\documentclass[a4paper,12pt,authoryear,twoside]{elsarticle}

\pdfoutput=1

\usepackage{latexsym}
\usepackage{amssymb}
\usepackage{amsmath}
\usepackage[varg]{txfonts}
\usepackage{mathrsfs}
\usepackage{upgreek}
\usepackage[small]{caption}
\usepackage[]{subfig}
\usepackage{verbatim}
\usepackage{array}
\usepackage{color}
\usepackage{graphicx}
\DeclareMathAlphabet{\mathpzc}{OT1}{pzc}{m}{it}
\usepackage{hyperref}
\usepackage{multirow}

\begin{document}
\pagestyle{empty}

\begin{frontmatter}

\title{Single Cooper pair transfer in stable and in exotic nuclei}
\author[1,2,3]{G. Potel\corref{}}
\ead{gregory.potel@gmail.com}
\author[1,2]{A. Idini}
\ead{andrea.idini@mi.infn.it}
\author[4]{F. Barranco}
\ead{barranco@us.es}
\author[2]{E. Vigezzi}
\ead{enrico.vigezzi@mi.infn.it}
\author[1,2,5]{\mbox{R. A. Broglia}\corref{cor1}}
\ead{broglia@mi.infn.it}
\address[1]{Dipartimento di Fisica, Universit\`{a} di Milano,
Via Celoria 16, 20133 Milano, Italy.}
\address[2]{INFN, Sezione di Milano Via Celoria 16, 20133 Milano, Italy.}
\address[3]{Departamento de Fisica Atomica, Molecular y Nuclear, Universidad de Sevilla, Facultad de Fisica, Avenida Reina Mercedes s/n, Sevilla, Spain}
\address[4]{Departamento de Fisica Aplicada III, Universidad de Sevilla, Escuela Superior de Ingenieros,
Sevilla, 41092 Camino de los Descubrimientos s/n,
Spain.}
\address[5]{The Niels Bohr Institute, University of Copenhagen, Blegdamsvej 17,
2100 Copenhagen {\O}, Denmark.}

\cortext[cor1]{Corresponding author. Tel:(+39)3388959875}

\thispagestyle{empty}
\begin{abstract}
Arguably, one of the greatest achievement of many--body physics has been that of developing the tools for a complete description and a thorough understanding of superconductivity in metals. At the  basis of it one finds BCS theory and the Josephson effect. The first recognized the central role played by the appearance of a macroscopic coherent field usually viewed as a condensate of strongly overlapping Cooper pairs, the quasiparticle va\-cuum. The second made it clear that a true gap is not essential for such a state of matter to exist, but rather a finite expectation value of the pair field. Consequently, the specific probe to study the superconducting state is Cooper pair tunneling. Tunneling experiments not only gave a measurable physical reality to the phase of the pair wavefunction, the Josephson current being phase dependent. It also provided detailed information concerning the origin and strength of the pairing force. In fact, once it was understood that Cooper pair tunneling was mainly the result of the individual, successive transfer of each of the partners of the pair induced by the mean field potential, and a realistic tunneling Hamiltonian was worked out, it was possible to quantitatively probe the correlations giving rise to Cooper pair binding and condensation. As a consequence, tunneling experiments helped at eliminating any uncertainty concerning the electron--electron, electron--plasmon and electron--phonon interactions which are at the basis of the origin of pairing in metals. The resulting, unified description which treated at the same level of physical and computational accuracy both tunneling processes (reaction) and the electron--electron interaction screened and renormalized by medium polarization effects (structure), also meant the end of superconductivity as a wide open, speculative field and the beginning of a thoroughly quantitative ``exact'' era, with uncertainties below the 10\% level.

From this vantage point of view it is not difficult to argue that important progress in the understanding of pairing in atomic nuclei will arise from a systematic, quantitative study of two--particle transfer reactions on drip line, exotic, halo nuclei (like for example $^{11}$Li), stabilized by the pairing correlations associated with a single Cooper pair, as well as on many (but still few)--Cooper pair systems like e.g. the Sn--isotopes, setting equal emphasis on the structure as well as on the reaction aspects of the process.
Time seems ripe for such a study, in keeping with the fact that one now knows how to correlate pairs of nucleons taking properly into account the interplay between bare and medium po\-la\-ri\-za\-tion (induced) nuclear pairing interactions. Also how to calculate the absolute value of the two--particle transfer cross sections taking properly into account the full non--locality of the Cooper pairs, as well as the multistep (successive, simultaneous and non--orthogonality) contributions to it. The above expectation is strongly supported by the results emerging from the analysis of a broad sample of two--nucleon transfer data.

In particular, from the analysis of recent data from $(p,t)$ reactions on $^{11}$Li and Sn--isotopes, carried out making use of a unified nuclear field theoretical description of structure and reaction mechanisms, which testify to the fact that theory is now able to provide an overall account of the experimental findings, in particular of the absolute two--particle transfer cross section, within experimental errors and without adjusting any free parameter. This is also true when the multi--step theory of two--particle transfer together with detailed, microscopic, nuclear structure wavefunctions is applied to $(t,p)$ and $({}^{16}$O,${}^{18}$O) data associated with the \mbox{${}^{206}$Pb$(gs)$ $\rightleftarrows {}^{208}$Pb$(gs)$} processes.
\end{abstract}


\end{frontmatter}

\clearpage
\tableofcontents

\clearpage

\section{Introduction}
Few years ago physicists celebrated all around the world the 50th anniversary of BCS theory (\cite{Bardeen:57a}, \cite{Bardeen:57b}, see also \cite{Cooper:11}). The success of such a theory is not so much, or better not only, the fact that it gave  the definitive solution to one of the most spectacular phenomenon of all of physics --permanent (super)currents estimated to be stable for $10^{10^{10}}$ y-- , but that it provided a paradigm for the phenomenon of spontaneous symmetry breaking and associated emergent properties (\cite{Anderson:94}). This paradigm demonstrated to be successful in a variety of fields starting from solid state physics and extending to nuclear and particle physics, field theory and astrophysics.

Bohr, Mottelson and Pines developed, in the summer of 1957 (\cite{Bohr:58}), the basis of the theory of nuclear superfluidity which was eventually applied to the description of the nuclear structure, in particular to the calculation of the moment of inertia of deformed nuclei (\cite{Belyaev:59}) and of quadrupole vibrations of spherical and of deformed nuclei (\cite{Bayman:60a}, \cite{Bayman:60b}, \cite{Kisslinger:63}, \cite{Bes:63}, \cite{Soloviev:65}, \cite{Bes:69}; see also \cite{Ring:80}).

In subsequent years the consequences of the phenomenon of spontaneous breaking of gauge symmetry in nuclei was investigated. In keeping with the fact that generalized rigidity in gauge space constitutes the most basic of the associated emergent properties, two-particle transfer reactions are the specific probe to study the individual members of the associated pairing rotational and vibrational  bands (\cite{Bohr:64}, \cite{Bes:66}, \cite{Bohr:75}, \cite{Brink:05}), a feat which is totally out of reach in the case of tunneling between two superconductors in condensed matter (\cite{Josephson:62}).

Due to the fact that the number of Cooper pairs participating in the nuclear condensate is small, 5--10 in the case of typical superfluid nuclei like the Sn--isotopes, one can study the phenomenon in terms of specific orbitals, some of which play an essential role in the transfer process (hot orbitals). For this purpose, \textbf{absolute} differential cross sections $d \sigma ((A+a) \longrightarrow ((A+2)+(a-2)))/d \Omega$ must be measured as well as calculated (\cite{Yoshida:62}, \cite{Ascuitto:69}, \cite{Ascuitto:71}, \cite{Glendenning:63}, \cite{Glendenning:65}, \cite{Bjerregaard:66}, \cite{Glendenning:68}, \cite{Bayman:71}, \cite{Broglia:72a}, \cite{Broglia:73}, \cite{Charlton:76}, \cite{Hashimoto:78}, \cite{Takemasa:79}, \cite{Bayman:82}, \cite{Maglione:85}, \cite{Igarashi:91}, \cite{Becha:97}, \cite{Tanihata:08}, \cite{Potel:10}).

Much has been learned concerning pairing correlations in finite many--body systems by studying metallic grains at low temperatures (c.f. \cite{Anderson:59}, \cite{Perenboom:81}, \cite{Lauritzen:93}, \cite{Farine:99}, \cite{Farine:02}), atomic clusters (\cite{Palstra:95}, \cite{Gunnarsson:97}, \cite{Gunnarsson:04}, \cite{Broglia:04b}) as well as by using semiclassical approximations (see e.g. \cite{Bengtsson:80}, \cite{Kucharek:89}, \cite{Broglia:04a} and references therein). Also by studying the manifestation of Berry phase in rotating nuclei (\cite{Nikam:87a}; see also \cite{Broglia:86}), as well as of pairing fluctuations and phase--transitions as a function of spin in highly rotating nuclei (see e.g. \cite{Bengtsson:79}, \cite{Bernath:93},\cite{Shimizu:89} and refs. therein).
New perspectives in the study of pairing correlations among nucleons are being opened through the study of exotic nuclei lying along the drip lines in general, and of halo nuclei in particular. Insight into these systems through two-particle transfer reactions (\cite{Lenske:98}, \cite{Khan:04}, \cite{Matsuo:10}), and also with break--up (\cite{Barranco:93},\cite{Bertsch:98}, \cite{Aumann:05}) and high--energy knock--out reactions (\cite{Hansen:03}, \cite{Tostevin:04}, \cite{Tostevin:07}) is expected to shed light on the relative role the bare and the induced pairing interactions play in regions of very low density (see e.g. Fig. 2 of \cite{Richter:93},  and Fig. 3.21 of \cite{Heyde:98}, Broglia, private communication), but also on BEC of dilute Fermi gases (see e.g. \cite{Pethick:09} and references therein), where single--pair transfer experiments are not possible.

In fact, and in keeping with the insight provided by the results of tunneling experiments concerning pairing in metals(\cite{Josephson:62}, \cite{Anderson:69}, \cite{Josephson:69}, \cite{Scalapino:69},  \cite{McMillan:69}, \cite{Esaki:73}, \cite{Giaever:73}, \cite{Josephson:73}, \cite{Nambu:91}), much is expected to be learned concerning the effective, strongly renormalized interaction responsible for the presence of Cooper pairs in the nuclear medium, by measuring the two--particle differential cross sections and strength functions (energy dependence of the tunneling phenomenon).
Within this scenario, the two--particle transfer reactions carried out at Ganil (\cite{Keeley:07a}), \cite{Chatterjee:08}, TRIUMF (\cite{Tanihata:08},\cite{Ball:11}) and Dubna (\cite{Golovkov:08}) as well as those in the planning stage which eventually will be studied,
are expected to expand in an important way the frontiers of our knowledge con\-cer\-ning quasispin pair alignment (\cite{Anderson:58}, \cite{Bohr:89}) and dynamical pair correlations in nuclei.

The paper is organized as follows. Section 2 provides a qualitative discussion of two--nucleon transfer reactions, while section 3 presents the associated formalism needed to carry out detailed calculations. In section 4 a unified nuclear structure--reaction mechanism analysis of the experiment $^{11}$Li($^1$H,$^3$H)$^9$Li (\cite{Tanihata:08}) on the single--Cooper pair, exotic, halo nucleus $^{11}$Li is discussed in detail, setting equal emphasis on the structure as well as on the reaction aspects of the analysis. This analysis provides clear evidence for the role the exchange of collective vibrations between Cooper pair partners plays in stabilizing these building blocks of nuclear superfluidity.
Section 5 deals with a similar analysis carried out in connection with the reaction $^{122}$Sn($p,t$)$^{120}$Sn (\cite{Guazzoni:99}) while in Section 6, a $v_{low-k}$ shell model analysis of the $^{112}$Sn($p,t$)$^{110}$Sn (\cite{Guazzoni:06}) is reported. In other words, Sections 5 and 6 present a detailed analysis of the insight two--nucleon transfer reaction can provide in the study of pairing rotational bands.
Section 7 reports on the microscopic calculation of the absolute two--neutron transfer reaction cross section associated with  $^{206}$Pb($t,p$)$^{208}$Pb (gs), thus providing an example of a quantitative analysis of the excitation of a member of a pairing vibrational band (\cite{Bes:66}). Making use of the same nuclear structure input, the heavy ion reaction $^{208}$Pb($^{16}$O,$^{18}$O)$^{206}$Pb (\cite{Bayman:82}) is discussed in Section 8.
The conclusions, which are collected in Section 9 are very simple to state: the present work marks, arguably, the beginning of a completely new era in the study of nuclear structure with two--nucleon transfer reactions, in which the previous qualitative and semi--quantiative arguments are now placed on a sound basis.

\section{Two--nucleon transfer reactions}
\begin{figure}
\begin{center}
\includegraphics*[width=10cm,angle=0]{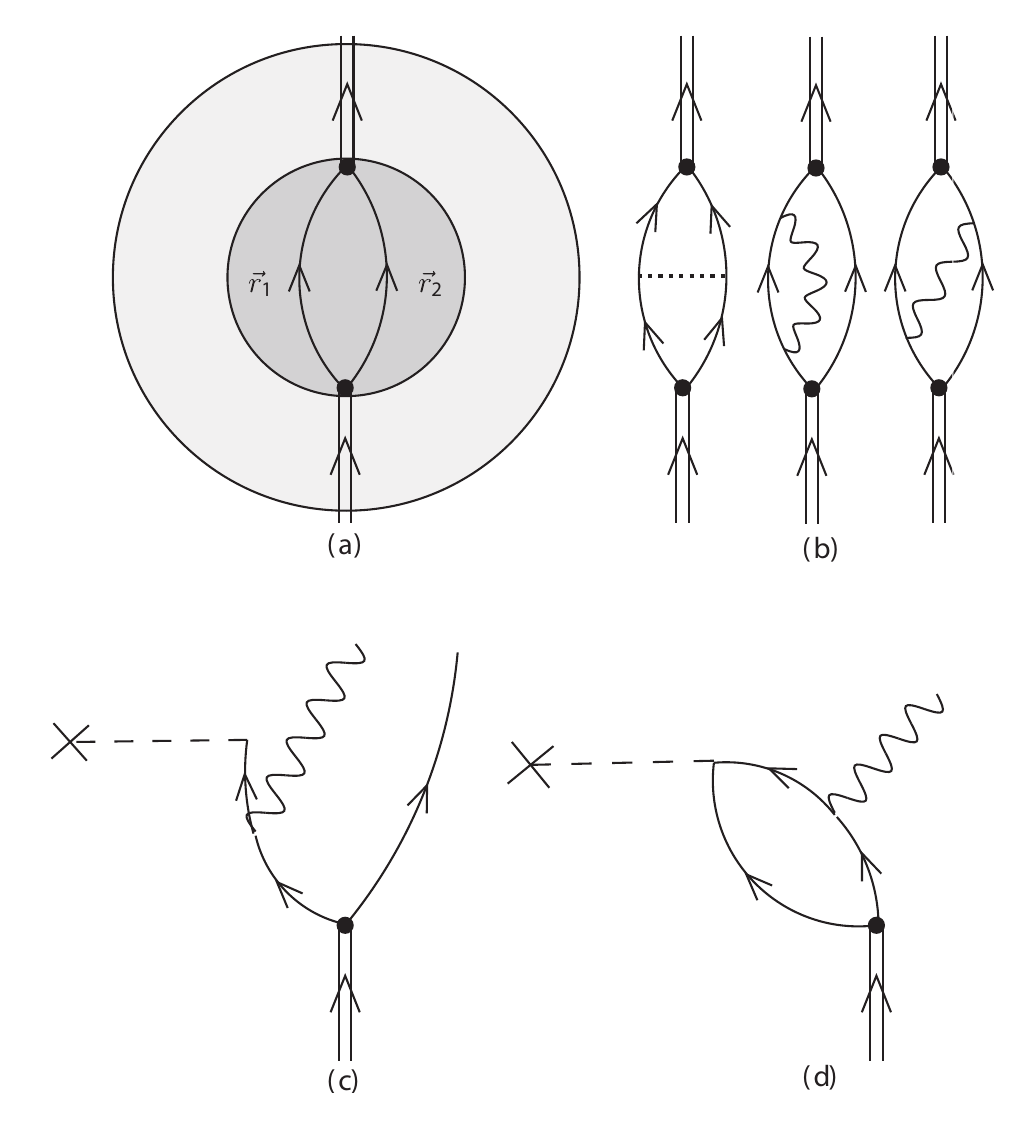}
\end{center}
\caption[Schematic representation of a two--particle correlated state]{(a) Schematic representation of a two--particle correlated state (pair vibration: pair addition mode). This object (Cooper pair; arrowed double line) is highly non--local, the fermions participating in it, and represented by a single arrowed line feeling the presence of the partner within distances inversely proportional to their correlation energy (light grey area) a fact which, in the case of halo exotic nuclei, leads to an increase of the effective nuclear radius of the systems by about 50\% as compared with the systematics. This may not sound much, if it were not for the fact that nuclear matter is highly incompressible and that small changes in the nuclear radius may imply nuclear instability. The mean single--particle field (dark grey area) can be viewed as an external field confining each of the member of Cooper pairs individually whose partners are correlated over distances considerably larger than nuclear dimensions, (b) the bare NN interaction (dotted horizontal line) correlating pairing vibrations can be renormalized through the interweaving of the nuclear pair with collective density, spin and isospin vibrations (wavy lines) making use of the particle--vibration coupling mechanism in terms of self--energy and vertex corrections, (c) direct one--particle pick--up can excite a 2$p$--1$h$ like state while (d) two--particle pick--up may lead to a collective 1$p$--1$h$ excited final state.}
\label{fig10}
\end{figure}

At the origin of the BCS many--body theory of superconductivity one finds Cooper's solution of a system composed of two--electrons lying on top of the Fermi surface and interacting through an attractive force, all the other electrons playing a role only through the exclusion principle (\cite{Cooper:56}). The ground states of $^{210}$Pb and of $^{11}$Li, in which two neutrons move around the cores  $^{208}$Pb and  $^{9}$Li respectively, provide nuclear embodiments of Cooper's model (see Fig. \ref{fig10}).

\subsection{The qualitative picture}
\begin{figure}[]
\begin{center}
\includegraphics*[width=12.5cm,angle=0]{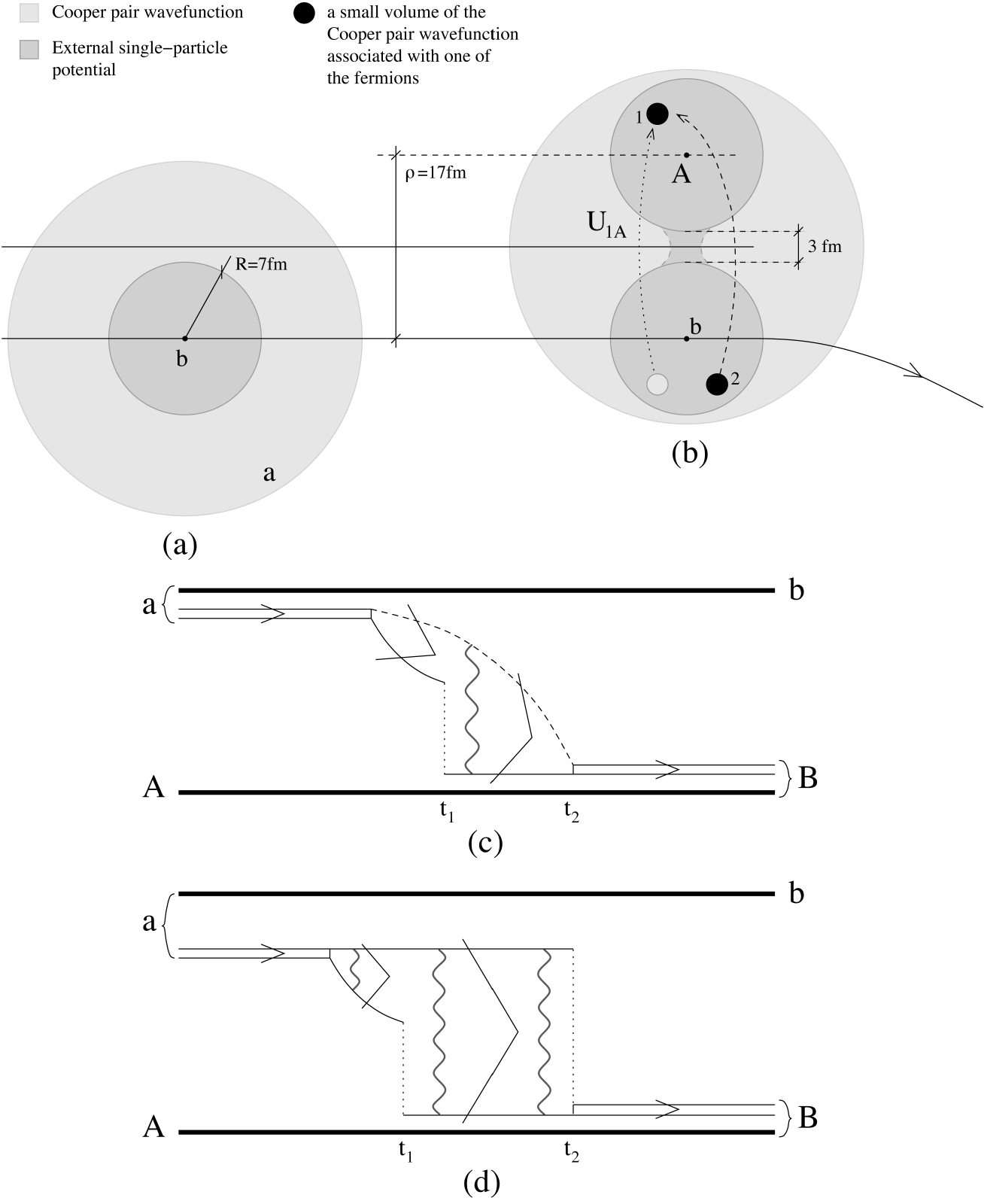}
\end{center}
\caption[Schematic representation of the transfer of two correlated nucleons]{\footnotesize{Schematic representation of the transfer of two correlated nucleons in first--order (c) and in second--order perturbation theory (d). In the first case a nucleon is transferred through the single--particle potential $U_{1A}$ acting at time $t_1$ (dotted vertical line). The second nucleon follows suit (dashed curve) through the non--orthogonality of the wavefunctions belonging to different nuclei and, at time $t_2$, it can be considered as a nucleon belonging to the nucleus $B$. In the second case, one nucleon is acted upon at $t=t_1$ by the single--particle field generated by the core $A$ ($U_{1A}$, dotted vertical line), and upon the second one at $t=t_2$. In all cases the correlation between the two nucleons is maintained throughout in keeping with the fact that $\xi\approx 30$ fm. The wavy lines in both (c) and (d) indicate the correlations existing between the members of the Cooper pair, entity which is represented by two--particle arrowed lines (pair addition mode). The light grey area in (a) indicates the (virtual) extension of the Cooper pair wavefunction, the darker one the actual extension of the single--particle wavefunctions in the external single--particle field $U_{1b}$. When the ion $a$ comes to the closest distance of approach from the target ion $A$ (see (b)), the Cooper pair wavefunction turns from virtual to real extending now over essentially a length equal to $\xi$. Transfer will receive (important) contributions over distances of $\xi$ , as schematically shown with the help of a small vo\-lume of the Cooper pair wavefunction (solid circle). Of course, the transfer integral $\langle\varphi_{Cooper}^{(A)}|U_{1A}|\varphi_{Cooper}^{(b)}\rangle$ receives contributions also from small distances, namely from all the dark grey area.}}\label{fig11}
\end{figure}

Much is known concerning two--nucleon pairing correlations of $|^{210}\text{Pb (gs)} \rangle$, a system which has been studied in detail also in terms of two--nucleon transfer reactions (see e.g. \cite{Broglia:73} and refs. therein). In this pair addition mode the two nucleons are correlated over a distance $\xi = \hbar \mathpzc{v}_F/2 E_{corr}$, where $E_{corr}$ plays the role of the pairing gap for open shell, superfluid, nuclei. In the case of  $^{210}$Pb, $E_{corr}\approx 1.2$ MeV. Thus $\xi = 25$ fm. Of course, if the two nucleons are subject to an external field (the central potential generated by e.g. the $^{208}$Pb core), they cannot move away from each other more than 14 fm (see Fig. \ref{fig10}), in keeping with the fact that the radius of $^{208}$Pb is $\approx$ 7 fm. On the other hand, in a heavy ion reaction with e.g. impact parameter 17 fm (see Fig. \ref{fig11} (a) and (b)), the central single--particle potential acting on one of the two nucleons to be transferred is much stronger than typical values of the pairing field. It will thus be this potential responsible for the transfer of one partner of the Cooper pair at time $t_1$ and of the second one at time $t_2$ (see Fig. \ref{fig11} (d)). And this two--step process will take place without loss of (pairing) correlation between the two nucleons. In other words, the Cooper pair is equally well formed at $t<t_1$ and $t>t_2$ (where the relative distance between the two neutrons is always less than 15 fm), than at $t_1<t<t_2$ where this distance can be much larger $(\approx 24)$ fm. A similar argument applies to the discussion of the origin of the non--orthogonality contribution (see Fig. \ref{fig11} (c)).

\section{The calculation of absolute cross sections}
In connection with the microscopic description of the tunneling process associated with the Josephson effect, (\cite{Cohen:62}) one can use the Hamiltonian
\begin{equation}\label{eq1_20}
H=H_1+H_2+\sum_{k,q} T_{kq}(a_{k\uparrow}^{\dagger} a_{q\uparrow}^{}+ a_{-q\downarrow}^\dagger a_{-k\downarrow}^{})+h.c. ,
\end{equation}
where  $H_1$ and $H_2$ are the separate Hamiltonians of the two superconductors on each side of the barrier; $T_{kq}$ is the exponentially small tunneling matrix element from state $k$ on one side to state $q$ on the other, and the relationship of phases shown is required by time reversal symmetry.
One of the many procedures for arriving at (\ref{eq1_20}) is to find sets of single--particle functions for each side of the barrier separately, in the absence of the potential of the other metal: then one eliminates the nonorthogonality effect by perturbation theory.
Let us now see how the above considerations translate in the case of the transfer of pairs of nucleons in a nuclear reaction, both in the case of a heavy ion and of light ion reactions.

\subsection{Semiclassical calculations (heavy ions)}
A very accurate and physical intuitive picture of heavy ion reactions, in particular of two--particle transfer reactions,
\begin{equation}\label{eq1_21}
    a(=b+2)+A \longrightarrow b+B(=A+2),
\end{equation}
described by the total hamiltonian
\begin{equation}\label{eq1_22}
\begin{split}
 H& = T_{aA}+H_a+H_A+V_{aA},\\
    & = T_{bB}+H_b+H_B+V_{bB},
\end{split}
\end{equation}
is obtained by assuming that the relative motion of the centers of mass of the two ions can be described, in both entrance $\alpha(\equiv(a,A))$ and exit $\beta(\equiv(b,B))$ channels, classically (\cite{Broglia:04a}).

In (\ref{eq1_22}) $T_{aA}$ is the kinetic energy of relative motion, $H_a$ and $H_A$ the Hamiltonians describing the intrinsic degrees of freedom of nuclei $a$ and $A$ respectively, while $V_{aA}$ is the effective interaction between the nucleons in $a$ and the nucleons in $A$. Similar notation has been used to describe the corresponding ions in the exit channel.

All the information concerning the process (\ref{eq1_21}) is obtained by solving the time--dependent Schr\"{o}dinger equation
\begin{equation}\label{eq1_23}
 i\hbar \frac{\partial \psi}{\partial t}= H \psi,
\end{equation}
with the initial condition that the nuclei $a$ and $A$ are in their ground states, and that the relative motion is described by a narrow wavepacket $\chi(\vec r_\beta-\vec R_\beta(t))$ of rather well--defined impact parameter and velocity. Let us expand $\psi$ on the channel wavefunctions $\psi_\beta=\psi_m^b(\xi_b)\psi_n^B(\xi_B)\exp(i\delta_\beta)$, according to
\begin{equation}\label{eq1_24}
    \begin{split}
       \psi& =\sum_\beta c((r_\beta-R_\beta),t) \psi_\beta(t)e^{-iE_\beta t/\hbar},\\
         & =\sum_\beta a_\beta(t) \chi(\vec r_\beta-\vec R_\beta(t),t) \psi_\beta(t)e^{-iE_\beta t/\hbar}.
     \end{split}
\end{equation}
Upon inserting (\ref{eq1_24}) in (\ref{eq1_23}), one obtains the coupled equations
\begin{equation}\label{eq1_25}
i\hbar \sum_\beta \dot {a}_\beta(t) \langle \psi_\xi|\psi_\beta\rangle_{\vec R_{\xi\gamma}}e^{-iE_\beta t / \hbar}= \sum_\gamma \langle \psi_\xi|V_\gamma-U_\gamma(r_\gamma)|\psi_\gamma\rangle _{\vec R_{\xi \gamma}}a_\gamma(t)e^{-iE_\beta t / \hbar},
\end{equation}
which, together with the condition $a_\gamma(-\infty)=\delta(\gamma,\alpha)$, allows to calculate $a_\beta$ and thus the transfer reaction cross section, proportional to $|a_\beta|^2$. The sub--index on the matrix elements indicate that in the integration over the degrees of freedom of the two nuclei, the average center--of--mass coordinate $\vec r_{\beta \gamma}=\tfrac{1}{2}(\vec r_\beta+\vec r_\gamma)$ should be identified with the average classical coordinate, that is,
\begin{equation}\label{eq1_26}
\vec r_{\beta \gamma}\rightarrow \vec R_{\beta \gamma} =\tfrac{1}{2}(\vec R_\beta+\vec R_\gamma).
\end{equation}
Of notice that
\begin{equation}\label{eq1_27}
\langle \psi_\gamma|V_\gamma-U_\gamma|\psi_\gamma\rangle=0,
\end{equation}
defines the ion--ion potential $U_\gamma$ to be the expectation value of the interaction $V_\gamma$ in the $\gamma$--channel.

A characteristic feature of the coupled equations (\ref{eq1_25}) is the presence of the overlap $\langle \psi_\xi|\psi_\beta\rangle$ on the left hand side. If $\xi$ and $\beta$ describe two channels of the same partition, e.g. $\xi=\beta'$, the prime indicating excited states of the nuclei $b$ and $B$, the overlap matrix is diagonal, namely,
\begin{equation}\label{eq1_28}
g(\vec R)=\langle \psi_\beta'|\psi_\beta\rangle_{\vec R}=\delta(\beta',\beta).
\end{equation}
If $\xi$ and $\beta$ describe different partitions, the overlap $\langle \psi_\xi|\psi_\beta\rangle$ is different from zero in the region where the densities of the two nuclei overlap (non--orthogonality).

The coupled equations (\ref{eq1_25}) can be written in a more compact way by introducing the adjoint channel wavefunction
\begin{equation}\label{eq1_29}
\omega_\xi=\sum_\gamma g^{-1}_{\xi \gamma} \psi_\gamma,
\end{equation}
fulfilling the orthogonality relation
\begin{equation}\label{eq1_30}
(\omega_\xi,\psi_\beta)=\delta(\xi,\beta).
\end{equation}
Solving (\ref{eq1_25}) in second order perturbation theory one obtains
\begin{equation}\label{eq1_31}
a_\beta(t)=(a_\beta(t))_{(0)}+(a_\beta(t))_{(1)}+(a_\beta(t))_{(2)},
\end{equation}
where
\begin{equation}\label{eq1_32}
(a_\beta(t))_{(0)}=\delta(\beta,\alpha),
\end{equation}
\begin{equation}\label{eq1_33}
(a_\beta(t))_{(1)}=\frac{1}{i \hbar}\int_{-\infty}^{t} \langle \omega_\beta|V_\alpha-U_\alpha|\psi_\alpha\rangle_{\vec R_{\alpha\beta}} e^{i(E_\beta-E_\alpha)t'/\hbar}dt',
\end{equation}
and
\begin{equation}\label{eq1_34}
\begin{split}
  (a_\beta(t))_{(2)} & =\left(\frac{1}{i \hbar}\right)^2 \sum_\gamma \int_{-\infty}^{t} \langle \omega_\beta|V_\gamma-U_\gamma|\psi_\gamma\rangle_{\vec R_{\beta\gamma}(t')} e^{i(E_\beta-E_\gamma)t'/\hbar}dt' \\
    & \times\int_{-\infty}^{t'} \langle \omega_\gamma|V_\alpha-U_\alpha|\psi_\alpha\rangle_{\vec R_{\gamma\alpha}(t'')} e^{i(E_\gamma-E_\alpha)t''/\hbar}dt''.
\end{split}
\end{equation}
The state vectors $|\omega\rangle$ have to include the non--orthogonality effects between channels $\beta, \gamma$ and $\alpha$. To second order one finds
\begin{equation}\label{eq1_35}
\begin{split}
  \omega_\beta =&  \psi_\beta-\sum_{\gamma\neq\beta}\langle \psi_\gamma|\psi_\beta\rangle_{\vec R_{\gamma\beta}}\psi_\gamma\\
    & +\sum_{\gamma\neq\beta,\alpha}\langle \psi_\alpha|\psi_\gamma\rangle_{\vec R_{\alpha\gamma}}
    \langle \psi_\gamma|\psi_\beta\rangle_{\vec R_{\gamma\beta}}\psi_\alpha.
\end{split}
\end{equation}
Out of the very many second--order processes, we shall here only discuss the second--order effects in the two nucleon transfer reaction $A+a (=b+2 \,\text{nucleons}) \rightarrow B(=A+2 \,\text{nucleons})+b$, where the intermediate channel $\gamma$ corresponds to the one--nucleon transfer channel $F (=A+1 \,\text{nucleons})+f(=b+1 \,\text{nucleons}).$

Inserting $\omega_\beta$ in (\ref{eq1_33}) and in (\ref{eq1_34}) one finds that the two--nucleon transfer amplitudes can be written as
\begin{equation}\label{eq1_36}
a(\infty)=(a_\beta)_{(1)}+(a_\beta)_{orth}+(a_\beta)_{succ}
\end{equation}
up to second order of perturbation theory. The associated differential cross section is obtained by multiplying the elastic cross section calculated in terms of the deflection function by $|a(\infty)|^2$.

 The different quantities appearing in (\ref{eq1_36}) are
\begin{equation}\label{eq1_37}
(a_\beta)_{(1)}=\frac{1}{i\hbar}\int_{-\infty}^{\infty} dt\langle \psi_\beta|V_\alpha-U_\alpha|\psi_\alpha\rangle_{\vec R_{\beta\alpha}} e^{i(E_\beta-E_\alpha)t/\hbar},
\end{equation}
\begin{equation}\label{eq1_38}
\begin{split}
(a_\beta)_{succ}=& \left(\frac{1}{i\hbar}\right)^2 \sum_{\gamma \neq (\beta,\alpha)} \int_{-\infty}^{\infty} dt\langle \psi_\beta|V_\gamma-U_\gamma|\psi_\gamma\rangle_{\vec R_{\beta\gamma}(t)} e^{i(E_\beta-E_\gamma)t/\hbar}\\
& \times \int_{-\infty}^{t} dt'\langle \psi_\gamma|V_\alpha-U_\alpha|\psi_\alpha\rangle_{\vec R_{\gamma\alpha}(t')} e^{i(E_\gamma-E_\alpha)t'/\hbar},
\end{split}
\end{equation}
and
\begin{equation}\label{eq1_39}
\begin{split}
(a_\beta)_{orth}=& -\frac{1}{i\hbar} \sum_{\gamma \neq (\beta,\alpha)} \int_{-\infty}^{\infty} dt\langle \psi_\beta|\psi_\gamma\rangle_{\vec R_{\beta\gamma}(t)}\\
& \times \langle \psi_\gamma|V_\alpha-U_\alpha|\psi_\alpha\rangle_{\vec R_{\gamma\alpha}} e^{i(E_\beta-E_\alpha)t/\hbar}.
\end{split}
\end{equation}
The first term $(a_\beta)_{(1)}$ describes the simultaneous transfer of two nucleons. The two--step successive transfer is described by $(a_\beta)_{succ}$, while $(a_\beta)_{orth}$ is a second--order contribution arising from the non--orthogonality of the wavefunctions associated with the different channels considered.

It is of notice that the main contribution to (\ref{eq1_37}) arises from the single particle potential in channel $\alpha$ and the overlap between the single--particle wave functions of a nucleon in $a(=b+2)$ and in $B(=A+2)$. The contribution of the two--body pairing interaction, also present in (\ref{eq1_37}) leads to a very small contribution.

In keeping with the above discussion, in the independent particle model, that is in the case in which there are no correlation between nucleons \mbox{$\sum_{\gamma} |\psi_\gamma\rangle\langle\psi_\gamma|=1$},  (\ref{eq1_39}) cancels exactly (\ref{eq1_37}). The transfer reaction is then described as a purely successive transfer by the amplitude (\ref{eq1_38}), as expected. In the opposite limit of very strong correlations between nucleons one expects that two--particle transfer occurs essentially as a simultaneous transfer process. This can be seen rewriting (\ref{eq1_37})--(\ref{eq1_39}) in the post--prior representation. In this representation, at variance from the prior--prior representation of (\ref{eq1_38}), the non--orthogonality term gets absorbed in the successive term which now reads

\begin{equation}\label{eq1_54}
\begin{split}
(\bar a_{\beta})_{(2)}=& \left(\frac{1}{i\hbar}\right)^2 \sum_{\gamma} \int_{-\infty}^{\infty} dt\langle \psi_\beta|V_\beta-U_\beta|\psi_\gamma\rangle_{\vec R_{\beta\gamma}} e^{i(E_\beta-E_\gamma)t/\hbar}\\
& \times \int_{-\infty}^{t} dt' \langle \psi_\gamma|V_\alpha-U_\alpha|\psi_\alpha\rangle_{\vec R_{\gamma\alpha}} e^{i(E_\gamma-E_\alpha)t'/\hbar}.
\end{split}
\end{equation}
The expression (\ref{eq1_37}) remains identical because of the post--prior symmetry. In the case in which the two particles to be transferred have a very strong mutual in\-te\-rac\-tion $V_{12}$, $(\bar a_{\beta})_{(2)}$ becomes very small. This is because in the intermediate state one has to break a pair, an event which becomes less and less likely as $V_{12}$ increases. Because in (\ref{eq1_54}) the transfer potential does not contain $V_{12}$, the effect mentioned above implies that $(\bar a_\beta)_{(2)}\rightarrow 0$ as $V_{12}\rightarrow\infty$. Now, in actual situations $V_{12}\ll\langle V \rangle$, the first picture (i.e. $a_{(1)}\approx-(a)_{orth}$) applies, and two--particle transfer process can be essentially viewed as a successive transfer process. In terms of the transfer amplitude (\ref{eq1_36}), the differential cross section can be written as
\begin{equation}\label{eq1_58}
\frac{d\sigma}{d\Omega}=|f(\theta)|^2,
\end{equation}
the scattering amplitude being
\begin{equation}\label{eq1_57}
f(\theta)=\frac{1}{2k}\sum_l(2l+1)\exp[2i(\sigma_l+\delta_l)]a_{l}(\infty)P_l(\cos \theta),
\end{equation}
where $\sigma_l$ and $\delta_l$ are the Coulomb and nuclear elastic phase shifts.

\subsection{Quantal calculations (heavy and light ion reactions)}
Similar calculations can be carried still for the case of heavy ion reactions, fully quantum mechanically. In the case of light ions this requirement, for energies of few MeV per nucleon, is a must.
In what follows we shall exemplify the workings of the closely interweaved structure--reaction formalism presented above to probe, through two particle transfer reactions, pairing correlations in atomic nuclei. We shall discuss two examples in detail, namely the $^{11}$Li($p,t$)$^9$Li and the $^{122}$Sn($p,t$)$^{120}$Sn($gs$) reaction processes, where a unified picture of both structure and reactions is available. We shall also comment on the reactions $^{112}$Sn($p,t$)$^{110}$Sn($gs$), $^{206}$Pb($t,p$)$^{208}$Pb($gs$) and $^{208}$Pb+$^{16}$O$\rightarrow { }^{206}$Pb+$^{18}$O.

In second order DWBA we need to calculate the simultaneous ($T^{(1)}$),  successive ($T^{(2)}_{succ}$) and non--orthogonal $(T^{(2)}_{NO})$ contributions to the transition amplitude between the initial $({j_i})^2$ and final $({j_f})^2$ single--particle states (\cite{Bayman:82}), namely,
\begin{subequations}
\begin{align}\label{eq1_40}
T^{(1)}(j_i,j_f)&=2\sum_{\sigma_1 \sigma_2}\int d\mathbf{r}_{fF}d\mathbf{r}_{b1}d\mathbf{r}_{A2}[\Psi^{j_f}(\mathbf{r}_{A1},\sigma_1)\Psi^{j_f}(\mathbf{r}_{A2},\sigma_2)]^{0*}_0 \chi^{(-)*}_{bB}(\mathbf{r}_{bB})\\
\nonumber & \times v(\mathbf{r}_{b1})[\Psi^{j_i}(\mathbf{r}_{b1},\sigma_1)\Psi^{j_i}(\mathbf{r}_{b2},\sigma_2)]^{0}_0 \chi^{(+)}_{aA}(\mathbf{r}_{aA}),
\end{align}
\begin{align}\label{eq1_41}
T^{(2)}_{succ}(j_i,j_f)&=2\sum_{K,M}\sum_{\substack{\sigma_1 \sigma_2\\\sigma'_1 \sigma'_2}}
\int d\mathbf{r}_{fF}d\mathbf{r}_{b1}d\mathbf{r}_{A2}[\Psi^{j_f}(\mathbf{r}_{A1},\sigma_1)\Psi^{j_f}(\mathbf{r}_{A2},\sigma_2)]^{0*}_0 \\ \nonumber &\times \chi^{(-)*}_{bB}(\mathbf{r}_{bB})
 v(\mathbf{r}_{b1})[\Psi^{j_f}(\mathbf{r}_{A2},\sigma_2)\Psi^{j_i}(\mathbf{r}_{b1},\sigma_1)]^{K}_M \\
\nonumber & \times \int d\mathbf{r}'_{fF}d\mathbf{r}'_{b1}d\mathbf{r}'_{A2}G(\mathbf{r}_{fF},\mathbf{r}'_{fF})
 [\Psi^{j_f}(\mathbf{r}'_{A2},\sigma'_2)\Psi^{j_i}(\mathbf{r}'_{b1},\sigma'_1)]^{K}_M\\
\nonumber  &\times \frac{2\mu_{fF}}{\hbar^2}v(\mathbf{r}'_{f2})
[\Psi^{j_i}(\mathbf{r}'_{A2},\sigma'_2)\Psi^{j_i}(\mathbf{r}'_{b1},\sigma'_1)]^{0}_0 \chi^{(+)}_{aA}(\mathbf{r}'_{aA}),
\end{align}
\begin{align}\label{eq1_42}
T^{(2)}_{NO}(j_i,j_f)&=2\sum_{K,M}\sum_{\substack{\sigma_1 \sigma_2\\\sigma'_1 \sigma'_2}}
\int d\mathbf{r}_{fF}d\mathbf{r}_{b1}d\mathbf{r}_{A2}[\Psi^{j_f}(\mathbf{r}_{A1},\sigma_1)\Psi^{j_f}(\mathbf{r}_{A2},\sigma_2)]^{0*}_0 \\ \nonumber &\times \chi^{(-)*}_{bB}(\mathbf{r}_{bB})
 v(\mathbf{r}_{b1})[\Psi^{j_f}(\mathbf{r}_{A2},\sigma_2)\Psi^{j_i}(\mathbf{r}_{b1},\sigma_1)]^{K}_M \\
\nonumber & \times \int d\mathbf{r}'_{b1}d\mathbf{r}'_{A2}
 [\Psi^{j_f}(\mathbf{r}'_{A2},\sigma'_2)\Psi^{j_i}(\mathbf{r}'_{b1},\sigma'_1)]^{K}_M\\
\nonumber  &\times
[\Psi^{j_i}(\mathbf{r}'_{A2},\sigma'_2)\Psi^{j_i}(\mathbf{r}'_{b1},\sigma'_1)]^{0}_0 \chi^{(+)}_{aA}(\mathbf{r}'_{aA}).
\end{align}
\end{subequations}
In these expressions, the spatial and spin coordinates of the two transferred nucleons are explicitly referred to with the subscripts 1 and 2. The subscripts $A$ and $b$ indicate the core to which the position of each of the nucleons are referred to. The vectors $\mathbf{r}_{aA}$, $\mathbf{r}_{bB}$ and $\mathbf{r}_{fF}$ are the relative motion coordinates in the initial, final and intermediate channels respectively. The transition potential responsible for the transfer of the pair is, in the \emph{post} representation,
\begin{equation}\label{eq1_43}
    V_\beta=v_{bB}-U_{\beta},
\end{equation}
where $v_{bB}$ is the interaction between the nuclei $B$ and $b$, and $U_{\beta}$ is the optical potential in the final channel. We make the assumption that $v_{bB}$ can be decomposed into a term containing the interaction between the cores $A$ and $b$ and the potential describing the interaction between $b$ and each of the transferred nucleons, namely
\begin{equation}\label{eq1_44}
    v_{bB}=v_{bA}+v_{b1}+v_{b2},
\end{equation}
where $v_{b1}$ and $v_{b2}$ is the same mean field potential we have used to define the single--particle wavefunctions of the neutrons in the nucleus $a$. The transition potential is
\begin{equation}\label{eq1_45}
    V_\beta=v_{bA}+v_{b1}+v_{b2}-U_{\beta}.
\end{equation}

Assuming that $\langle \beta |v_{bA}|\alpha \rangle \simeq \langle \beta |U_{\beta}|\alpha \rangle $ (i.e, assuming that the matrix element of the core--core interaction between the initial and final states is very similar to the matrix element of the real part of the optical potential), one obtains the final expression of the transfer potential in the \emph{post} representation,
\begin{equation}\label{eq1_45x}
    V_\beta\simeq v_{b1}+v_{b2}.
\end{equation}
This last approximation seems reasonable when dealing with heavy ion reactions in which there is no charge transfer, but more care has to be exerted when dealing with  reactions in which light ions are involved.

To calculate the total pair transfer amplitude, a sum of the contributions associated with each mean field contribution, labeled by the quantum numbers ($j_i,j_f$) and weighted with the correspondent two--nucleon spectroscopic amplitude $B_j$, is to be carried out leading to
\begin{equation}\label{eq1_55}
    T_ {2NT}=\sum_{j_fj_i}B_{j_f}B_{j_i}\left(  T^{(1)}(j_i,j_f)+T^{(2)}_{succ}(j_i,j_f)-T^{(2)}_{NO}(j_i,j_f)\right).
\end{equation}
The quantity $B_j\equiv B(j=0;j,j)$ is a special realization of the two--nucleon spectroscopy amplitude

\begin{equation}\label{eq61}
B(J;j_1,j_2)=\sum_{M,M_i} \langle J_i \; M_i \; J M|J_f \; M_f\rangle \langle \Psi_{J_f M_f}|P^\dagger(j_1,j_2;J M)|\Psi_{J_i M_i}\rangle,
\end{equation}
where
\begin{equation}\label{eq54}
P^\dagger(j_1,j_2;J M)=N\sum_m \langle j_1 \; m \;j_2 \;M-m|J \; M\rangle \;a^\dagger_{j_1m}a^\dagger_{j_2M-m} ,
\end{equation}
is the (renormalized) pair creation operator.
In other words,$B(J;j_1,j_2)$ is the amplitude of finding in the $|A+2;J_f,M_f>$ nuclear state, two nucleons moving in the single--particle orbitals $j_1$ and $j_2$ and coupled to angular momentum $J$, on top of the state $|A;J_i,M_i>$, coupled to total angular momentum $(J,J_i)J$.
Of notice that in Eq. (\ref{eq1_55}) the nuclear structure information which is essentially all contained in the amplitudes $B_{ij}$, is closely interweaved with the reaction amplitudes. This is the reason why the absolute value of two--nucleon transfer cross sections can display large enhancements as compared to pure configuration cross sections, thus revealing the coherence of (Cooper) pair correlations resulting from the pairing interaction. Eq. (\ref{eq1_55}) also testifies to the fact that quantitatively accurate description of pair transfer requires to treat on par both structure and reaction aspects of the process. Within this scenario Eq. (\ref{eq1_55}) provides another circumstantial evidence strongly supporting the fact that structure and reactions are but two aspects of the same many--body physics.

The differential cross section associated with the two--particle transfer amplitudes discussed above can be written as
\begin{equation}\label{eq1_56}
    \frac{d\sigma}{d\Omega}=\frac{\mu_i\mu_f}{(4\pi\hbar^2)^2}\frac{k_f}{k_i}|T_ {2NT}|^2,
\end{equation}
where $\mu_i,\mu_f$ are the reduced masses in entrance and exit channels respectively, while $k_f,k_i$ are the corresponding relative momenta.

\section{The ${}^{11}\textrm{Li}\left( {}^{1}\textrm{H}, {}^{3}\textrm{H} \right) {}^{9}\textrm{Li}$ reaction: pairing in exotic halo light nuclei}

In halo nuclei, some of the constituent neutrons or protons venture beyond the
drop's surface and form a misty cloud or halo. Not surprisingly, these
extended nuclei behave very differently from ordinary (``normal'') nuclei
lying along the stability valley in the chart of nuclides. In
particular, they are larger than normal nuclei of the same mass number, and
they interact with them with larger cross sections as well.
In the case of $^{11}$Li, the best studied drip line exotic halo nucleus,
the last two neutrons are very weakly bound.
Consequently, these neutrons need very little energy to move away
from the nucleus. There they can remain in their ``stratospheric'' orbits,
spreading out and forming a tenuous halo. If one neutron is taken away from
$^{11}$Li, a second neutron will come out immediately, leaving behind the
core of the system, the ordinary nucleus
$^{9}$Li. This result testifies to the fact that pairing, plays
 a central role in the stability of $^{11}$Li.

The basic experimental facts which characterize $^{11}$Li (\cite{Shulgina:09} and refs. therein) and which are also
of  particular relevance in connection with pairing in this system are: a)
$^9 _3$Li$_6$ and $^{11}_3$Li$_8$ are stable, $^{10}_3$Li$_7$ is not, b) the
two-neutron separation energy in $^{11}$Li is only $S_{2n}=378\pm5$; $369.15\pm0.65$
MeV, (\cite{Bachelet:08}; \cite{Smith:08}) as compared with values of 10 to 30 MeV in stable nuclei,
 c) $^{10}$Li displays s- and p-wave resonances at low energy, their centroids
lying within the energy range 0.1-0.25 MeV and 0.5-0.6 MeV
respectively (\cite{Zinser:95}) while these orbitals are well bound in nuclei
of the same mass lying along the stability valley,
 d) the mean square radius of $^{11}$Li, $\langle r^2 \rangle^{1/2}=3.55\pm 0.10$; $3.27\pm 0.24$; $3.12 \pm 0.06$ fm (\cite{Kobayashi:89}, \cite{Al-Khalili:96} , \cite{Hansen:96}, \cite{Shulgina:09}) is very large as compared to the value 2.32$\pm$0.02 fm
of the $^9$Li core, and testifies to the fact that the neutron halo must have
a large radius ($\approx$6-7 fm),
 e) the momentum distribution of the halo neutrons is found to be exceedingly
narrow, its FWHM being equal to $\sigma_{\bot}=48\pm 10$ MeV/c for the
 (perpendicular) distribution observed in the case of the break up of
$^{11}$Li on $^{12}$C, a value which is of the order of one fifth of that
measured
during the break up of normal nuclei (\cite{Kobayashi:93,Tanihata:96})
 f) the ground state of $^{11}$Li is a mixture of configurations where the two
 halo nucleons move around the $^9$Li core in $s^2-$ and $p^2-$configurations
with almost equal weight (\cite{Aoi:97,Simon:99}) while the wavefunctions of nuclei displaying two valence nucleons, although being strongly mixed are, as a rule, dominated by a single two-particle configuration.

Before discussing the sources of pairing correlations in $^{11}$Li, the single--particle resonant spectrum of $^{10}$Li has to be treated. Below we shall follow the Nuclear Field Theory (NFT) description of these subject (see e.g. \cite{Bes:76a},\cite{Mottelson:76},\cite{Bortignon:77}), mainly following \cite{Barranco:01}. The basis of (bare) single-particle states used was determined by calculating the eigenvalues  and eigenfunctions of a nucleon moving in the mean-field of the $^9$Li core, for which a Saxon-Woods potential parametrized following \cite{Bohr:69}, Vol I, Eqs. (2-181,2-182); \cite{Bortignon:98}, Eq.(3.48) was used.

\begin{figure}
\centerline{\includegraphics*[width=.63\textwidth,angle=0]{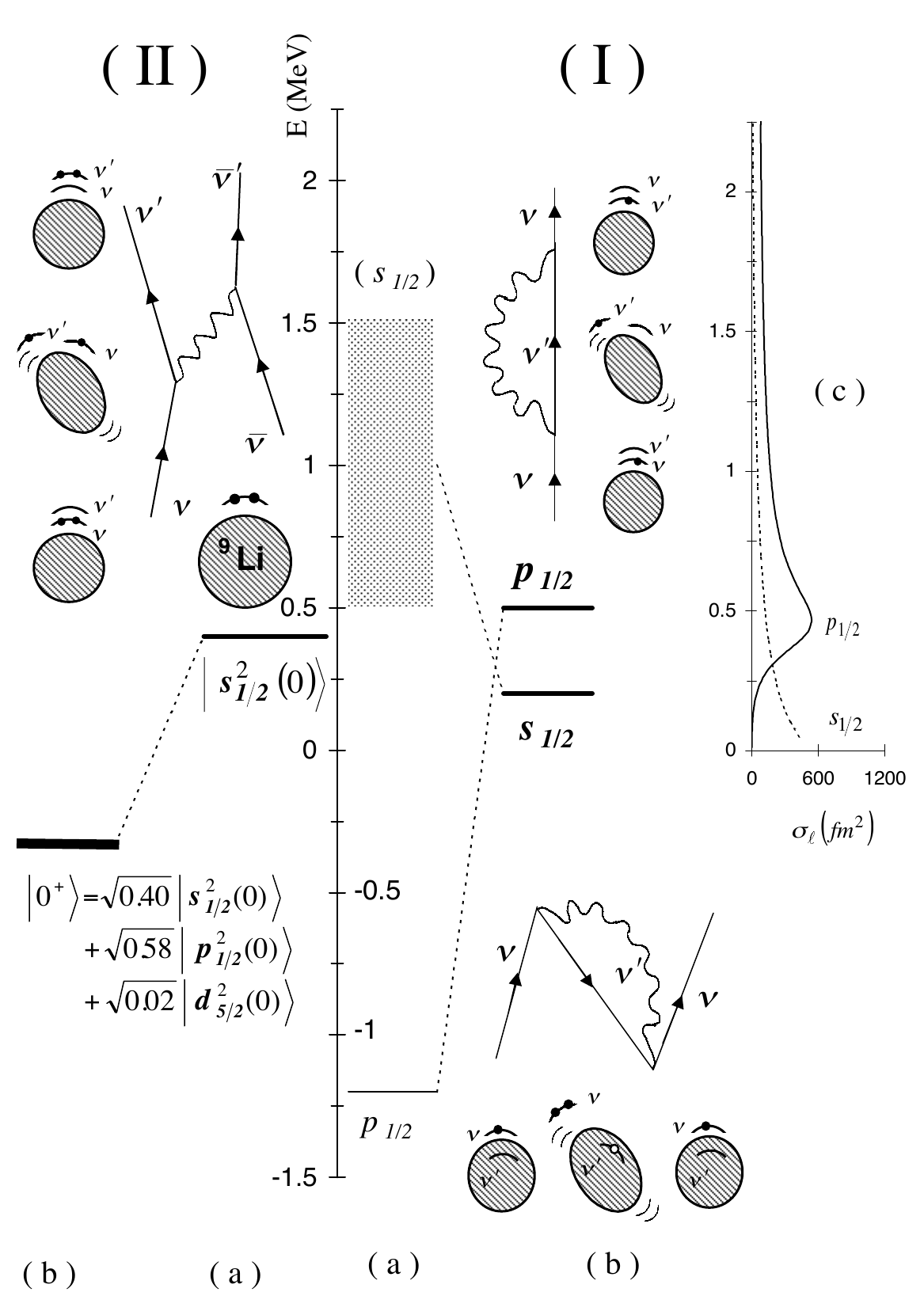}}
\caption{\protect\footnotesize (I) {\bf Single-particle neutron resonances in $^{10}$Li}.
 In (a) the position of the levels $s_{1/2}$ and $p_{1/2}$ calculated
 making use of mean field theory is shown (hatched area and
thin horizontal line respectively).
The coupling of a single-neutron
(upward pointing arrowed line) to a vibration (wavy line) calculated
making use of the Feynman diagrams displayed in (b)
(schematically depicted
also
in terms  of either solid dots (neutron) or open circles (neutron hole)
moving in a single-particle level around or in
the $^9$Li core (hatched area)), leads to conspicuous shifts in the energy
centroid of the
$s_{1/2}$ and $p_{1/2}$ resonances (shown by thick horizontal lines) and
eventually to an inversion in their sequence.
In (c) we show the calculated partial cross section
$\sigma_l$ for neutron elastic scattering off $^9$Li.
(II) {\bf The two-neutron system $^{11}$Li}.
We show in (a) the mean-field picture
of $^{11}$Li, where two neutrons (solid dots) move in time-reversal states
 around the core $^9$Li (hatched area) in the $s_{1/2}$ resonance leading to
an unbound $s^2_{1/2}(0)$ state where the two neutrons are coupled to zero angular momentum. The exchange of vibrations between the two neutrons shown in the upper part of the figure leads to a density dependent interaction which, added to the nucleon-nucleon interaction, correlates the two--neutron system leading
to a bound state $|0^+ \rangle$, where the two neutrons move with probability 0.40, 0.58 and 0.02
in the two-particle configurations $s^2_{1/2}(0)$,  $p^2_{1/2}(0)$ and
$d^2_{5/2}(0)$ respectively.}
\label{Li11_fig1}
\end{figure}

The continuum states of this  potential  were calculated by solving the problem in a box of radius equal to 40 fm, chosen so as to make the results associated with $^{10}$Li and $^{11}$Li discussed below, stable.  While mean field theory predicts the orbital $p_{1/2}$ to be lower than the $s_{1/2}$ orbital (cf. Fig. \ref{Li11_fig1}, I(a)),
experimentally the situation is reversed.
Similar parity inversions have
been observed in other isotones of $_3^{10}$Li$_7$, like e.g.
$^{11}_4$Be$_7$. Shell model calculations testify to the fact that the
effect of core excitation, in particular of quadrupole type, play a
central role in this invertion (\cite{Sagawa:93}, \cite{Gori:04}, cf. also \cite{Vinh:95}).
In keeping with this result,
the effect the coupling of the $p_{1/2}$ and $s_{1/2}$
orbitals of $^{10}$Li to quadrupole vibrations of
the $^9$Li core has on the properties of the 1/2$^+$ and 1/2$^-$ states
of this system has been studied (monopole and dipole vibrations display no low-lying strength
and their coupling to the single-particle
states of $^{10}$Li lead to negligible contributions).
The vibrational states of $^9$Li were calculated by diagonalizing,
in the random phase approximation (RPA), a multipole-multipole separable
interaction taking into account the contributions arising from the excitation
of particles into the continuum states.
The self-consistent value for the coupling strength was adopted,
because a calculation in the neighbor nucleus $^{10}$Be yields
good agreement with  the experimentally known
transition probability of the quadrupole low-lying vibrational state
(\cite{Raman:87},\cite{AjzenbergSelove:88}).

In the calculation of the renormalization effects of the single-particle resonances of $^{10}$Li due to the coupling to vibrational states one considered not only the effective-mass-like diagrams (upper-part graph of Fig. \ref{Li11_fig1}, I(b), so called polarization processes, cf. e.g. \cite{Mahaux:85}) leading to attractive (negative) contributions to the single-particle energies, but also those couplings leading to Pauli
principle (repulsive) correction processes associated with diagrams containing two--particles, one--hole and a vibration in the intermediate
states (lower-part diagram of Fig. \ref{Li11_fig1}, I(b), associated with the exchange between valence nucleons and those participating in vacuum zero point fluctuations, typical of Lamb shift--like process). Because of such Pauli correction
processes, the $p_{1/2}$ state experiences an upward shift in energy,
arising from the coupling of this orbital to the $p_{3/2}$ hole-state through
quadrupole vibrational states, in keeping with the fact that the ($p_{1/2}
p_{3/2}^{-1}$) particle-hole excitation constitutes an important component of
the quadrupole vibration wavefunction (cf. Table 1).
\begin{table}
\begin{center}
\begin{tabular}{|c|c|c|c|c|c|c|}
\hline
\small &  \small $1p_{3/2}^{-1}\;1p_{1/2}$ & \small $1p_{3/2}^{-1}\;8f_{7/2}$ & \small $1p_{3/2}^{-1}\;9f_{7/2}$ & \small $1s_{1/2}^{-1}\;3d_{5/2}$ & \small $1p_{3/2}^{-1}\;p_{1/2} (\pi)$ & \small $1s_{1/2}^{-1}\;1d_{5/2}(\pi)$ \\
\hline
$X_{ph}$ & 1.02 & 0.07 &  0.08 &  0.07 & 0.15 & 0.09 \\
$Y_{ph}$ & 0.28 & 0.05 &  0.06 &  0.06 & 0.09 & 0.07 \\
\hline
\end{tabular}
\caption{
RPA wavefunction of the collective low-lying quadrupole vibration of  $^{9}$Li
(X and Y are the forwardsgoing and backwardsgoing amplitudes respectively).
The energy of this state is $E_{2+}$ = 3.3 MeV.
}
\end{center}
\label{Li11_RPA1}
\end{table}
As a consequence, the $p_{1/2}$
state becomes unbound, turning into a low-lying resonance with centroid
$E_{res} \approx 0.5$MeV.
Due to the coupling to the vibrations the $s-$states are instead shifted
downwards. In fact, in this case there are essentially no (repulsive)
contributions arising from the Pauli correction processes. On the other
hand (attractive) effective-mass-like processes with intermediate states consisting
of one particle
plus a vibrational state of the type
($d_{5/2} \times 2^+$) lead to a virtual state with  $E_{virt} =
0.2 $ MeV (cf. Fig. \ref{Li11_fig1}, I(b)). The above results provide an overall
account of the $s-$ and $p-$resonances observed experimentally.
The important difference between the distribution of the  single-particle
strength associated with the resonant state $p_{1/2}$ and the virtual
state $s_{1/2}$ can be observed in Fig. \ref{Li11_fig1}, I(c), where the partial cross section
$\sigma_l$ for neutron elastic scattering off $^9$Li is shown. While
$\sigma_p$ displays  a clear peak at 0.5 MeV, $\sigma_s$ is a smoothly
decreasing function of the energy. A small increase in the
depth of the potential felt by the $s-$neutron will lead to a (slightly)
bound state, hence the name of virtual (cf. e.g. \cite{Landau:81}).

Let us now discuss the mechanism by which the Cooper pair neutron halo binds to the ${}^9$Li core to give rise to ${}^{11}$Li. While in the infinite system the existence of a bound state
of the (Cooper) pair happens for an arbitrarily weak interaction
\cite{Cooper:56},
 in the nuclear case this phenomenon takes place only if the strength
of the nucleon-nucleon potential is larger than a critical value, value connected
with the discreteness of the nuclear spectrum. In fact, calculations carried out
making use of a particularly successful parametrization of the (bare)
potential (Argonne potential, \cite{Wiringa:84}),
show that the nuclear forces
are able to bind Cooper pairs in open shell nuclei (leading to sizable
pairing gaps (1-2 MeV) \cite{Barranco:97}),
but not in closed shell nuclei, the most
important contributions to the nucleon-nucleon (pairing) interaction
arising from high multipole components of the force (\cite{Belyaev:59}).

The situation
is however quite different for the "open  shell" nucleus $^{11}$Li, as in
this case the bare nucleon-nucleon interaction is not able to bind
the two last neutrons to the $^9$Li ``core``. In fact, diagonalizing the
Argonne potential  in the basis of two-particle states $| nlj \times n'lj (0)>$
coupled to angular momentum zero, does not lead to a bound state. The low-lying states resulting from the diagonalization of the  Argonne $v_{14}$
nucleon-nucleon force are essentially dominated by one of the configurations $| s_{1/2}^2 (0)>, |p_{1/2}^2(0)>$ or
$ | d_{5/2}^2 (0)>$. In fact, the Argonne interaction
produces almost no  mixing between $s-$, $p-$waves and $d-$waves,
but essentially it only lowers the energy of
the unperturbed (resonant) configurations $s^2_{1/2}(0)$ and
$p^2_{1/2}(0)$ by about 80 keV without giving rise to a bound
system. It is of notice that the $d^2_{5/2}(0)$ configurations
are essentially not shifted by the bare NN--interaction. This result is very different from that obtained in nuclei
lying along the stability valley where typical pairing correlation energies
are of the order of 1.5 MeV.
Making use of the same single-particle levels and of the same
matrix elements
of the nucleon-nucleon potential in connection with the BCS equations does not
lead to a solution but to the trivial one of zero pairing gap  ($\Delta_{\nu} =0,
U_{\nu}V_{\nu}$ = 0).
At the basis of this negative result is the fact, already mentioned above,
that the most important
single-particle states allowed to the halo neutrons of $^{11}$Li to correlate
are the $s_{1/2}$, $p_{1/2}$ and $d_{5/2}$ orbitals. Consequently the two
neutrons are not able, in this  low-angular momentum phase space, to profit
fully from the strong force-pairing interaction.

\begin{figure}
\centerline{\includegraphics*[width=.73\textwidth,angle=0]{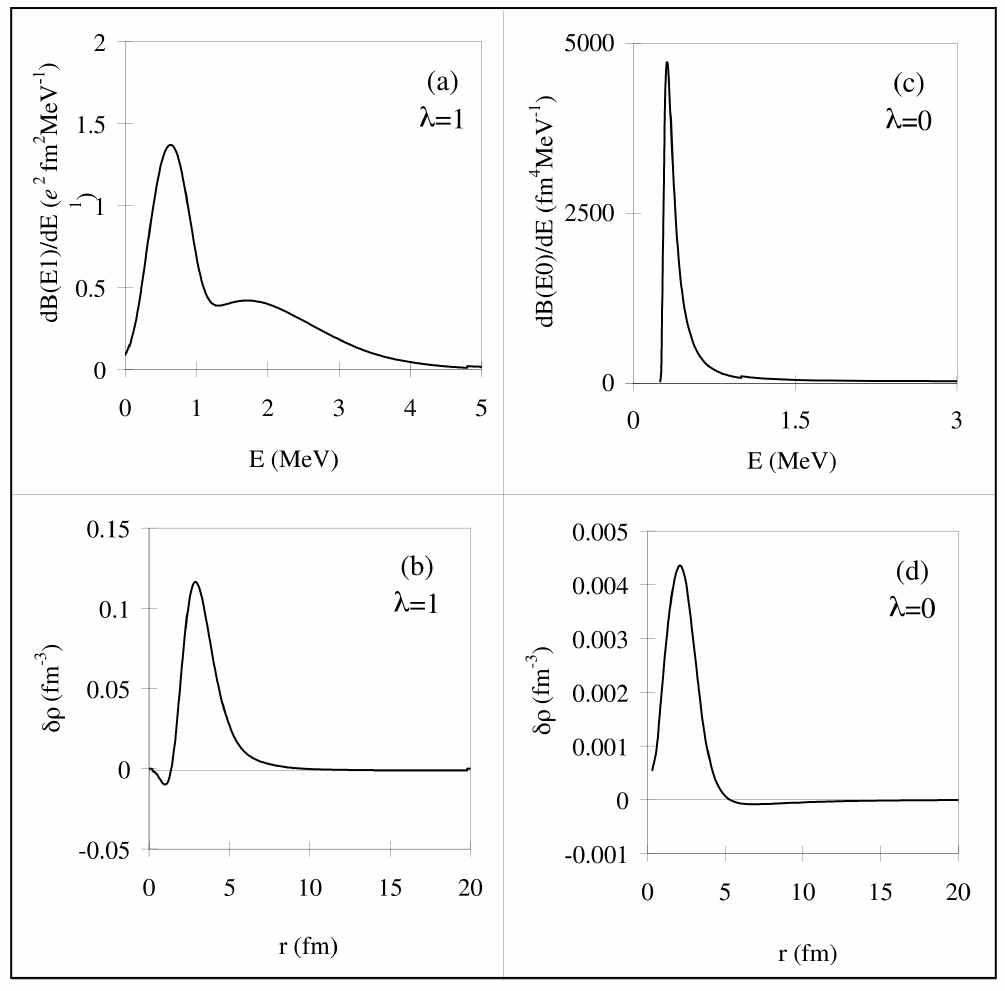}}
\caption{\protect\footnotesize
Dipole and  monopole linear response functions and transition
densities of $^{11}$Li calculated in the RPA.
}\label{Li11_fig2}
\end{figure}

Because of this result and those of \cite{Barranco:99},
and in keeping with the fact that
$^{11}$Li displays low-lying collective vibrations (\cite{Sackett:93,Zinser:97,Nakamura:06}),
it is fair to expect that the exchange of these vibrations between the two outer neutrons of $^{11}$Li is the main source of pairing available
to them to correlate. The L=0,1, and 2-vibrational spectrum of $^{11}$Li (see Tables 1, 2 and 3) needed to calculate
the matrix elements of this induced interaction was determined in much
the same way as in $^9$Li, that is making use of the RPA.
\begin{table}
\begin{center}
\begin{tabular}{|c|c|c|c|c|c|c|}
\hline
\small& \small $1p_{3/2}^{-1}\;1p_{1/2}$ & \small $2s_{1/2}^{-1}\;5d_{3/2}$ & \small $1p_{1/2}^{-1}\;6p_{3/2}$ & \small $2s_{1/2}^{-1}\;3d_{5/2}$ & \small $2s_{1/2}^{-1}\;5d_{5/2}$ & \small $1p_{3/2}^{-1}\;1p_{1/2}(\pi)$ \\
\hline
$X_{ph}$  & 0.824 &  0.404  &  0.151 &  0.125 & 0.126 & 0.16 \\
$Y_{ph}$ & 0.119 & 0.011 &  -0.002 &  -0.049 & -0.011 & 0.07 \\
\hline
\end{tabular}
\caption{
RPA wavefunction of the collective low-lying quadrupole phonon in $^{11}$Li,
of energy  $E_{2+}$ =5.05 MeV, which contributes to the induced interaction in Fig.\ref{Li11_fig1}.II. All listed amplitudes, with the exception of the one displayed in the last column, correspond to neutron transitions.
}
\end{center}
\label{Li11_RPA2}
\end{table}
\begin{table}
\begin{center}
\begin{tabular}{|c|c|c|c|c|c|c|c|}
\hline
& $1p_{1/2}^{-1}\;2s_{1/2}$ & $1p_{1/2}^{-1}\;3s_{1/2}$ & $1p_{1/2}^{-1}\;4s_{1/2}$ & $1p_{1/2}^{-1}\;1d_{3/2}$ & $1p_{3/2}^{-1}\;5d_{5/2}$ & $1p_{3/2}^{-1}\;6d_{5/2}$ & $1p_{3/2}^{-1}\;7d_{5/2}$ \\
\hline
$X_{ph}$ & 0.847 & -0.335 & 0.244 &  0.165 & 0.197 & 0.201 & 0.157 \\
$Y_{ph}$ & 0.088 &  0.060 & 0.088 &  0.008 & 0.165 & 0.173 & 0.138 \\
\hline
\end{tabular}
\caption{
RPA wavefunction of the strongest low-lying dipole vibration of
$^{11}$Li, ($E_{1-}$ =0.75 MeV),
and  contributing most importantly to the pairing induced interaction
(Fig.\ref{Li11_fig1}.II). All the listed amplitudes refer to neutron transitions.
}
\end{center}
\label{Li11_RPA3}
\end{table}
The soft dipole response  is shown in Fig. \ref{Li11_fig2}(a).
The low-lying quadrupole response is concentrated in a single peak,
whose wavefunction is shown in Table 2.
The monopole response exhausting 94\% of the EWSR
is displayed in Fig. \ref{Li11_fig2}(c).
All the resulting vibrational
states were coupled to the single-particle states making use
of the corresponding
transition densities (form factors, cf. Fig. \ref{Li11_fig2}(b) and \ref{Li11_fig2}(d))
and associated particle--vibration coupling strengths.

The calculations of the effects of the exchange of soft-dipole vibrational phonons includes the core dynamics in a fashion which is, to a large
extent, equivalent to the calculations reported
(within this context see also \cite{Esbensen:97}, \cite{Broglia:02b})
within the framework of a three-body cluster model. In fact, in
this model the soft dipole mode arises from the vibrations of the two weakly
bound neutrons with respect to the $^9$Li core. Within this model
an interaction
correlating the two neutrons arises from a cross term of the recoil kinetic
energy of the core. In the Appendix of \cite{Esbensen:97} this term is shown to be
equivalent to a separable dipole-dipole interaction of the type used in our
calculations. To be noted however, that the cluster model does not include
the dynamical renormalization effects of the  single-particle motion
(in particular s- and p-motion) arising from
the coupling of single-particle motion to
quadrupole vibrations of the $^9$Li core. In fact, in the cluster model
the core is assumed to be inert.

Allowing the two outer neutrons of $^{11}$Li to both exchange phonons (induced
interaction, Fig. \ref{Li11_fig1}, II(a)), as well as to emit and later reabsorb
them (self-energy correction, Fig.\ref{Li11_fig1}, I(b)), leads to a bound
(Cooper) pair, the lowest eigenstate of the associated
secular matrix being \mbox{$E_{gs}=-0.270$ MeV}. This result is mostly due to the exchange of the low-lying dipole
vibrations shown in Fig.\ref{Li11_fig2}(a) with associated wavefunction collected in
Table 3. Adding the nucleon-nucleon Argonne potential to the induced interaction, one obtains \mbox{$E_{gs} = -0.330$} MeV, and thus a two-neutron
separation energy quite close to the experimental value.
 Measured from the unperturbed energy
 of a pair of neutrons in the lowest state calculated for $^{10}$Li, namely
the $s$-resonance
($E_{unp}=2E_{s_{1/2}}=$ 400 keV, cf. Fig.\ref{Li11_fig1},I(b)),
it leads to a pairing correlation
 energy $E_o=E_{unp}-E_{g.s.}=$ 0.730 MeV (cf. Fig.\ref{Li11_fig1}, II(b)).

From the associated two-particle ground state wavefunction
$\Psi_0(\vec r_1, \vec r_2)(\equiv \langle \vec r_1, \vec r_2 |0^+ \rangle)$,
one obtains a momentum distribution
(whose FWHM is $\sigma_{\bot}=56$ MeV/c, for $^{11}$Li on $^{12}$C)
 and ground state occupation
probabilities of the two-particle states $s_{1/2}^2$(0), $p_{1/2}^2$(0) and
$d_{5/2}^2(0)$
 (0.40, 0.58 and 0.02
respectively, cf. Fig. \ref{Li11_fig1},II(b))
which provide an overall account of the experimental findings.
The radius of the associated single-particle distribution is 7.1 fm.
 Adding to this density
 that of the core nucleons one obtains the total density
of $^{11}$Li (see discussion below as well as Fig. \ref{Li11_fig5}). The associated mean square radius (3.9 fm) is
slightly larger than the experimental value.

\begin{figure}
\centerline{\includegraphics*[width=.63\textwidth,angle=0]{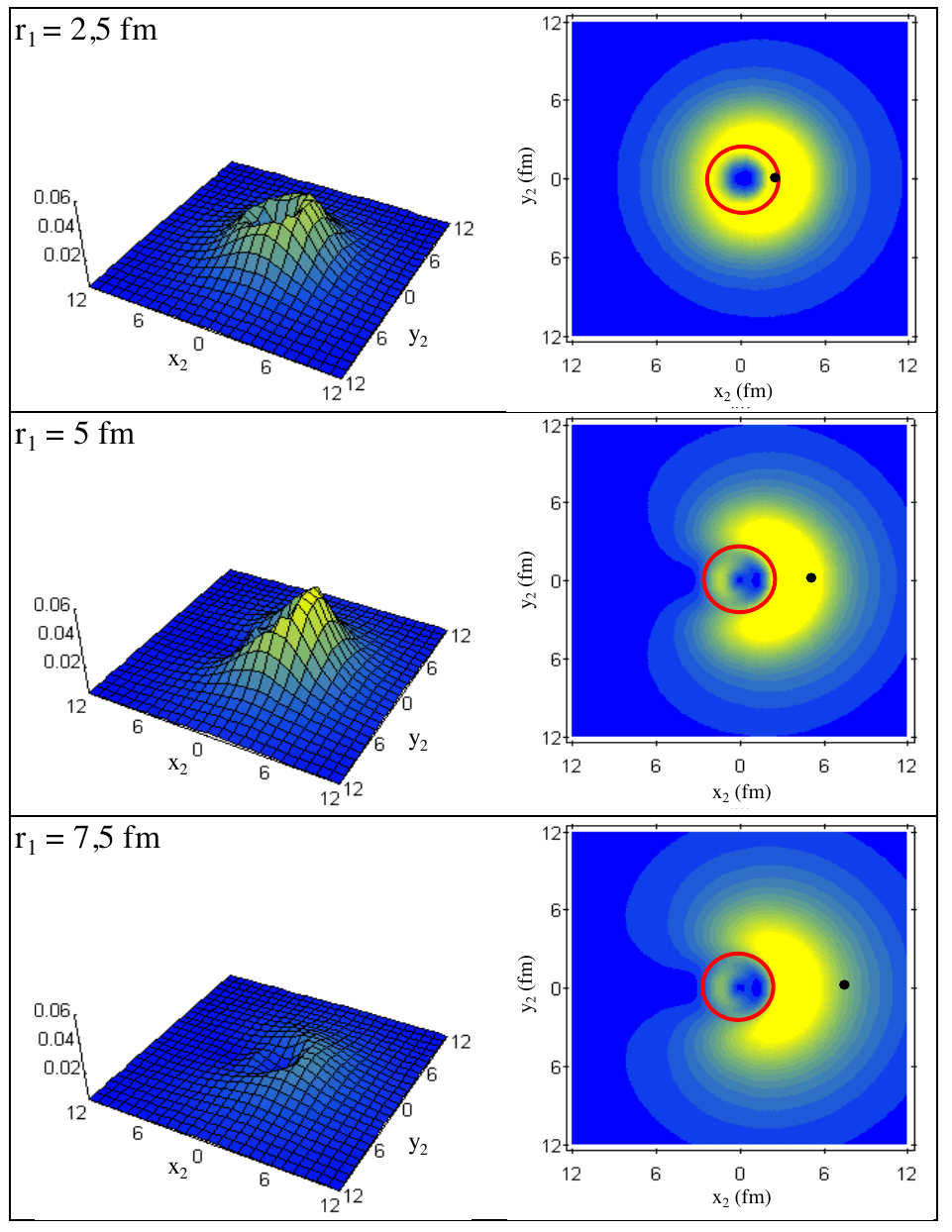}}
\caption{\protect\footnotesize
{\bf Spatial structure of two-neutron Cooper pair}.
The modulus squared wavefunction
$|\Psi_0 (\vec r_1,\vec r_2)|^2=|\langle \vec r_1,\vec r_2|0^+ \rangle|^2$
 (cf. Fig. \ref{Li11_fig1}, II (b)) describing the motion of the two halo
neutrons around the
$^9$Li core (normalized to unity and multiplied by 16$\pi^2r_1^2r_2^2)$ is
displayed as a function of the cartesian coordinates
$x_2 = r_2 \cos(\theta_{12})$
and $y_2=r_2 \sin(\theta_{12})$ of particle 2, for fixed value of
the position of particle 1
($r_1$=2.5, 5, 7.5 fm) represented in the right panels by a solid dot, while
the core $^9$Li is shown as a red circle. The numbers appearing on the $z$-axis
of the three-dimensional plots displayed on the left side of the figure are
in units of fm$^{-2}$.
}\label{Li11_fig3}
\end{figure}

\begin{figure}
\centerline{\includegraphics*[width=.63\textwidth,angle=0]{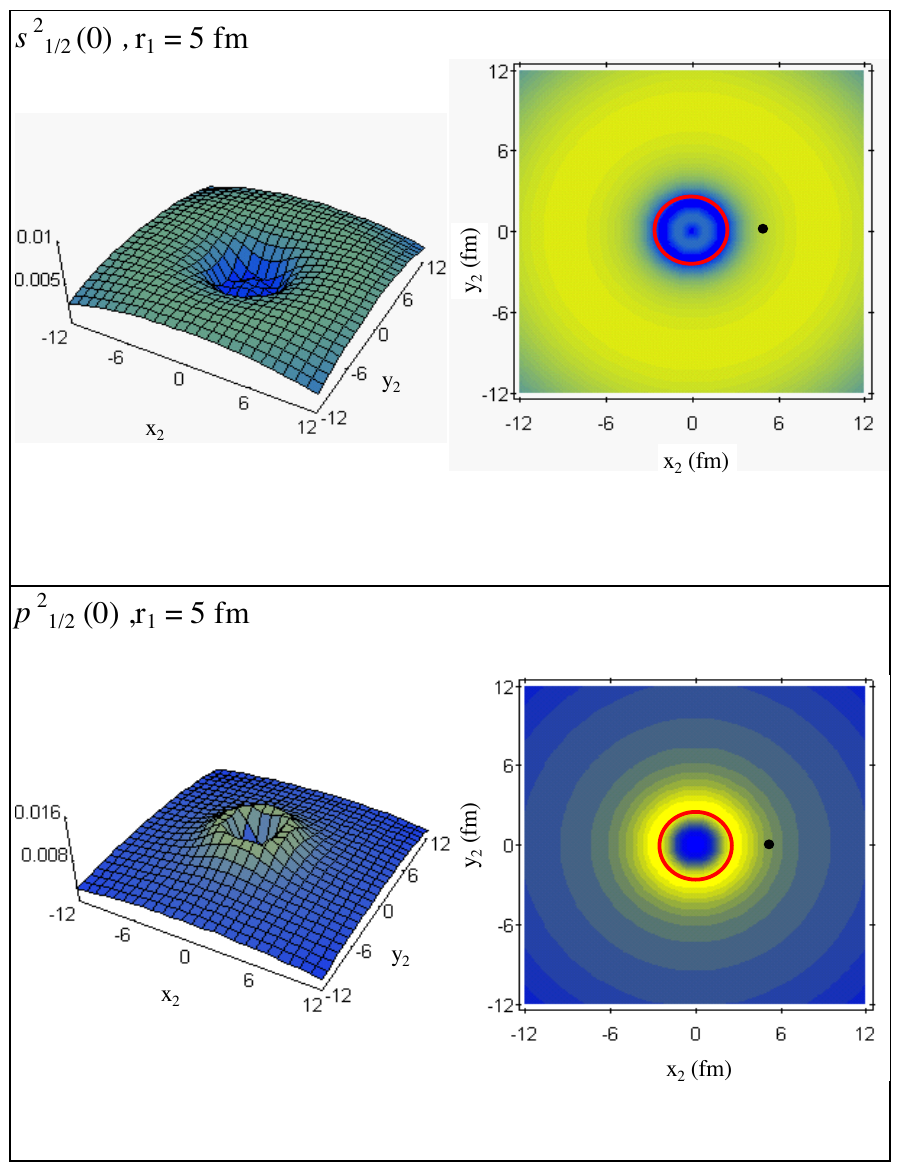}}
\caption{\protect\footnotesize
{\bf Spatial distribution of the pure two-particle configurations
 $ \mbox{\boldmath $s_{1/2}^2$}$ (0)
and $ \mbox{\boldmath $p_{1/2}^2$}$(0)}
as a function of the $x$- and  $y$-coordinates of particle
 2, for a fixed
value of the coordinate  of particle 1 ($r_1$=5 fm). For more details cf.
 caption to Fig. \ref{Li11_fig3}.
}\label{Li11_fig4}
\end{figure}

\begin{figure}
\centerline{\includegraphics*[width=.83\textwidth,angle=0]{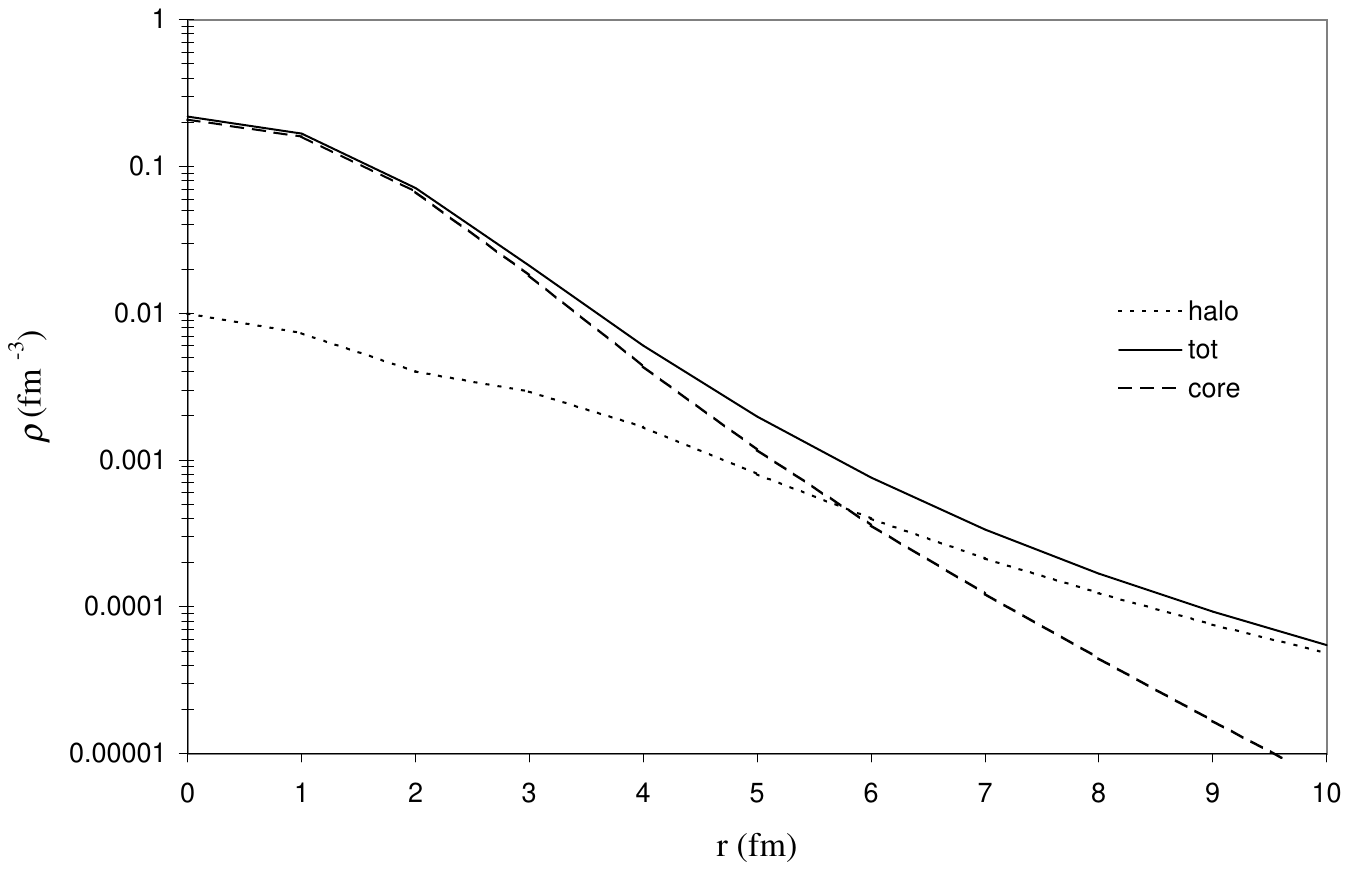}}
\caption{\protect\footnotesize
{\bf Calculated single--particle density of ${}^{11}$Li} (thick continuous curve). The contributions of the density of the core ${}^9$Li and of the halo neutrons are also shown.
}\label{Li11_fig5}
\end{figure}

The spatial structure of the Cooper pair described by the wavefunction
$\Psi_0(\vec r_1, \vec r_2)$ is displayed in Fig. \ref{Li11_fig3}.
The mean square radius of the center of mass of the two neutrons is
$\langle r_{cm}^2 \rangle^{1/2}$ = 5.4 fm.
This result testifies to the importance the correlations have
in collecting the small (enhanced) amplitudes of the uncorrelated
two-particle configuration $s_{1/2}^2(0)$ in the region between 4 to 5
fm, region in which the $p_{1/2}^2(0)$, helped by the centrifugal barrier,
displays a somewhat larger concentration (cf. Fig. \ref{Li11_fig4}).
From the above results, it emerges that the exchange of vibrations
between the least bound neutrons leads to a (density-dependent) pairing
interaction acting essentially only outside the core (cf. also \cite{Bertsch:91}).
Of notice that the long wavelength behaviour of these vibrations
is connected with  the excitation of the neutron halo,  the large size of which
not only makes the system easily polarizable but provides also the
elastic medium through which the loosely bound neutrons exchange vibrations with
each other.
Within this context, see the calculated single--particle density of ${}^{11}$Li (thick continuous curve) displayed in Fig. \ref{Li11_fig5}, resulting from the summed contributions of the density of the ${}^9$Li core (dashed curve) and of the halo neutrons (dotted curve).

%

From the above narrative it emerges that essentially all of the experimental facts characterizing ${}^{11}$Li can be explained, even quantitatively, in terms of a simple scenario: the formation of a neutron Cooper pair which is held together to the $^9$Li core by exchanging low--frequency phonons corresponding to long--wavelength vibrations of the system. Such a pairing mechanism is clearly reflected in the calculated ground state wavefunction of  $^{11}$Li,
\begin{equation}\label{eq1}
    |^{11}\text{Li}(gs);3/2^-\rangle = |\tilde 0\rangle_\nu \otimes |1 p_{3/2}(\pi)\rangle,
\end{equation}
where $\pi$ and $\nu$ indicate proton and neutron degrees of freedom respectively, while $|\tilde 0\rangle_\nu$ indicates the halo neutron Cooper pair wavefunction, that is,
\begin{equation}\label{eq2}
    |\tilde 0\rangle_\nu = | 0\rangle +  \alpha |(p_{1/2},s_{1/2})_{1^-}\otimes 1^-;0\rangle+\beta|(s_{1/2},d_{5/2})_{2^+}\otimes 2^+;0\rangle,
\end{equation}
with
\begin{equation}\label{eq5}
 \alpha\approx 0.7, \quad \text{and} \quad \beta\approx 0.1,
\end{equation}
and
\begin{equation}\label{eq4}
    | 0\rangle=0.45 |s_{1/2}^2(0)\rangle + 0.55 |p_{1/2}^2(0)\rangle+0.04|d_{5/2}^2(0)\rangle,
\end{equation}
the states $| 1^-\rangle$ and $| 2^+\rangle$ being the (RPA) states describing the dipole pigmy resonance of $^{11}$Li and  the quadrupole vibration of $^{11}$Li.
 The intrinsic non--observability of virtual processes (like the exchange of collective vibrations between Cooper pair partners leading to the second and third components of the state $|\tilde 0\rangle_\nu$) is a fact. However, in those cases in which the experimental tool exists which specifically probes the phenomenon under study, one can force the virtual processes of interest to become real. In this way one could, for example, hope to observe the collective vibrations of $^{11}$Li and of  $^{9}$Li correlating the two--halo neutrons, with the help of a two--particle transfer process, specific probe of pairing in nuclei.

In what follows it is shown that the experiment $^1$H($^{11}$Li,$^{9}$Li)$^3$H recently carried out at TRIUMF by \cite{Tanihata:08}, provides direct evidence of such process.

\begin{figure*}
\centerline{\includegraphics*[width=.99\textwidth,angle=0]{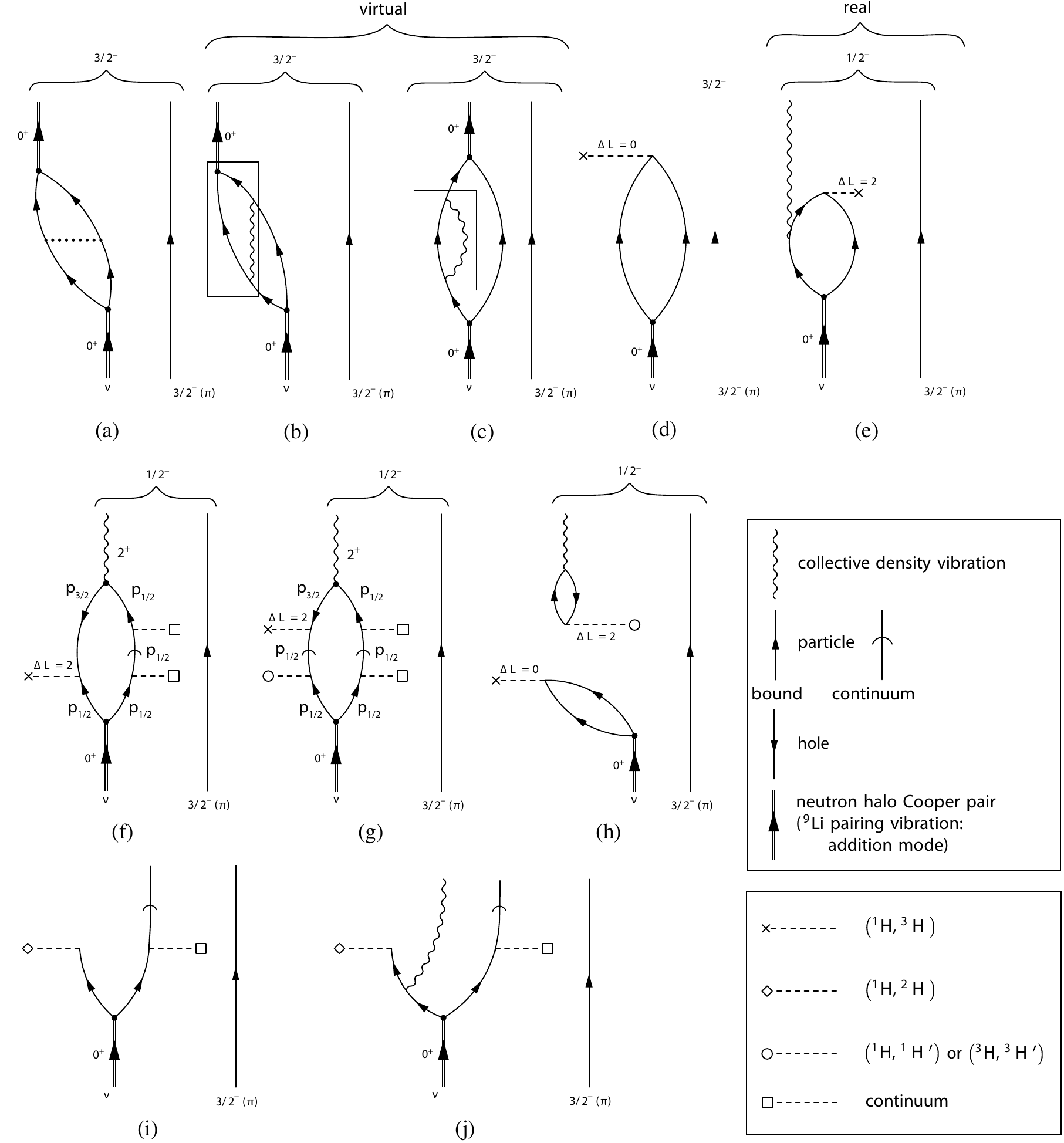}}
\caption{\protect\footnotesize Representative Nuclear Field Theory--Feynman diagrams associated with correlation process ((a),(b),(c)) and with one-- and two--particle pick--up reactions ((i),(j) and (d),(e) respectively) of the halo neutrons of $^{11}$Li (Cooper pair, indicated in terms of a double arrowed line). Also shown are the possible diagrams associated with other channels (breakup and inelastic) populating the $1/2^-$ (2.69 MeV) state: f) one of the neutrons is picked up (the other one going into the continuum, i.e. breaking up from the $^9$Li core) together with a neutron from the $p_{3/2}$ orbital of the $^9$Li core  leading eventually to the excitation of the $1/2^-$ final state ($2^+$ density mode (wavy line) coupled to the $p_{3/2}(\pi)$), g) the proton field acting once breaks the Cooper pair forcing one of the halo neutrons to populate a $p_{1/2}$ continuum state (the other one follows suit), while acting for the second time picks up one of the neutrons moving in the continuum and another one from those moving in the $p_{3/2}$ orbital of $^9$Li eventually leaving the core in the quadrupole mode of excitation. In (h) the  two--step transfer to the $^9$Li ground state plus the inelastic final channel process exciting the $(2^+\otimes p_{3/2}(\pi))_{1/2^-}$ state is shown.}\label{fig1}
\end{figure*}

To convey the details of such an analysis the NFT--Feynman diagrams used to calculate the wavefunction in Eq. (\ref{eq1}) generalized to deal also with reaction processes (\cite{Broglia:04a}), have been used (Fig.\ref{fig1}).

 From the diagrams displayed in Figs. \ref{fig1} (b) and \ref{fig1} (c), it is easy to understand how the virtual propagation of collective vibrations (in the present case $1^-$ and $2^+$ vibrations) can be forced to become a real process: by transferring one or two units of angular momentum in a two--neutron pick up process. In particular, the correlation mechanism displayed in Figs. \ref{fig1} (b) and \ref{fig1} (c) predicts a direct excitation of the quadrupole multiplet of $^9$Li (see Fig. \ref{fig1}(e), see also \cite{Brink:05} Fig. 11.6). On the other hand, if the two--neutron pick--up process takes place before the virtual excitation of the vibrational mode, the ground state of $^9$Li is populated (Fig. \ref{fig1} (d)).


The $1/2^-$ (2.69 MeV) first excited state of $^9$Li can also be excited through a break up process in which one (see Fig. \ref{fig1}(f)), or both neutrons (see Fig. \ref{fig1}(g)) are forced into the continuum for then eventually one of them to fall into the $1p_{3/2}$ orbital of $^9$Li and excite the quadrupole vibration of the core, in keeping with the fact that the main RPA amplitude of this state is precisely $X(1p^{-1}_{3/2},1p_{1/2})(\approx 1)$ (see Table 1). The remaining channel populating the first excited state of $^9$Li is associated with an inelastic process (see Fig. \ref{fig1}(h)): two--particle transfer to the ground state of $^{9}$Li and Final State (inelastic scattering) Interaction (FSI) between the outgoing triton and $^{9}$Li in its ground state, resulting in the inelastic excitation of  the $1/2^-$ state. It was shown in \cite{Potel:10} that the probabilities $p_l=|S_l^{(c)}|^2$ associated with each of the processes discussed above, where the amplitude $S_l^{(c)}$ is related to the total cross section associated with each of the channels $c$  by the expression (\cite{Satchler:80}, \cite{Landau:81})
\begin{equation}\label{eq6}
    \sigma_c=\frac{\pi}{k^2}\sum_l(2l+1)|S_l^{(c)}|^2,
\end{equation}
are small. Consequently, the interference between the contributions of the processes mentioned above to the differential cross section were taken into account in \cite{Potel:10} making use of second order perturbation theory, instead of a coupled channel treatment (see e.g. \cite{Ascuitto:69},\cite{Tamura:70},\cite{Khoa:04}, \cite{Rodriguez:09}, \cite{Keeley:07b} and refs. therein).

\begin{figure}
\centerline{\includegraphics*[width=.8\textwidth,angle=0]{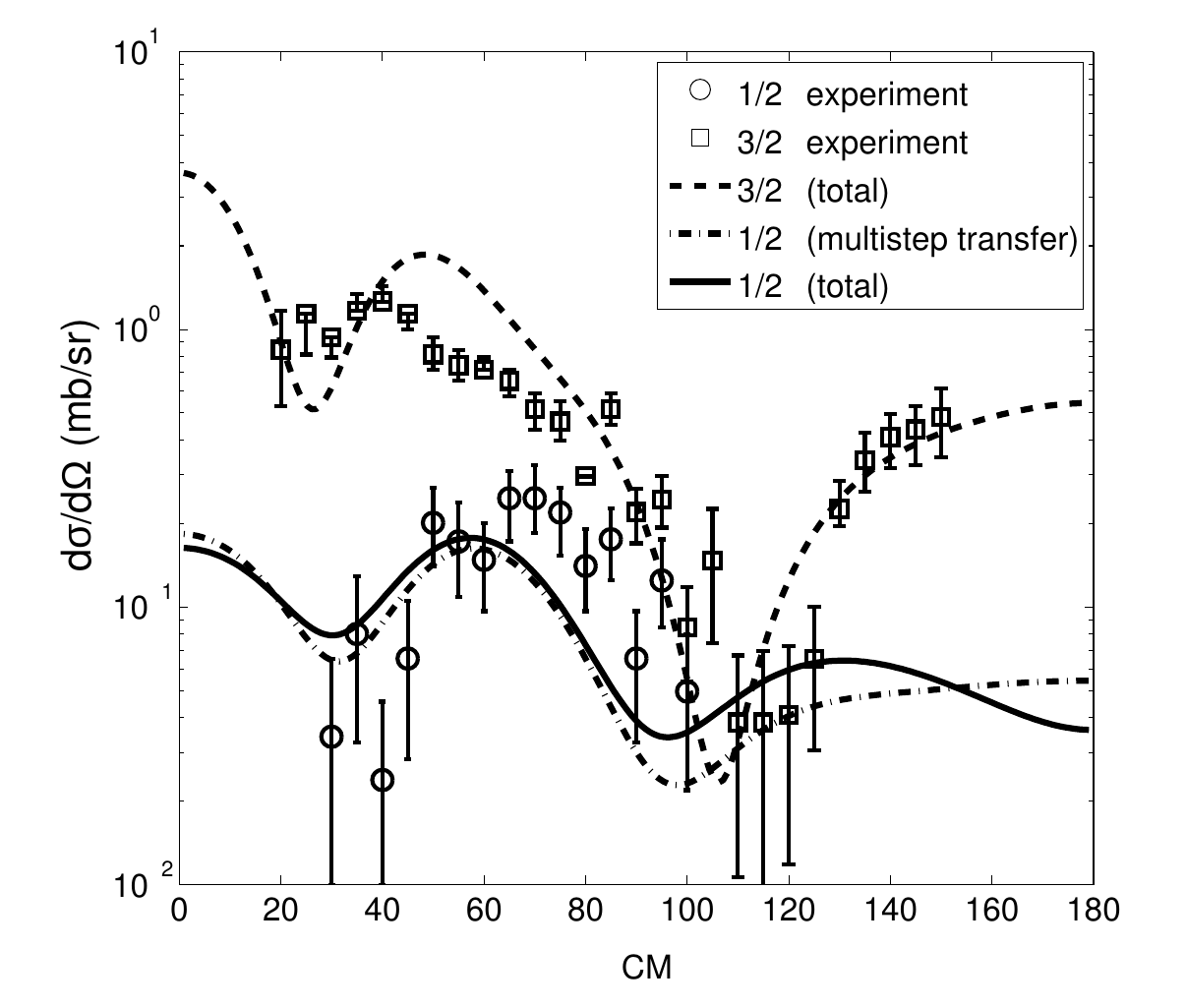}}
\caption{\protect\footnotesize Experimental (\cite{Tanihata:08}) and theoretical differential cross sections (including multistep transfer as well as breakup and inelastic channels).  of the
$^1$H($^{11}$Li,$^9$Li)$^3$H  reaction populating the ground state ($3/2^-$) and the first excited state ($1/2^-$; 2.69 MeV) of $^{9}$Li. Also shown (dash--dotted curve) is the differential cross section associated with this state but taking into account only multistep transfer. The optical potentials used are from \cite{An:06}and \cite{Tanihata:08}.}\label{fig2}
\end{figure}

\begin{figure}
\centerline{\includegraphics*[width=.8\textwidth,angle=0]{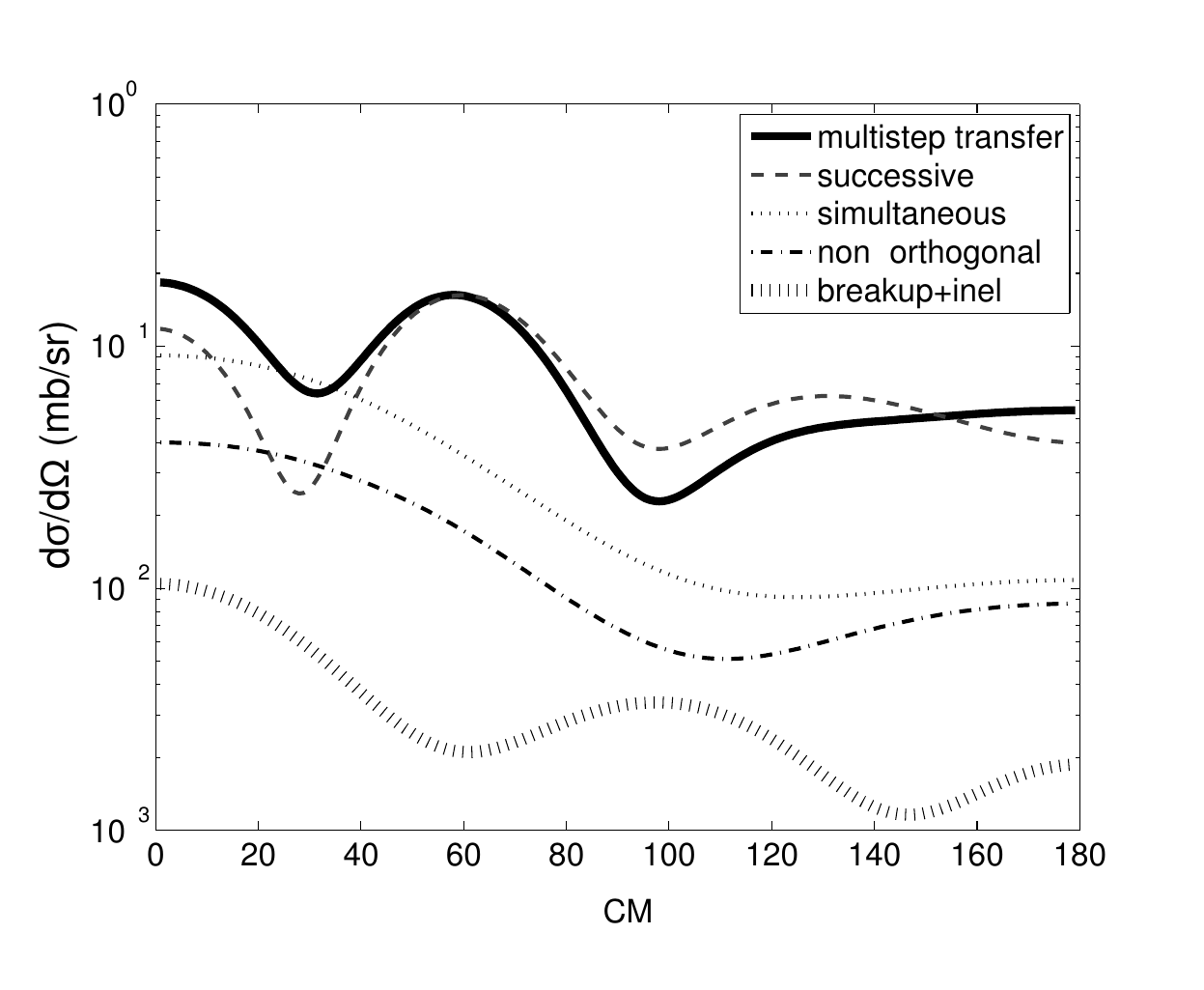}}
\caption{Successive, simultaneous and non-orthogonality contributions (prior representation)
to the  $^1$H($^{11}$Li,$^9$Li)$^3$H differential cross section
associated with the $1/2^-$ state
of $^9$Li, displayed in Fig. \ref{fig2}. Also shown is the (coherent) sum of the breakup and inelastic channel contributions.}\label{fig3}
\end{figure}

 Making use of the elements discussed above, multistep transfer, breakup and inelastic channels were calculated. In all the calculations the proton--neutron potential involved in the transfer was parametrized according to \cite{Tang:65} with a depth adjusted so as to reproduce the experimental deuteron binding energy. This is also true concerning the calculation of the $(t,p)$ and $(p,t)$ results displayed in Sections 5--7. The results are shown in Figs. \ref{fig2} and \ref{fig3} and in Table \ref{tab2}. Theory provides an overall account of the experimental findings. In particular, in connection with the $1/2^-$ state, this result essentially emerges from cancellations and coherence effects taking place between the three terms contributing to the multistep two--particle transfer cross section (see Fig. \ref{fig3}), tuned by the nuclear structure amplitudes associated with the process shown in Fig. \ref{fig1}(e) as well as Eqs. (\ref{eq1})--(\ref{eq4}). In fact, and
as shown in Figs. \ref{fig2} and \ref{fig3}, the contributions of inelastic and break up processes (Figs. \ref{fig1}(f),(g) and (h) respectively) to the population of the $1/2^-$ (2.69 MeV) first excited state of $^9$Li are negligible as compared with the process depicted in Fig. \ref{fig1}(e). In the case of the breakup channel (Figs. \ref{fig1}(f) and \ref{fig1}(g)) this is a consequence of the low bombarding energy of the $^{11}$Li beam (inverse kinematics), combined with the small overlap between continuum (resonant) neutron $p_{1/2}$ wavefunctions and  bound state wavefunctions. In the case of the inelastic process (Fig. \ref{fig1}(h)), it is again a consequence of the relative low bombarding energy. In fact, the adiabaticity parameters $\xi_C,\xi_N$ (see eqs. (IV.12) and (IV.14) of \cite{Broglia:05c}) associated with Coulomb excitation and inelastic excitation in the t+$^9$Li channel are larger than 1, implying an adiabatic cutoff. In other words, the quadrupole mode is essentially only polarized during the reaction but not excited. The situation is quite different in the case of the virtual process displayed in Fig. \ref{fig1} (e). Being this an off--the--energy shell process, energy is not conserved, and adiabaticity plays no role. It would be very interesting to challenge the results of the above calculations with eventual data for the same reaction at higher incident energy.

It is worth mentioning that the final states observed in the two neutron pick--up process can, in principle, also be populated in a one--particle pick--up process (see Figs. \ref{fig1}(i) and \ref{fig1}(j)).

\begin{figure}
\centerline{\includegraphics*[width=.8\textwidth,angle=0]{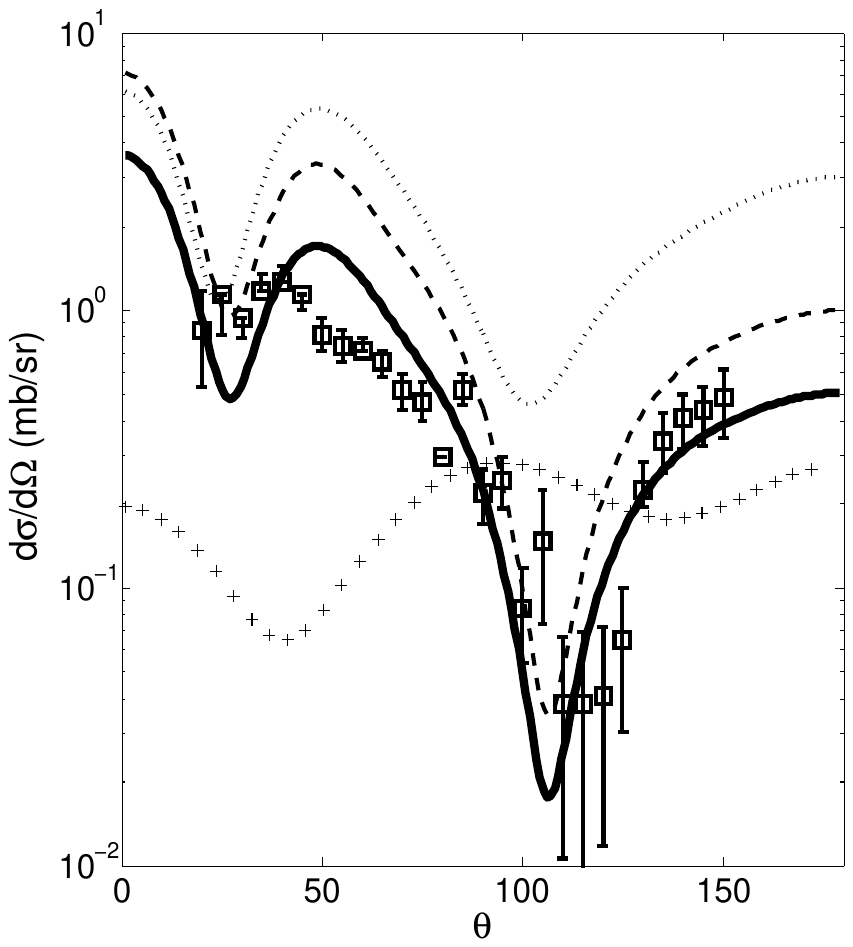}}
\caption{Differential cross section associated with the reaction ${}^{11}\textrm{Li}\left( {}^{1}\textrm{H}, {}^{3}\textrm{H} \right) {}^{9}\textrm{Li}(gs)$ calculated making use of: a) the wavefunction (\ref{eq1})--(\ref{eq4}) (continuous solid curve), b) the wavefunction $|0> = 0.63|s_{1/2}^{2}(0)>+0.77|p_{1/2}^{2}(0)>+0.06|d_{5/2}^{2}(0)>$, i.e. like (\ref{eq4}) but normalized not to 0.51 but to 1, tantamount of not including ground state correlations (dashed continuous curve), c) $s_{1/2}^{2}$ pure configuration (dotted continuous curve), d) $p_{1/2}^{2}$ pure configuration (cross continuous curve). The experimental data is also reported (see Fig. \ref{fig2}).}
\label{Li11_fig6}
\end{figure}

While the direct excitation of the $1/2^{-}$,2.69 MeV state of ${}^{9}$Li carries out an important message, namely evidence for phonon mediated pairing interaction in nuclei and can rightly be considered a milestone in two--nucleon transfer studies of nuclear superfluidity, the absolute value of the ${}^{11}\textrm{Li}\left( {}^{1}\textrm{H}, {}^{3}\textrm{H} \right) {}^{9}\textrm{Li}(gs)$ cross section contains also much information concerning nuclear pairing, as can be seen from Fig. \ref{Li11_fig6} and Table \ref{Li11_tab6} (cf. also \cite{Tanihata:08}). It clearly emerges from these results that the ground state correlations associated with the exchange of phonons between the two halo neutrons can change the absolute ground state cross section by a factor of 2, an effect that is experimentally confirmed.

\begin{table}
\begin{center}
\begin{tabular}{|c|c|c|c|}
\cline{3-4}
\multicolumn{2}{c|}{ } &  \multicolumn{2}{c|}{$\sigma$($^{11}$Li(gs)  $\to$ $^9$Li (i)) $(m$b)}\\
\hline
i & $\Delta L$ & Theory & Experiment \\
\hline
gs ($3/2^-$)& 0 & 6.1 &  5.7 $\pm$ 0.9\\
\hline
 2.69 MeV $(1/2^-)$ & $\hspace{1.6cm} 2\quad\left\{\begin{array}{l}
                        (\beta=0.1)\\
                        (\beta=0)
                      \end{array}\right.$
  & $\begin{array}{c}
                        0.7 \\
                        0.05
                      \end{array} $& 1.0 $\pm$ 0.36\\
\hline
\end{tabular}
\caption{\protect\footnotesize Integrated two-neutron differential transfer cross sections, in the angular range \mbox{20$^{\circ}$--150$^{\circ}$} in which the observation has been made, associated with the ground state (gs ($3/2^-$))
and with the first excited state (2.69 MeV; $1/2^-$) of $^9$Li in comparison with the data (\cite{Tanihata:08}). In the case of the $1/2^-$ state two calculations have been carried out. One making use of the microscopic wavefunction given in Eqs. (\ref{eq1})--(\ref{eq4}). A second one in which it is (arbitrarily) assumed that $\beta=0$ (see Eq. (\ref{eq2})). That is, that the only processes populating the first excited state of $^9$Li are associated with breakup and inelastic channels (see also Fig. \ref{fig3}).}\label{tab2}
\end{center}
\end{table}

\begin{table}
\begin{center}
\begin{tabular}{|c|c|}
\cline{2-2}
\multicolumn{1}{c|}{ } &  $\sigma$($^{11}$Li(gs)  $\to$ $^9$Li (gs)) $(m$b)\\
\hline
exp.                     &  $5.7 \pm 0.9$ \\
\hline
GSC ${}^{a)}$            &  $6.1$         \\
\hline
no GSC ${}^{b)}$         &  $12.1$         \\
\hline
$s_{1/2}^{2}$ ${}^{c)}$  &  $23.2$         \\
\hline
$p_{1/2}^{2}$ ${}^{d)}$  &  $2.2$         \\
\hline
\end{tabular}
\caption{\protect Absolute cross sections resulting from the integration of the calculated differential cross sections reported in Fig. \ref{Li11_fig6} as well as of the associated experimental data, integrated in the angular range \mbox{20$^{\circ}$--150$^{\circ}$}.
For the meaning of a), b), c) and d) we refer to the caption of Fig. \ref{Li11_fig6}.}
\label{Li11_tab6}
\end{center}
\end{table}



\section{The ${}^{122}\textrm{Sn}(p,t){}^{120}$Sn (gs) reaction: pairing in superfluid nuclei}

\begin{figure}
\begin{center}
\includegraphics[width=0.65\textwidth]{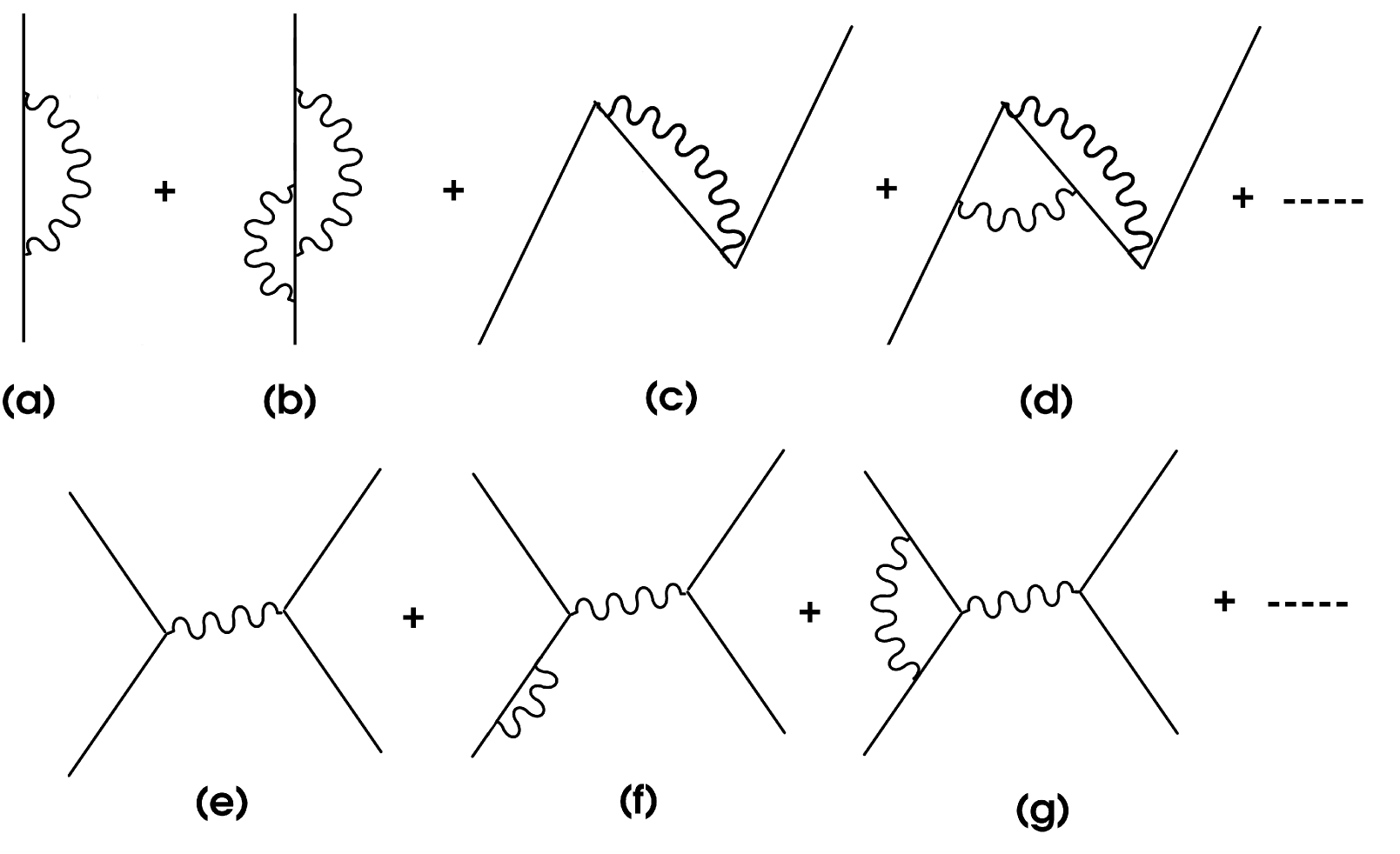}
\end{center}
\caption{
Renormalization processes arising from the particle vibration coupling
phenomenon. A line indicates quasiparticles obtained from BCS theory,
making use of the mean field single-particle states and the nucleon-nucleon $v_{14}$
Argonne potential. The wavy line indicates the vibrational states.}
\label{Sn122_fig1}
\end{figure}

\begin{figure}
\begin{center}
\includegraphics[width=0.65\textwidth]{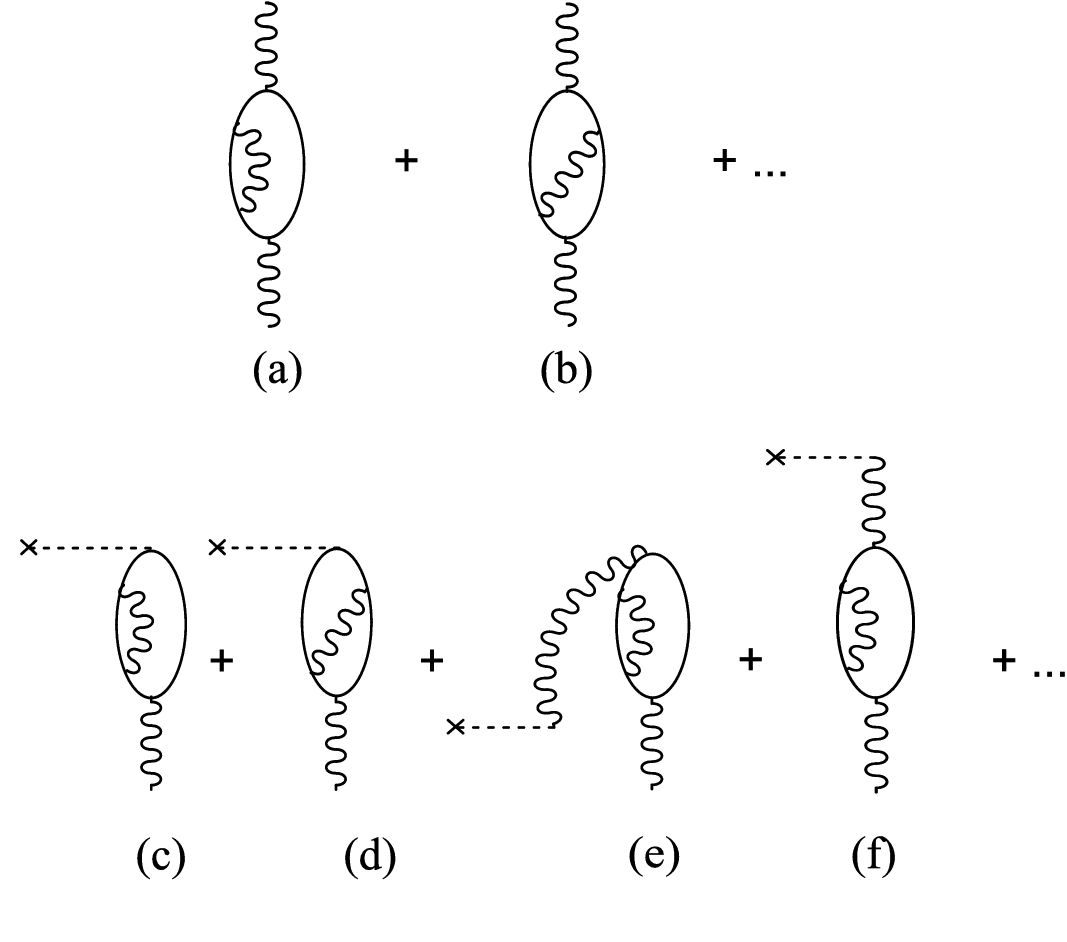}
\end{center}
\caption{
Most relevant processes taken into account in the renormalization of the
energy of the phonon (a-b) and of the associated  transition strength (c-f). }
\label{Sn122_fig2}
\end{figure}

A consistent fraction of the pairing gap (30--50\%) in medium heavy nuclei is due to a long-range pairing force arising from the
exchange of low-lying collective vibrations (see e.g. \cite{Barranco:99}, \cite{Barranco:04} and \cite{Pastore:08}). This seems to be also consistent with recent studies of pairing pro\-per\-ties of nuclear ground states making use of lowest--order contributions of chiral three--nucleon interactions (\cite{Hebeler:09}, \cite{Duguet:10}).
Such ambitious studies involve a number of steps, namely: 1) determination of the mean field from the low momentum plus 3N interactions $(v_{NN+3N})_{\textrm{low--}k}$, 2) calculation of the associated linear response, taking into account the coupling to $2p-2h$ states (see Fig. \ref{Sn122_fig2}), 3) determination, making use of these elements and of e.g. nuclear field theory diagramatic techniques, of the dressed single--particles (polarization and correlation processes), as well as of the induced pairing interaction $v_\textrm{ind}$ including vertex renormalization effects (see Fig.\ref{Sn122_fig1}), 4) use $(v_{NN+3N}({}^1S_0))_{\textrm{low--}k} + v_\textrm{ind}$ to calculate the anomalous density and $(v_{NN+3N})_{\textrm{low--}k} - (v_{NN+3N}({}^1S_0))_{\textrm{low--}k}$ to correct the normal density, 5) restart the whole process until convergence is achieved.

While such a program, although being within reach and eventually forthcoming has not yet been implemented, a number of important results have been obtained by using Skyrme interactions in the $p-h$ sector (mean field), and Gogny (or $v_{14}$ Argonne potential as in the present case, see below) in the $pp$ sector. In keeping with these results, one can argue that a quantitative description of pairing in nuclei can likely be attained by correlating pairs of nucleons through the bare nucleon-nucleon potential and the exchange of collective surface vibrations.  In what follows, we provide evidence for such a scenario in the case of typical superfluid nuclei, namely $^{119}$Sn, $^{120}$Sn and  $^{121}$Sn.

The formalism used to carry out the calculations, is based on the Dyson equation
(\cite{Terasaki:02a} and \cite{Terasaki:02b}). It can describe on equal footing the dressed one-particle state $\tilde a $  of an
odd nucleon renormalized by the (collective) response of all the other nucleons (Figs. \ref{Sn122_fig1}(a)-(d)),
the renormalization of the energy $\hbar \omega_{\nu}$ (Figs. \ref{Sn122_fig2}(a)-(b)) and of the transition probability $B(E\lambda)$ (Figs. 2(c)-2(f))
of the collective vibrations of the even system where the number of nucleons remains constant (correlated particle-hole excitations), and the induced interaction due to the exchange of collective vibrations between pairs of nucleons, moving in time reversal states close to the Fermi energy (Figs. \ref{Sn122_fig1}(e)-(g)), including both self-energy and vertex correction processes. Within this framework,  the self-consistency existing between the dynamical deformations of the density and of the potential sustained by "screened" particle-vibrations coupling vertices leads to renormalization effects which make  finite (stabilize)  the collectivity and the self-interaction of the elementary modes of nuclear excitation.
In particular of the low-lying surface vibrational modes. Such a scenario provides an accurate description of many seemingly unrelated experimental findings, in terms of very few (theoretically calculable) parameters, namely: the $k-$mass $m_k$ (\cite{Mahaux:85}) and  the particle vibration coupling vertex $h(ab\nu)$, associated to the process in which a quasiparticle changes its state of motion from the unperturbed quasiparticle state $a$ to $b$, by absorbing or emitting a vibration $\nu$ (\cite{Bohr:75},\cite{Mottelson:76},\cite{Bertsch:83} and \cite{Bortignon:98}).

The Dyson equation describing the renormalization of a quasiparticle $a$, due to this variety of couplings (see Fig.\ref{Sn122_fig1}(a)--(d)) can be written as (\cite{VanderSluys:93})

\begin{eqnarray}
\left [
\left( \begin{matrix}
E_a  &  0 \cr
0  &  -E_a \cr   \end{matrix} \right )
+
\left( \begin{matrix}
\Sigma_{11}(\tilde E_a)   &  \Sigma_{12}(\tilde E_a) \cr
 \Sigma_{12}(\tilde E_a)  &  \Sigma_{22}(\tilde E_a) \cr   \end{matrix} \right )
 \right ]
 \left ( \begin{matrix}
\tilde x_a \cr
\tilde y_a \cr \end{matrix}
\right )  =
{\tilde E_a}
\left ( \begin{matrix}
\tilde x_a \cr
\tilde y_a \cr \end{matrix}
\right ), \quad \quad \quad \quad \quad \quad \quad \quad
\label{Sn122_eq1}
\end{eqnarray}
where $\Sigma_{ii}$ and
$\Sigma_{ij} , (i \neq j) $ are the normal and abnormal self-energies.
The quantities $E_a$ denote the quasiparticle energies obtained from a
previous diagonalization of the bare nucleon-nucleon potential within
the framework of  the generalized Bogoliubov-Valatin transformation. A Skyrme interaction (Sly4 parametrization,
with $m_k \approx 0.7 m $ \cite{Chabanat:97}), is solely used to determine the properties of the  bare single-particle states. The collective vibrations in the particle-hole channel were determined by diagonalizing a separable multipole--multipole interaction (\cite{Bohr:75}, \cite{Ring:80}). In the particle-particle (pairing) channel the interactions used were the bare nucleon-nucleon $v_{14}$ Argonne potential (\cite{Wiringa:84}) and the exchange of collective vibrations.

\begin{figure}
\begin{center}
\includegraphics[width=1.0\textwidth]{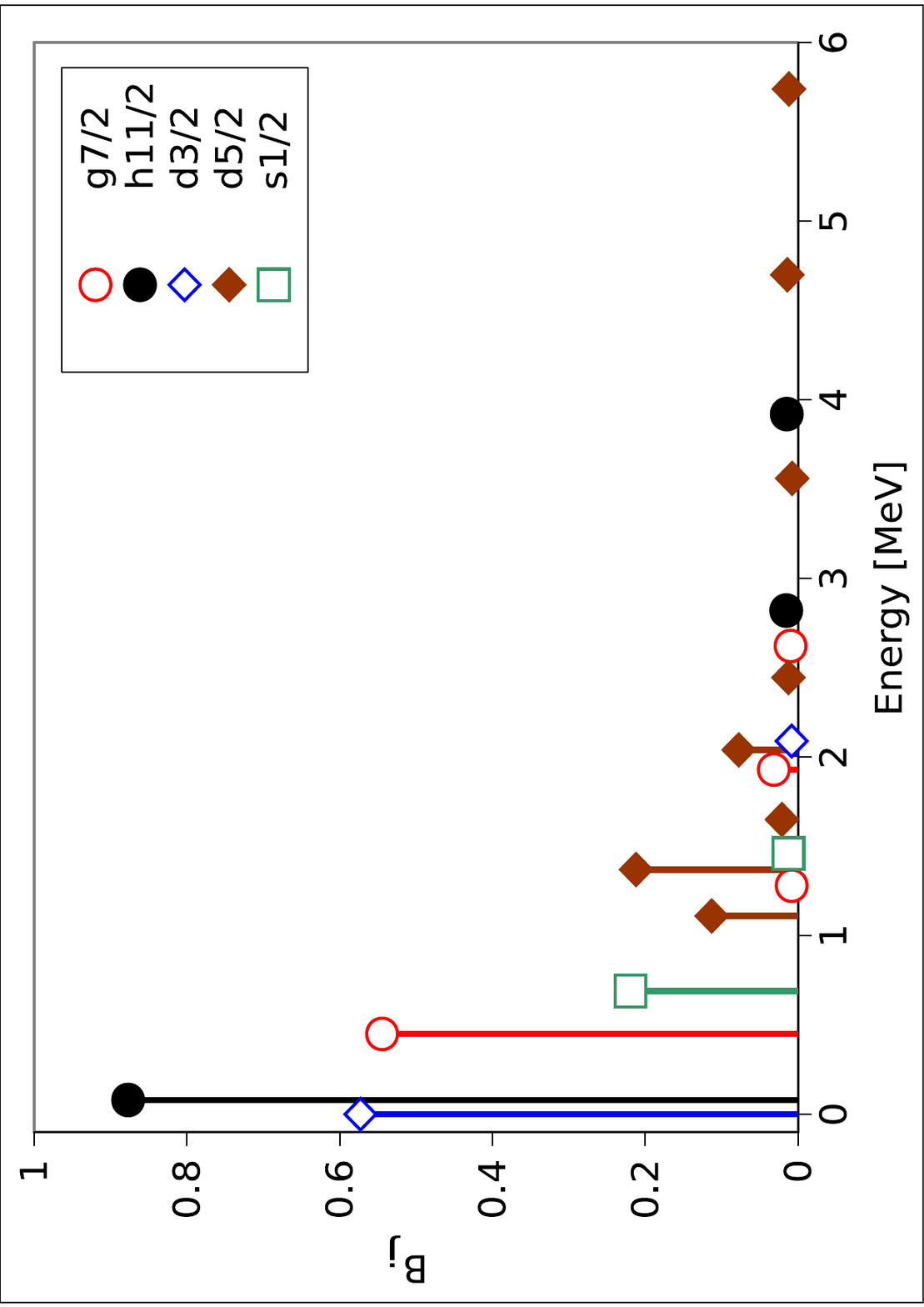}
\end{center}
\caption{Two--nucleon transfer spectroscopic amplitudes $B_j = \sqrt{j+1/2} \tilde{u_j} \tilde{v_j}$ (see also Table \ref{Sn122_tab1}), where $\tilde{u_j}=\tilde{x_j}u_j-\tilde{y_j}v_j$ and $\tilde{v_j}=\tilde{x_j}v_j+\tilde{y_j}u_j$.}
\label{Sn122_fig3}
\end{figure}

Eq. (\ref{Sn122_eq1}) is solved iteratively, and simultaneously for all the involved quasiparticle states. At each iteration step, the original quasiparticle states $a$ become fragmented over the different eigenstates $\tilde a $, a fragmentation which also affects the two--particle transfer spectroscopic amplitudes (see Table \ref{Sn122_tab1} and \ref{Sn122_fig3})


\begin{table}[h]
\begin{tabular}{|c|c|c|}
\hline
state & $E_{qp}$ & $B_j$ \\ \hline
$ 2 d_{3/2}$ & 0 & 0.57 \\ \hline
$ 1 h_{11/2}$ & 0.08 & 0.88 \\ \hline
$ 1 g_{7/2}$ & 0.45 & 0.54 \\ \hline
$ 3 s_{1/2}$ & 0.69 & 0.22 \\ \hline
$ 2 d_{5/2}$ & 1.11 & 0.12 \\ \hline
$ 1 g_{7/2}$ & 1.28 & 0.008 \\ \hline
$ 2 d_{5/2}$ & 1.37 & 0.21 \\ \hline
$ 3 s_{1/2}$ & 1.46 & 0.013 \\ \hline
$ 2 d_{5/2}$ & 1.65 & 0.021 \\ \hline
$ 1 g_{7/2}$ & 1.93 & 0.032 \\ \hline
$ 2 d_{3/2}$ & 2.02 & 0.008 \\ \hline
$ 2 d_{5/2}$ & 2.04 & 0.077 \\ \hline
$ 2 d_{5/2}$ & 2.57 & 0.013 \\ \hline
$ 1 g_{7/2}$ & 2.58 & 0.01 \\ \hline
$ 1 h_{11/2}$ & 2.82 & 0.015 \\ \hline
$ 2 d_{5/2}$ & 3.56 & 0.008 \\ \hline
$ 1 h_{11/2}$ & 3.92 & 0.015 \\ \hline
$ 2 d_{5/2}$ & 4.70 & 0.014 \\ \hline
$ 2 d_{5/2}$ & 5.74 & 0.012 \\ \hline
$ 2 d_{5/2}$ & 8.14 & 0.008 \\
\hline
\end{tabular}
\caption{Two--particle transfer spectroscopic amplitudes $B_j=\sqrt{j+1/2} \tilde{u_j} \tilde{v_j}$ (see also Fig. \ref{Sn122_fig3}) for the valence orbitals of superfluid nuclei around ${}^{120}$Sn. In the first and second columns the nlj single--particle quantum numbers and the quasiparticle energies are displayed respectively.}
\label{Sn122_tab1}
\end{table}

\begin{figure}
\begin{center}
\includegraphics[width=0.65\textwidth]{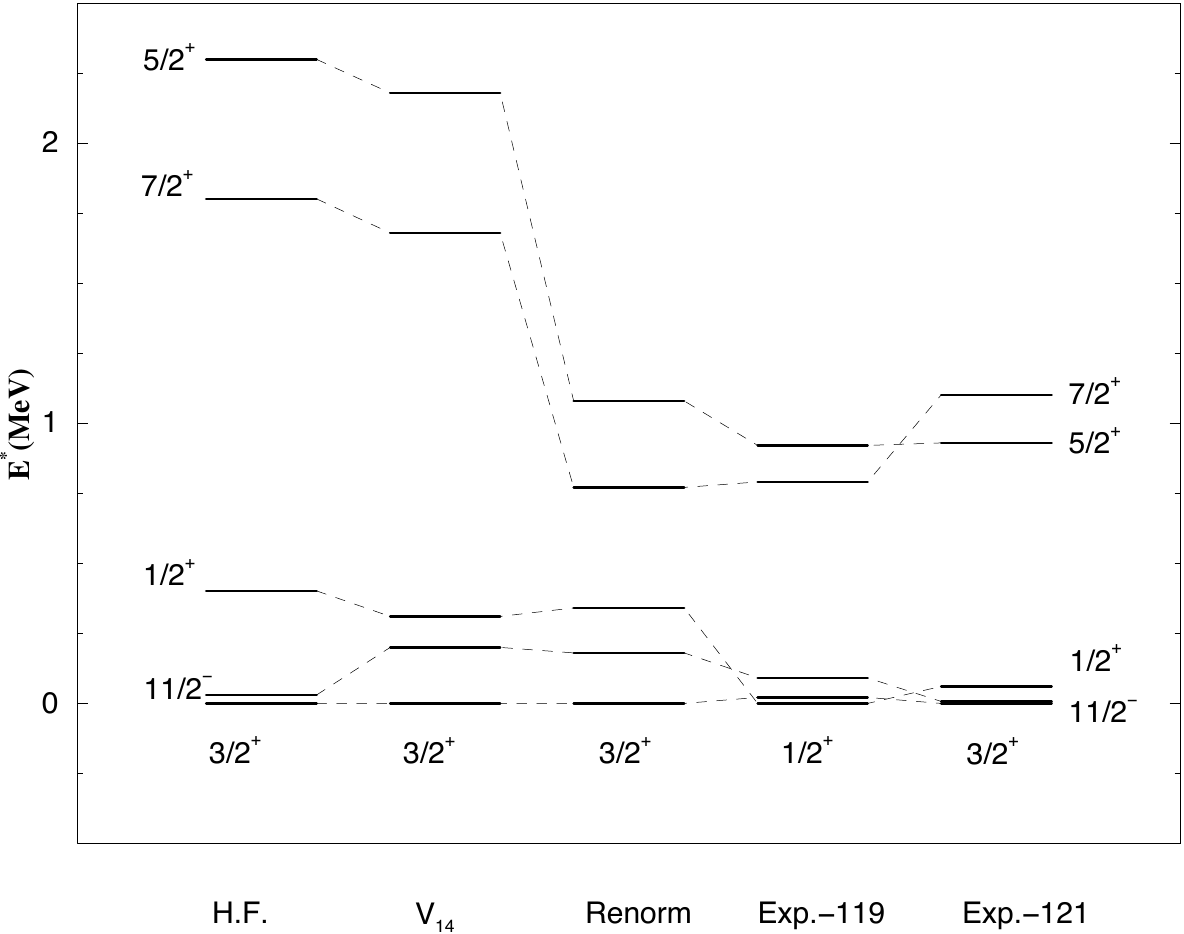}
\end{center}
\caption{
The spectra of the lowest quasiparticle states in $^{120}$Sn  calculated
using Hartree-Fock theory, BCS with the Argonne $v_{14}$ potential,
and after renormalization, are
compared to the experimental levels in the odd neighbouring nuclei $^{119}$Sn
and $^{121}$Sn.}
\label{Sn122_fig3b}
\end{figure}

\begin{figure}
\begin{center}
\includegraphics[width=0.65\textwidth]{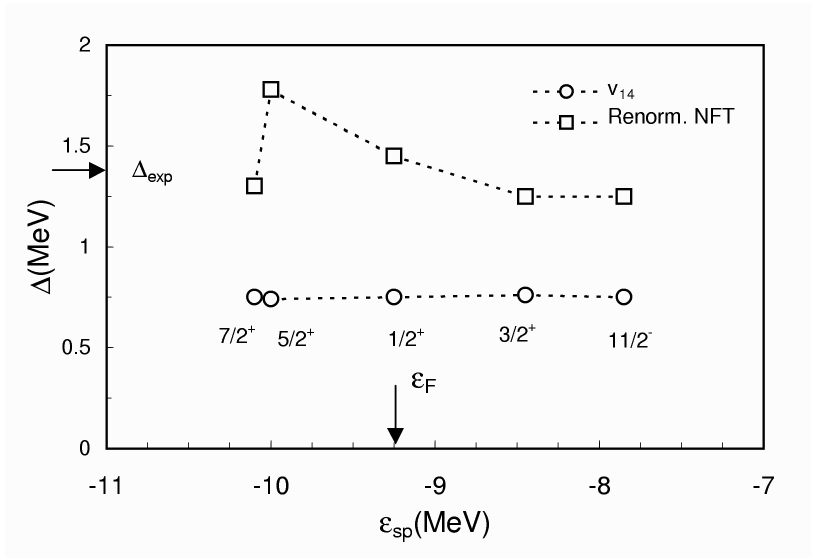}
\end{center}
\caption{
The state-dependent pairing gap for the levels close to the Fermi energy
obtained using BCS theory with the $v_{14}$ Argonne potential (circles)
is compared with the results obtained including renormalization effects
($\tilde{\Delta_j}=2\tilde{E}_j\tilde{u_j}\tilde{v_j}/(\tilde{u_j}^{2}+\tilde{v_j}^{2})$, \cite{Terasaki:02a},\cite{Schrieffer:64}).}
\label{Sn122_fig4}
\end{figure}

\begin{figure}
\begin{center}
\includegraphics[width=0.8\textwidth]{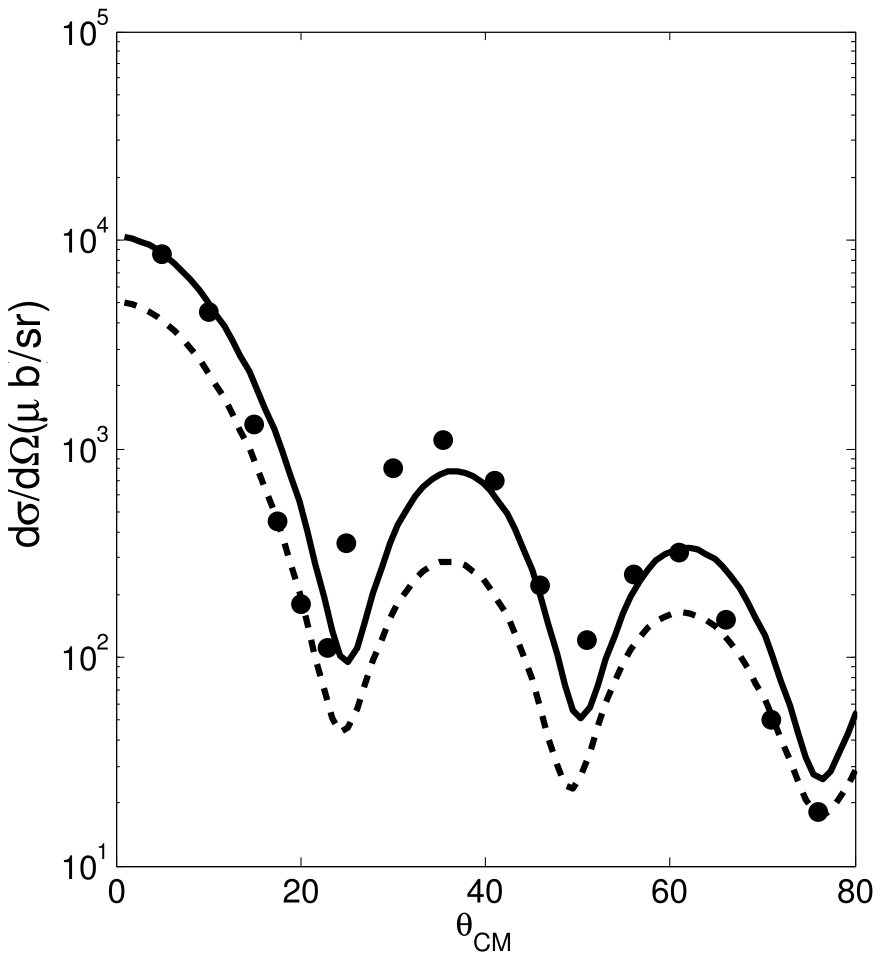}
\end{center}
\caption{Differential cross section associated with the ${}^{122}\textrm{Sn}(p,t){}^{120}\textrm{Sn}(gs)$ reaction at 26 MeV calculated making use of the spectroscopic amplitudes $B_j$ (see Table \ref{Sn122_tab1}) in comparison with the experimental data (\cite{Guazzoni:99}). The dashed line correspond to the calculation making use of just the NN $v_{14}$ potential without the induced interaction. The optical parameters used in the calculations are quoted in \cite{Guazzoni:99}.}
\label{Sn122_fig5}
\end{figure}

As seen from Fig. \ref{Sn122_fig3b},
Hartree-Fock theory is not able to account for the experimental quasiparticle energies of the low-lying states.
Diagonalizing the Argonne $v_{14}$ nucleon-nucleon potential in the
Hartree-Fock
basis, within the framework of the generalized Bogoliubov-Valatin approximation, the situation remains largely unchanged.
                                                                                                                                                                    This is consistent with the fact that $HF+v_{14}$ accounts for about half of the empirical pairing gap value ($\approx$ 1.4 MeV) obtained from the odd-even mass difference (see Fig. \ref{Sn122_fig4}). Solving (\ref{Sn122_eq1}) but this time taking into account also the induced pairing interaction arising from the exchange of vibrational mode one obtains the state dependent pairing gap labeled Renorm.NFT in Fig.\ref{Sn122_fig4} and the corresponding quasiparticle spectrum displayed in Fig.\ref{Sn122_fig3b}.
 Making use of the two-nucleon spectroscopic amplitudes associated with the full solution of Eq. (\ref{Sn122_eq1}) ($HF+v_{14}+$induced $\equiv$ Renorm.NFT) (see Table \ref{Sn122_tab1} and Fig. \ref{Sn122_fig3}), the absolute cross section associated with the ${}^{122}\textrm{Sn}(p,t){}^{120}\textrm{Sn}(gs)$ reaction at $26$MeV was calculated, making use of the optical parameter reported in \cite{Guazzoni:99}. The results in comparison with the experimental data are displayed in Fig. \ref{Sn122_fig5} and Table 6.

 Of notice that the relative importance of the bare and induced interaction contributions to the pairing gap depends, among other things, on the treatment of the $k$--dependence of $m_k$, a question which will be eventually solved through the use of $(v_{NN+3N})_{\textrm{low--}k}$. Within the present treatment of the pairing gap, one can turn the above argumentation upside down, and state that because the collectivity of vibrational modes arises from a coherent sum of two--quasiparticle states, it will be rather independent of details of the single--particle spectrum, and depend mainly on the average level density around the Fermi energy. Thus, this contribution is expected to be rather stable, provided the collective states show a weak isotopic dependence as it is the case for the Sn--isotopes.

%
%
%

\begin{table}[h]
\begin{center}
\begin{tabular}{|c|c|c|}
\cline{2-3}
\multicolumn{1}{c|}{}                  & \multicolumn{2}{c|}{$\sigma(\mu$b)}\\ \cline{2-3}
\multicolumn{1}{c|}{}                  & Theory & Experiment \\
\hline
Renorm.NFT       & 2466        & \multirow{2}{*}{$2505 \pm 376 (\pm 18.02)$}\\ \cline{1-2}
$v_{14}$         & 969         &\\
\hline
\end{tabular}
\end{center}
\caption{Absolute cross section in $\mu$b associated with the reaction ${}^{122}\textrm{Sn}(p,t){}^{120}\textrm{Sn}(gs)$ at an incident proton energy of 26 MeV integrated in the angular range $5^{\circ}$ to $75^{\circ}$. In the third column the experimental value is reported (\cite{Guazzoni:99}) together with the estimated systematic error of 15$\%$. The quantity quoted in parenthesis is the statistical error of the measurement. In the second column the theoretical predictions displayed corresponds to the full solution of the Dyson (see Eq. \ref{Sn122_eq1}) equation taking properly into account both the bare and the induced interaction (Renorm.NFT) as well as the solution associated with only the bare $v_{14}$ NN--potential (see also Figs. \ref{Sn122_fig3b} and \ref{Sn122_fig4}).}
\label{Sn122_tab2}
\end{table}

\section{The $^{112}$Sn($p,t$)$^{110}$Sn (gs) reaction}
\begin{figure}
\centerline{\includegraphics*[width=13cm,angle=0]{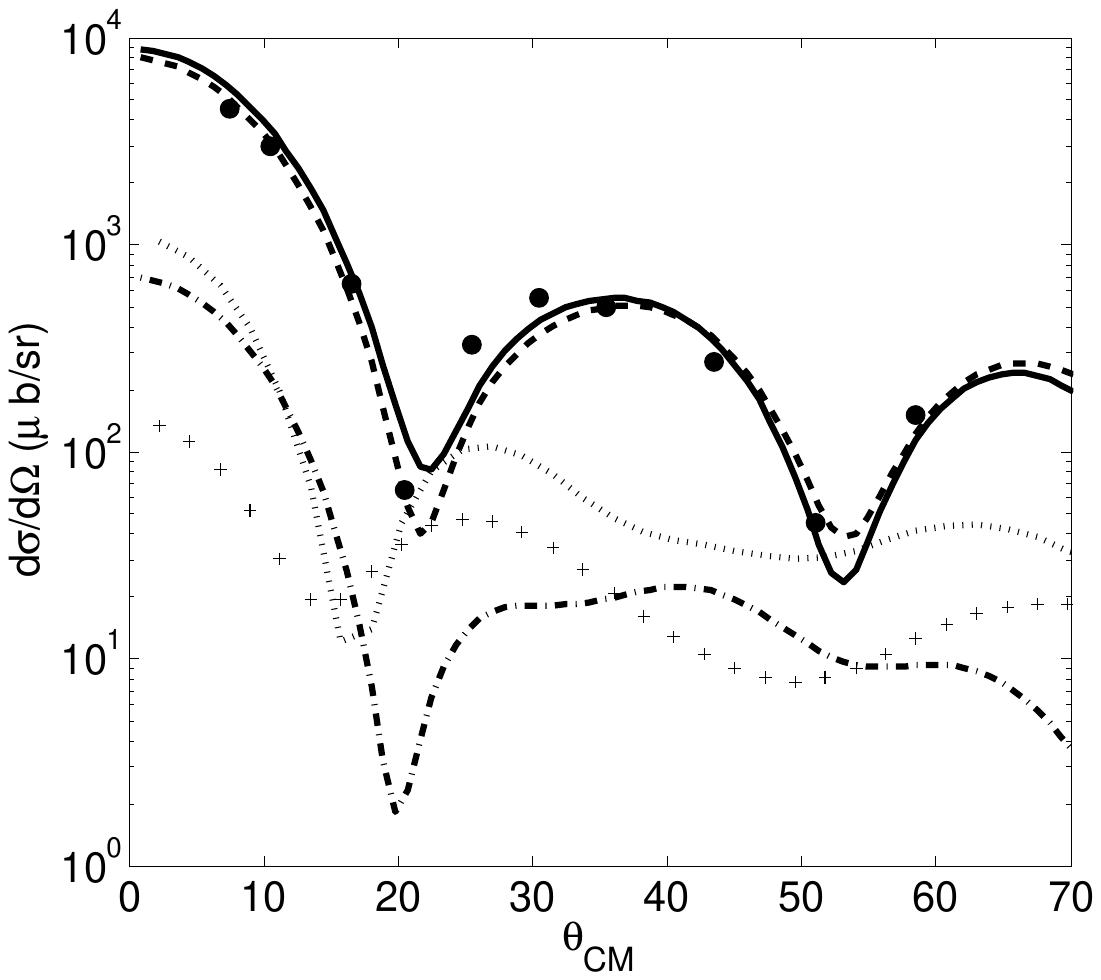}}
\caption{Differential cross section associated with the $^{112}$Sn($p,t$)$^{110}$Sn (gs) reaction at an incident proton energy of \mbox{26 MeV}. The dots represent the experimental data (\cite{Guazzoni:06}), the solid curve the total theoretical prediction. The other curves represent the successive (dashed), simultaneus (dotted-dashed), non--orthogonality (dotted) and simultaneus+non--ortogonality (crosses) contributions. The optical potential used in the calculation is quoted in this reference.}\label{fig18}
\end{figure}

In ref. \cite{Guazzoni:06} a study of the spectroscopy of $^{110}$Sn via the $(p,t)$ reaction on $^{112}$Sn at an incident proton energy of 26 MeV was reported. The data was analyzed in term of spectroscopic amplitudes obtained by diagonalizing  a renormalized low momentum potential in a shell model basis. This $v_{low-k}$ potential was derived from the CD--Bonn nucleon--nucleon potential (\cite{Machleidt:01}). It was then used to derive a two--body interaction to act between the valence neutrons  (see \cite{Kuo:71}, \cite{Suzuki:80}, \cite{Bogner:02}). It was assumed in (see \cite{Coraggio:04}) that $^{100}$Sn is a closed core, the valence nucleon being allowed to move on the $0g_{7/2},1d_{5/2},1d_{3/2},2s_{1/2}$ and $0h_{11/2}$ orbitals. The energy of these levels were taken from the results of an analysis of low--energy spectra of light odd Sn--isotopes. The resulting values of the two--nucleon transfer spectroscopic amplitudes were reported in Table IV of \cite{Guazzoni:06}; see also \cite{Covello:97} and \cite{Andreozzi:96} and the optical parameters reported in the same reference. Making use of these values we have calculated, taking into account successive, simultaneous and non--orthogonality contributions as explained above in the paper, the gs$\rightarrow$ gs absolute differential cross section. The results of the calculation in comparison with the experimental data are reported in Fig. \ref{fig18} and Table \ref{Sn112_tab1}.

\begin{table}[h]
\begin{center}
\begin{tabular}{|c|c|}
\hline
\multicolumn{2}{|c|}{$\sigma(\mu$b)}\\
\hline
Theory & Experiment \\
\hline
1301        & $1309 \pm 200 (\pm 14)$\\
\hline
\end{tabular}
\end{center}
\caption{Absolute cross section in $\mu$b associated with the reaction ${}^{112}\textrm{Sn}(p,t){}^{110}\textrm{Sn}(gs)$ at an incident proton energy of 26 MeV integrated in the angular range $6^{\circ}$ to $57.5^{\circ}$. In the second column the experimental value is reported (\cite{Guazzoni:06}) together with the estimated systematic error of 15$\%$. The error quoted in parenthesis is the statistical error of the measurement. In the first column the theoretical prediction is displayed.}
\label{Sn112_tab1}
\end{table}

\section{The $^{208}$Pb($t,p$)$^{206}$Pb (gs) reaction: pairing in normal nuclei}

The basic property associated with superfluid nuclei, is the existence of a finite value $\alpha_0=<0|P^{\dagger}|0>(=\sum_{\nu}u_{\nu}v_{\nu}$ within BCS theory) of the pair transferred operator $P^{\dagger}=\sum_{\nu>0}a^{\dagger}_{\bar{\nu}}a^{\dagger}_{\nu}$ ($P=\sum_{\nu>0}a_{\nu}a_{\bar{\nu}}$), in the (mean field) ground state, implying an interweaving of particle and hole degrees of freedom around the Fermi energy, as testified by the structure of quasiparticle excitations ($\alpha^{\dagger}_{\nu}=u_{\nu}a^{\dagger}_{\nu}-v_{\nu}a_{\bar{\nu}}$). This scenario implies that the superfluid nucleus defines a privileged orientation in gauge space. Fluctuation in particle number associated with the interaction among quasiparticle states, proportional to the field ($u^{2}_{\nu}+v^{2}_{\nu}$) restores gauge symmetry (see e.g. \cite{Bes:66} and \cite{Broglia:00}).

Around closed shell nuclei, while $\alpha_0=0$, the standard deviation $\sigma=\linebreak =(\sum_{int}<0|P^{\dagger}|int><int|P|0>)^{1/2}=(\sum_{int}|<int|P|0>|^{2})^{1/2}$ is, as a rule, large. In particular in the case in which $|0>=|{}^{208}Pb(gs)>$.
In other words, the $|{}^{206}Pb(gs)>$ can be viewed as a vibrational mode of $|{}^{208}Pb(gs)>$ which change particle number by 2. Similarly, the $|{}^{210}Pb(gs)>$ is interpreted as the pair addition mode of $|{}^{208}Pb(gs)>$.
Pairing vibrational bands around closed shell nuclei have been studied in detail, in particular that around ${}^{208}$Pb (see e.g. \cite{Flynn:72}, \cite{Broglia:73}).

Within the RPA approximation one can calculate the forwards-- and backwards--going amplitudes X and Y and thus the two--particle transfer spectroscopic amplitudes. In the case of the reaction $^{206}$Pb$(t,p)^{208}$Pb$(gs)$ (pair removal mode) one can write

\begin{equation}\label{Eq15}
 B_{nlj}=\sum_{i}X_{i}\delta_{nlj,i}-\sum_{k}Y_{k}\delta_{nlj,k}
\end{equation}
where $i$ label the states below N=126 shell closure (i.e. $3p_{1/2}$,$2f_{5/2}$,...) while $k$ those above (i.e. $2g_{9/2}$,$1i_{11/2}$,...). Making use of the RPA wavefunctions obtained following the prescription of e.g. \cite{Broglia:73} (also \cite{Broglia:67}), the B-coefficients listed in Table \ref{Tab_RPA} are worked out.
Of notice that the RPA wavefunction of the pair removal mode ($|{}^{206}Pb(gs)>$) is normalized according to $\sum_{i} X_{i}^2-\sum_k Y_{k}^2=1$, and that the second term in Eq.(\ref{Eq15}) is associated with ground state correlations. Neglecting it, and using as normalization condition $\sum_{i} X_{i}^2=1$, one obtains the so called Tamm--Dancoff approximation to the B--coefficients.

\begin{figure}
\centerline{\includegraphics*[width=13cm,angle=0]{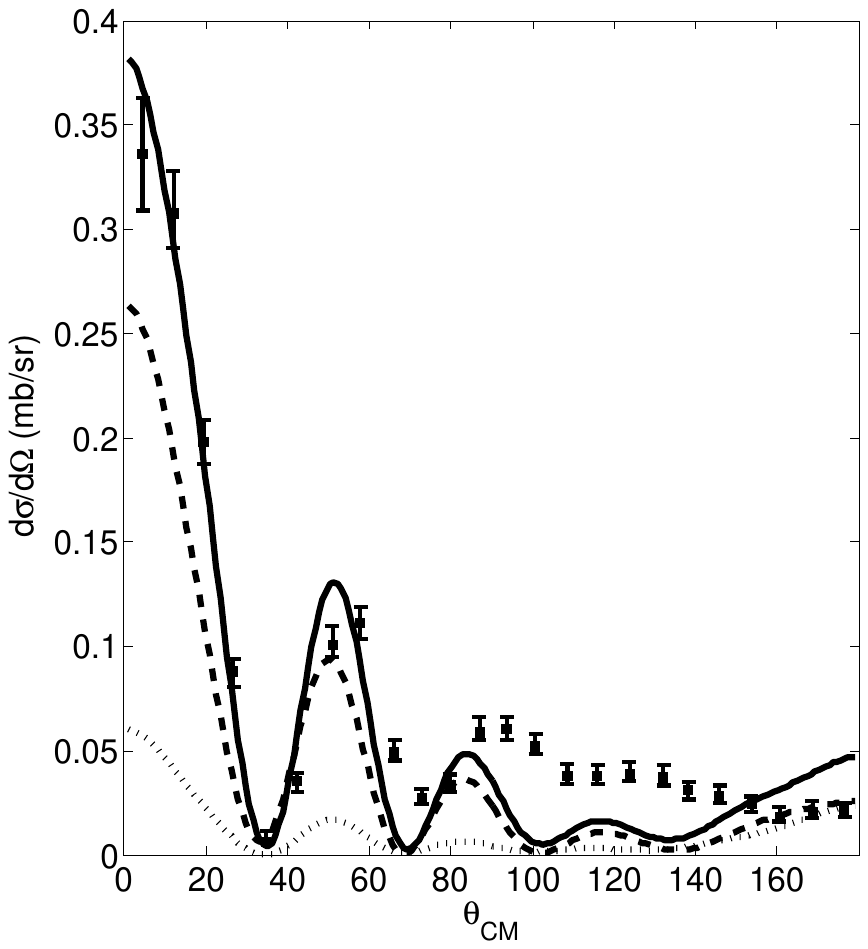}}
\caption{Differential cross section associated with the reaction $^{208}$Pb($t,p$)$^{206}$Pb (gs) at an incident triton energy of \mbox{12 MeV} (\cite{Bjerregaard:66}). The theoretical predictions (continuous curve) were worked out making use of the two--nucleon spectroscopic amplitudes displayed in Table \ref{Tab_RPA} taking into account ground state correlations (column labeled RPA of Table \ref{Tab_RPA}). The results neglecting this contributions (TD entry of Table \ref{Tab_RPA}) are shown with a dashed curve. Those calculated making use of the pure $p_{1/2}^{-2}(0)$ configuration are displayed with a dotted curve.  The optical parameters used in the calculations were determined from elastic scattering data and are quoted in \cite{Broglia:67}.}
\label{Fig_CS}
\end{figure}

Making use of these two--nucleon transfer spectroscopic amplitudes and of the optical parameters reported in the above two references, the absolute differential cross sections where calculated (see Fig.\ref{Fig_CS}). Integrating these cross sections in the angular range 4.5--176.5 degrees, one obtains the value reported in Table \ref{Tab_exp} in comparison with the experimental findings.

\begin{table}[h]
\begin{tabular}{|c|c c|}
\hline
              & \multicolumn{2}{c|}{$B_{nlj}$} \\ \hline
state $nlj$   & RPA  & (TD)   \\ \hline
$ 1 h_{9/2}$  & 0.15 & (0.14) \\
$ 2 f_{7/2}$  & 0.21 & (0.26) \\
$ 1 i_{13/2}$ & 0.29 & (0.28) \\
$ 3 p_{3/2}$  & 0.23 & (0.22) \\
$ 2 f_{5/2}$  & 0.32 & (0.31) \\
$ 3 p_{1/2}$  & 0.89 & (0.85) \\ \hline

$ 2 g_{9/2}$  & 0.18 &        \\
$ 1 i_{11/2}$ & 0.15 &        \\
$ 1 j_{15/2}$ & 0.13 &        \\
$ 3 d_{5/2}$  & 0.06 & (--)   \\
$ 4 s_{1/2}$  & 0.06 &        \\
$ 2 g_{7/2}$  & 0.10 &        \\
$ 3 d_{3/2}$  & 0.05 &        \\
\hline
\end{tabular}
\caption{Two--nucleon transfer spectroscopic amplitudes (Eq. (\ref{Eq15})) calculated taking into account ground state correlations (RPA) and neglecting them (TD) (see \cite{Broglia:67}).}
\label{Tab_RPA}
\end{table}

\begin{table}[h]
\begin{center}
\begin{tabular}{|c|c|c|}
       \cline{2-3}
\multicolumn{1}{c|}{}    &  \multicolumn{2}{c|}{ $\sigma (m$b)}\\
       \cline{2-3}
\multicolumn{1}{c|}{}    & Theory & Experiment \\
\hline
RPA &  0.52  &  \multirow{3}{*}{$0.68 \pm 0.20$} \\
\cline{1-2}
TD  &  0.34  &                                 \\
\cline{1-2}
$p_{1/2}^{-2}(0)$ & 0.08 & \\
\hline
\end{tabular}
\end{center}
\caption{Cross section associated with the $^{208}$Pb($t,p$)$^{206}$Pb$(gs)$ reaction at a triton bombarding energy of 12 MeV, integrated in the angular range 4.5--176.5 degrees. The systematic errors of the experimental data (\cite{Bjerregaard:66}) are estimated to be 30$\%$. The three theoretical entries correspond to the results including (RPA) and not including (TD) ground state correlations, as well as considering the ground state of ${}^{206}$Pb as a pure $p_{1/2}^{-2}(0)$ configuration.}
\label{Tab_exp}
\end{table}

\clearpage

\section{The $^{208}$Pb($^{16}$O,$^{18}$O)$^{206}$Pb (gs) reaction: heavy ion processes}

\begin{figure}
\centerline{\includegraphics*[width=13cm,angle=0]{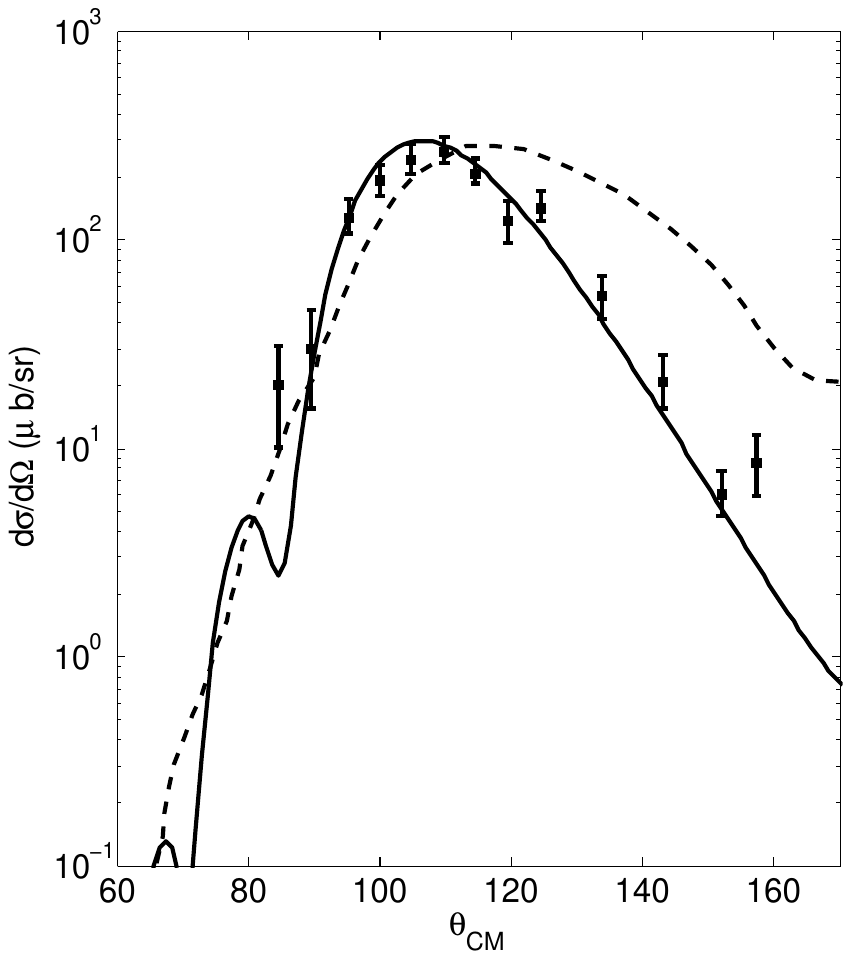}}
\caption{Differential cross section for the reaction $^{208}$Pb($^{16}$O, $^{18}$O)$^{206}$Pb at an energy of 86 MeV of the $^{16}$O in the laboratory frame, expressed in $\mu$b$/sr$. We present the results of the quantal (continuous line) and semiclassical (dashed line) calculations, along with the experimental data (dots), (\cite{Lilley:83}). The optical potentials used in the calculations are displayed in Table \ref{Tab2}.}
\label{fig17}
\end{figure}

We now turn our attention to the results obtained in the analysis of the heavy--ion reaction $^{208}$Pb($^{16}$O, $^{18}$O)$^{206}$Pb at 86 MeV $^{16}${O} bombarding energy. As the wavelength of the relative motion is relatively short at this energy ($\lambda\simeq0.8$ fm), the semiclassical scheme is applicable. In Fig. \ref{fig17} we display the corresponding results for the transition to the $^{206}$Pb ground state worked out making use of the $B_j$--coefficient displayed in Tables \ref{Tab_RPA} ($^{206}$Pb; RPA) and \ref{Tab_RPA2} ($^{18}$O (gs)) in comparison with the results worked out within the framework of a fully quantal theory, (second order DWBA formalism) and making use of the optical potential displayed in Table \ref{Tab2} in comparison with the experimental data (\cite{vonOertzen:83b},\cite{Lilley:83}, see also \cite{Bayman:82}).
 \begin{table}[h]
  \begin{center}
   \begin{tabular}{|c|c|}
    \hline
    $nlj$ & $B_{nlj}$ \\ \hline
     $1d_{5/2}$     &  0.89     \\
     $2s_{1/2}$     &  0.45     \\ \hline
   \end{tabular}
  \end{center}
  \caption{Two--particle transfer spectroscopic amplitudes associated with the transfer of two neutron between $|^{16}O$(gs)$>$ and $|^{18}O$(gs)$>$.}
  \label{Tab_RPA2}
 \end{table}

\begin{table}[h]
\begin{center}
\begin{tabular}{|c|c|c|c|c|c|}
\hline
$V$ & $r_V$ & $a_V$ & $W$ & $r_W$ & $a_W$\\ \hline
65 & 1.35 & 0.34 & 45 & 1.34 & 0.33\\
\hline
\end{tabular}
\end{center}
\caption[Parameters of the optical potentials]{Parameters of the optical potential used in the calculation of the reaction $^{208}$Pb($^{16}$O,$^{18}$O)$^{206}$Pb (\protect\cite{Bayman:82}).}\label{Tab2}
\end{table}

%

\begin{table}[h]
\begin{center}
\begin{tabular}{|c|c|c|}
\cline{2-3}
\multicolumn{1}{c|}{} & \multicolumn{2}{c|}{$\sigma(m$b)}\\
\cline{2-3}
\multicolumn{1}{c|}{} & Theory & Experiment \\
\hline
 Quantal             & 0.80   & \multirow{2}{*}{$0.76 \pm 0.23$}  \\
\cline{1-2}
 Semiclassical       & 1.02   & \\
\hline
\end{tabular}
\end{center}
\caption{Cross section associated with the reaction $^{208}$Pb($^{16}$O,$^{18}$O)$^{206}$Pb(gs) integrated in the angular range from 84.5 to 158.5 degrees. The systematic errors are estimated to be 30\%.}
\label{Tab4}
\end{table}

\section{Conclusions}
Examples of studies of pairing in nuclei with the help of two--nucleon transfer reaction have been discussed. They cover both single-- as well a many--Cooper pair systems. Those corresponding to the first group range from light, weakly bound, drip--line systems to highly stable, near closed shell nuclei lying along the stability valley. \mbox{Sn--isotopes} provide the embodiment of many (but still few) Cooper pair--open--shell systems lying along the valley of stability.

\begin{figure}
\centerline{\includegraphics*[width=13cm,angle=0]{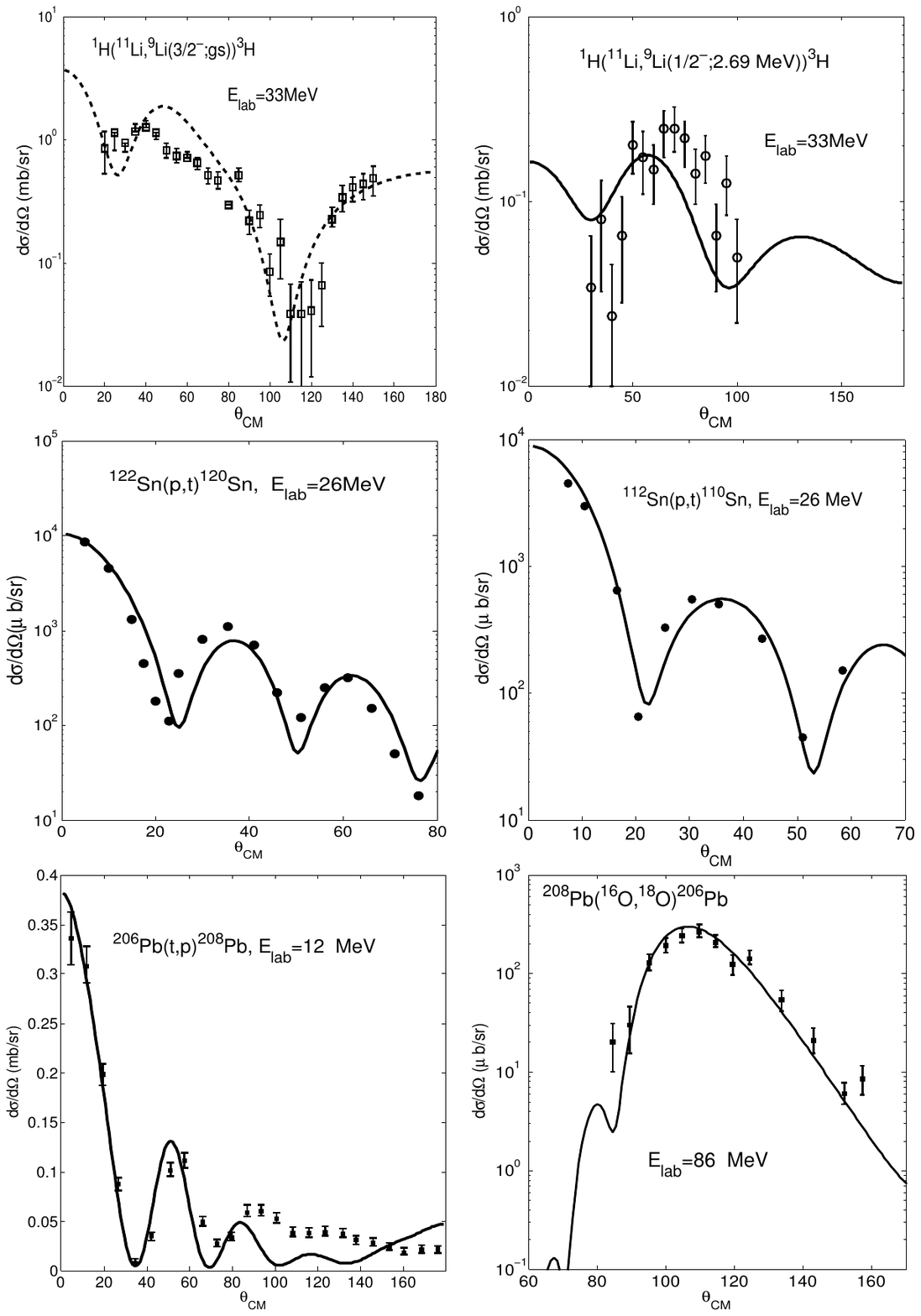}}
\caption{Summary of the absolute differential cross section predictions in comparison with the experimental data (see Figs. \ref{fig2}, \ref{Sn122_fig5}, \ref{fig18}, \ref{Fig_CS} and \ref{fig17}).}
\label{Fig_Last}
\end{figure}

As it emerges from the previous narrative and as can be seen from Fig. \ref{Fig_Last} and Table \ref{Tab_Last}, theoretical predictions reproduce the data within experimental errors without free parameters. This is a consequence of the use of reliable optical parameters for entrance, intermediate and exit channels and to the treatment, on equal footing, of the structure and of the reaction aspects of the phenomena under discussion. Within this scenario, it is only a question of time before the optical potential becomes routine part of the reaction--structure computational output/input.

It is well established that single Cooper pair transfer is the specific tool to probe pairing correlations in nuclei. This fact translates itself through structure--reaction calculations, in the fact that the absolute value of two--particle transfer cross sections are a result of the interweaving of a number of structure amplitudes and of single--particle reaction form factors. The interference of such contributions can lead to important amplifications of the physical effects which are at the basis of nuclear pairing. In particular, the relative role played by bare and by induced pairing interactions, as well as that played by ground state correlations, in connection with the structure and stability of nuclear Cooper pairs.

Particularly revealing examples of the validity of the above scenario are provided by:\\
1) the absolute cross section associated with the first excited state of \mbox{${}^{9}$Li} in the \mbox{${}^{11}\textrm{Li}\left( {}^{1}\textrm{H}, {}^{3}\textrm{H} \right) {}^{9}\textrm{Li}(1/2^{-};2.69$MeV)} reaction: a $10^{-2}$ probability of the component \linebreak $|(s_{1/2},d_{5/2})_{2^{+}} \otimes 2^{+}; 0>$ in the ${}^{11}$Li ground state leads to an order of magnitude increase of \mbox{$\sigma({}^{11}Li$(gs)$\rightarrow{}^{9}Li(1/2^{-}))$,} from a value of $5\times10^{-2}m$b to a value of $0.7m$b (exp. value $1.0 \pm 0.36 m$b), providing direct evidence of phonon mediated pairing in nuclei;\\
2) the change in the absolute cross section associated with the reaction \linebreak \mbox{${}^{122}\textrm{Sn}(p,t){}^{120}$Sn(gs)} from $969\mu$b to $2466\mu$b (exp.value $2505 \pm 376 \mu$b), when phonon mediated pairing is added to the bare NN--interaction in the calculation of the many Cooper pair ground state wavefunctions of the Sn--isotopes;\\
3) the absolute cross section change of the reaction  $^{208}$Pb($t,p$)$^{206}$Pb(gs) from $340 \mu$b to $520 \mu$b (exp. value $680 \pm 210 \mu$b), by including in the $|{}^{206}Pb$(gs)$>$ state, about 9\% of ground state correlations.

Arguably, the above results are likely to signal, if not the starting of the ``exact'' era of nuclear pairing studies, in any case the end of the qualitative one which was mainly based on relative two--particle transfer reactions cross section calculations.

We want to thank Ben Bayman for discussions concerning the construction of the two--particle transfer code. Financial support from the Ministry of Science and Innovation of Spain grants FPA2009--07653 and ACI2009--1056 are acknowledge by FB and GP and by FB respectively.

\begin{table}
 \begin{center}
  \begin{tabular}{|c|c|c|c|}
\cline{2-4}
\multicolumn{1}{c|}{}                                                                & \multicolumn{3}{|c|}{$\sigma($gs$\rightarrow$f)}              \\
\cline{2-4}
\multicolumn{1}{c|}{}                                                                & f        & Theory         & Experiment                        \\
\hline
\multirow{2}{*}{${}^{11}$Li$\left({}^{1}\textrm{H},{}^{3}\textrm{H}\right){}^{9}$Li} & gs       & 6.1 ${}^{a)}$  & $5.7 \pm 0.9$ ${}^{a)}$           \\
\cline{2-4}
                                                                                     & $1/2^{-}$& 0.7 ${}^{a)}$  & $1.0 \pm 0.36$ ${}^{a)}$          \\
\hline
${}^{122}\textrm{Sn}(p,t){}^{120}$Sn                                                 & gs       & 2466 ${}^{b)}$ & $2505 \pm 376 (\pm 18)$ ${}^{b)}$ \\
\hline
$^{112}$Sn($p,t$)$^{110}$Sn                                                          & gs       & 1301 ${}^{b)}$ & $1309 \pm 200 (\pm 14)$ ${}^{b)}$ \\
\hline
$^{208}$Pb($t,p$)$^{206}$Pb                                                          & gs       & 0.52 ${}^{a)}$ & $0.68 \pm 0.21$ ${}^{a)}$         \\
\hline
$^{208}$Pb($^{16}$O,$^{18}$O)$^{206}$Pb                                              & gs       & 0.80 ${}^{a)}$  & $0.76 \pm 0.18$ ${}^{a)}$         \\
\hline

  \end{tabular}
 \end{center}
 \caption{Summary of the absolute two--particle transfer cross sections predictions in comparison with the experimental data (see Tables \ref{tab2}, \ref{Sn122_tab2}, \ref{Sn112_tab1}, \ref{Tab_exp} and \ref{Tab4}). The superscripts ${}^{a)}$ and ${}^{b)}$ indicate that the cross sections are measured in $m$b and in $\mu$b respectively.}
 \label{Tab_Last}
\end{table}

\bibliographystyle{apalike}

\begin{thebibliography}{}

\bibitem[Ajzenberg-Selove, 1988]{AjzenbergSelove:88}
Ajzenberg-Selove, F. (1988).
\newblock Energy levels of light nuclei {A} = 5-10.
\newblock {\em Nucl. Phys. A}, 490:1.

\bibitem[Al-Khalili and Tostevin, 1996]{Al-Khalili:96}
Al-Khalili, J.~S. and Tostevin, J.~A. (1996).
\newblock Matter radii of light halo nuclei.
\newblock {\em Phys. Rev. Lett.}, 76:3903.

\bibitem[An and Cai, 2006]{An:06}
An, H. and Cai, C. (2006).
\newblock Global deuteron optical model potential for the energy range up to
  183 {M}e{V}.
\newblock {\em Phys. Rev. C}, 73:054605.

\bibitem[Anderson, 1958]{Anderson:58}
Anderson, P.~W. (1958).
\newblock Random--{P}hase {A}pproximation in the theory of superconductivity.
\newblock {\em Phys. Rev.}, 112:1900.

\bibitem[Anderson, 1959]{Anderson:59}
Anderson, P.~W. (1959).
\newblock Theory of dirty superconductors.
\newblock {\em J. Phys Chem. Solids}, 11:26.

\bibitem[Anderson, 1969]{Anderson:69}
Anderson, P.~W. (1969).
\newblock Superconductivity in the past and in the future.
\newblock In Parks, R.~D., editor, {\em Superconductivity}, volume~2, page
  1343, New York. Marcel Dekker, Inc.

\bibitem[Anderson, 1994]{Anderson:94}
Anderson, P.~W. (1994).
\newblock {\em A career in theoretical physics}.
\newblock World Scientific, Singapore.

\bibitem[Andreozzi et~al., 1996]{Andreozzi:96}
Andreozzi, F., Coraggio, L., Covello, A., Gargano, A., Kuo, T. T.~S., Li,
  Z.~B., and Porrino, A. (1996).
\newblock Realistic shell-model calculations for neutron deficient {S}n
  isotopes.
\newblock {\em Phys. Rev. C}, 54:1636.

\bibitem[Aoi et~al., 1997]{Aoi:97}
Aoi, N., Yoneda, K., Miyatake, H., Ogawa, H., Yamamoto, Y., Ideguchi, E.,
  Kishida, T., Nakamura, T., Notani, M., Sakurai, H., Teranishi, T., Wu, H.,
  Yamamoto, S.~S., Watanabe, Y., Yoshida, A., and Ishihara, M. (1997).
\newblock {$\beta$--spectroscopy of $^{11}$Li and $^{14}$Be --With a
  $\beta$--n--$\gamma$ triple coincidence method}.
\newblock {\em Nucl. Phys. A}, 616:181c.

\bibitem[Ascuitto and Glendenning, 1969]{Ascuitto:69}
Ascuitto, R.~J. and Glendenning, N.~K. (1969).
\newblock Inelastic processes in particle transfer reactions.
\newblock {\em Phys. Rev.}, 181:1396.

\bibitem[Ascuitto et~al., 1971]{Ascuitto:71}
Ascuitto, R.~J., Glendenning, N.~K., and S{\o}rensen, B. (1971).
\newblock Confirmation of strong second order processes in $(p, t)$ reactions
  on deformed nuclei.
\newblock {\em Phys Lett B}, 34:17.

\bibitem[Aumann, 2005]{Aumann:05}
Aumann, T. (2005).
\newblock Reactions with fast radioactive beams of neutron-rich nuclei.
\newblock {\em Eur. Phys. J A}, 26:441.

\bibitem[Bachelet et~al., 2008]{Bachelet:08}
Bachelet, C., Audi, G., Gaulard, C., Gu\'{e}naut, C., Herfurth, F., Lunney, D.,
  de~Saint~Simon, M., and Thibault, C. (2008).
\newblock {New Binding Energy for the Two-Neutron Halo of $^{11}$Li}.
\newblock {\em Phys. Rev. Lett.}, 100:182501.

\bibitem[Ball et~al., 2011]{Ball:11}
Ball, G., Buchmann, L., Davids, B., Kanungo, R., Ruiz, C., and Svensson, C.
  (2011).
\newblock Physics with reaccelerated radioactive beams at {TRIUMF-ISAC}.
\newblock {\em J. Phys. G}, 38:024003.

\bibitem[Bardeen et~al., 1957a]{Bardeen:57a}
Bardeen, J., Cooper, L.~N., and Schrieffer, J.~R. (1957a).
\newblock Microscopic theory of superconductivity.
\newblock {\em Phys. Rev.}, 106:162.

\bibitem[Bardeen et~al., 1957b]{Bardeen:57b}
Bardeen, J., Cooper, L.~N., and Schrieffer, J.~R. (1957b).
\newblock Theory of superconductivity.
\newblock {\em Phys. Rev.}, 108:1175.

\bibitem[Barranco et~al., 2001]{Barranco:01}
Barranco, F., Bortignon, P.~F., Broglia, R.~A., Col{\`{o}}, G., and Vigezzi, E.
  (2001).
\newblock The halo of the exotic nucleus $^{11}${Li}: a single {C}ooper pair.
\newblock {\em Europ. Phys. J. A}, 11:385.

\bibitem[Barranco et~al., 2004]{Barranco:04}
Barranco, F., Broglia, R.~A., Col\`{o}, G., Gori, G., Vigezzi, E., and
  Bortignon, P.~F. (2004).
\newblock Many-body effects in nuclear structure.
\newblock {\em Eur. Phys. J. A}, 21:57.

\bibitem[Barranco et~al., 1997]{Barranco:97}
Barranco, F., Broglia, R.~A., Esbensen, H., and Vigezzi, E. (1997).
\newblock Role of finite nuclei on the pairing gap of the inner crust of
  neutron stars.
\newblock {\em Physics Letters B}, 390:13.

\bibitem[Barranco et~al., 1999]{Barranco:99}
Barranco, F., Broglia, R.~A., Gori, G., Vigezzi, E., Bortignon, P.~F., and
  Terasaki, J. (1999).
\newblock Surface vibrations and the pairing interaction in nuclei.
\newblock {\em Phys. Rev. Lett.}, 83:2147.

\bibitem[Barranco et~al., 1993]{Barranco:93}
Barranco, F., Vigezzi, E., and Broglia, R.~A. (1993).
\newblock Nuclear effects in the break-up of $^{11}${L}i.
\newblock {\em Phys. Lett. B}, 319:387.

\bibitem[Bayman, 1960a]{Bayman:60a}
Bayman, B. (1960a).
\newblock A derivation of the pairing-correlation method.
\newblock {\em Nucl. Phys.}, 15:33.

\bibitem[Bayman, 1960b]{Bayman:60b}
Bayman, B.~F. (1960b).
\newblock {\em Seniority, Quasiparticles and Collective Vibrations, {Lecture}
  notes at the Palmer Physical Laboratory, Princeton University}.
\newblock {Unpublished}.

\bibitem[Bayman, 1971]{Bayman:71}
Bayman, B.~F. (1971).
\newblock Finite--range calculation of the two--neutron transfer reaction.
\newblock {\em Nucl. Phys. A}, 168:1.

\bibitem[Bayman and Chen, 1982]{Bayman:82}
Bayman, B.~F. and Chen, J. (1982).
\newblock One-step and two-step contributions to two-nucleon transfer
  reactions.
\newblock {\em Phys. Rev. C}, 26:1509.

\bibitem[Becha et~al., 1997]{Becha:97}
Becha, M.~B., Blyth, C.~O., Pinder, C.~N., Clarke, N.~M., Ward, R.~P., Hayes,
  P.~R., Pearce, K.~I., Watson, D.~L., Ghazarian, A., Cohler, M.~D., Thompson,
  I.~J., and Nagarajan, M.~A. (1997).
\newblock The $^{40}${C}a($t,p)^{42}${C}a reaction at triton energies near 10
  {MeV} per nucleon.
\newblock {\em Phys. Rev. C}, 56:1960.

\bibitem[Belyaev, 1959]{Belyaev:59}
Belyaev, S.~T. (1959).
\newblock Effect of pairing correlations on nuclear properties.
\newblock {\em Kgl. Danske Videnskab. Selskab, Mat.-fys. Medd.}, 31:No11.

\bibitem[Bengtsson and Frauendorf, 1979]{Bengtsson:79}
Bengtsson, R. and Frauendorf, S. (1979).
\newblock Quasiparticle spectra near the yrast line.
\newblock {\em Nucl. Phys. A}, 327(1).

\bibitem[Bengtsson and Schuck, 1980]{Bengtsson:80}
Bengtsson, R. and Schuck, P. (1980).
\newblock Semiclassical treatment of nuclear pairing properties.
\newblock {\em Phys. Lett. B}, 89:321.

\bibitem[Bernath et~al., 1993]{Bernath:93}
Bernath, M., Broglia, R.~A., Shimizu, Y.~R., and S\, J.~H. (1993).
\newblock Pairing phase transition and two-nucleon transfer reactions in
  rotating nuclei.
\newblock {\em Nucl. Phys. A}, 565(2).

\bibitem[Bertsch et~al., 1983]{Bertsch:83}
Bertsch, G.~F., Bortignon, P.~F., and Broglia, R.~A. (1983).
\newblock Damping of nuclear excitations.
\newblock {\em Rev. Mod. Phys.}, 55:287.

\bibitem[Bertsch and Esbensen, 1991]{Bertsch:91}
Bertsch, G.~F. and Esbensen, H. (1991).
\newblock Pair correlations near the neutron drip line.
\newblock {\em Ann. Phys.}, 209:327.

\bibitem[Bertsch et~al., 1998]{Bertsch:98}
Bertsch, G.~F., Hencken, K., and Esbensen, H. (1998).
\newblock Nuclear breakup of {B}orromean nuclei.
\newblock {\em Phys. Rev. C}, 57:1366.

\bibitem[B{\`{e}}s, 1963]{Bes:63}
B{\`{e}}s, D.~R. (1963).
\newblock Beta--vibrations in even nuclei.
\newblock {\em Nucl. Phys.}, 49:544.

\bibitem[B{\`{e}}s and Broglia, 1966]{Bes:66}
B{\`{e}}s, D.~R. and Broglia, R.~A. (1966).
\newblock Pairing vibrations.
\newblock {\em Nucl. Phys.}, 80:289.

\bibitem[B{\`{e}}s et~al., 1976]{Bes:76a}
B{\`{e}}s, D.~R., Broglia, R.~A., Dussel, G.~G., Liotta, R.~J., and
  Sof{\'{\i}}a, H.~M. (1976).
\newblock The nuclear field treatment of some exactly soluble models.
\newblock {\em Nucl. Phys. A}, 260:1.

\bibitem[B{\`{e}}s and Sorensen, 1969]{Bes:69}
B{\`{e}}s, D.~R. and Sorensen, R.~A. (1969).
\newblock The pairing--plus--quadrupole model.
\newblock {\em Adv. Nucl. Phys.}, 2:129.

\bibitem[Bjerregaard et~al., 1966]{Bjerregaard:66}
Bjerregaard, J.~H., Hansen, O., Nathan, O., and Hinds, S. (1966).
\newblock States of $^{208}${P}b from double triton stripping.
\newblock {\em Nucl. Phys.}, 89:337.

\bibitem[Bogner et~al., 2002]{Bogner:02}
Bogner, S., Kuo, T. T.~S., Coraggio, L., Covello, A., and Itaco, N. (2002).
\newblock Low momentum nucleon-nucleon potential and shell model effective
  interactions.
\newblock {\em Phys. Rev. C}, 65:051301(R).

\bibitem[Bohr, 1964]{Bohr:64}
Bohr, A. (1964).
\newblock In {\em {Comptes Rendus du Congr\`{e}s International de Physique
  Nucl\'{e}aire}}, volume~1, page 487. Centre National de la Recherche
  Scientifique.

\bibitem[Bohr and Mottelson, 1969]{Bohr:69}
Bohr, A. and Mottelson, B.~R. (1969).
\newblock {\em Nuclear Structure, Vol.I}.
\newblock Benjamin, New York.

\bibitem[Bohr and Mottelson, 1975]{Bohr:75}
Bohr, A. and Mottelson, B.~R. (1975).
\newblock {\em Nuclear Structure, Vol.II}.
\newblock Benjamin, New York.

\bibitem[Bohr et~al., 1958]{Bohr:58}
Bohr, A., Mottelson, B.~R., and Pines, D. (1958).
\newblock Possible analogy between the excitation spectra of nuclei and those
  of the superconducting metallic state.
\newblock {\em Phys. Rev.}, 110:936.

\bibitem[Bohr and Ulfbeck, 1988]{Bohr:89}
Bohr, A. and Ulfbeck, O. (1988).
\newblock Quantal structure of superconductivity, gauge angle.
\newblock {\em 1st Tops\o{}e Summer School, AEK Ris\o{}, Denmark
  (unpublished)}.

\bibitem[Bortignon et~al., 1998]{Bortignon:98}
Bortignon, P.~F., Bracco, A., and Broglia, R.~A. (1998).
\newblock {\em Giant Resonances}.
\newblock Harwood Academic Publishers, Amsterdam.

\bibitem[Bortignon et~al., 1977]{Bortignon:77}
Bortignon, P.~F., Broglia, R.~A., B{\`{e}}s, D.~R., and Liotta, R. (1977).
\newblock Nuclear field theory.
\newblock {\em Phys. Rep.}, 30:305.

\bibitem[Brink and Broglia, 2005]{Brink:05}
Brink, D. and Broglia, R.~A. (2005).
\newblock {\em Nuclear Superfluidity}.
\newblock Cambridge University Press, Cambridge.

\bibitem[Broglia, 1973]{Broglia:72a}
Broglia, R.~A. (1973).
\newblock The pairing model.
\newblock {\em Annals of Physics}, 80:60.

\bibitem[Broglia, 1986]{Broglia:86}
Broglia, R.~A. (1986).
\newblock Pairing phase transitions in nuclei.
\newblock In {\em International Summer School, Theory of Nuclear Structure and
  Reactions}, La Rabida, Huelva, Spain. World Scientific.
\newblock 133.

\bibitem[Broglia et~al., 2002]{Broglia:02b}
Broglia, R.~A., Barranco, F., Col\'{o}, G., Vigezzi, E., Bortignon, P.~F.,
  Gori, G., and Terasaki, J. (2002).
\newblock {Nuclear field theory description of the three-body system
  $^{11}$Li}.
\newblock In Norman, E., Schroeder, L., Wozniak, G., and Smith, A.~M., editors,
  {\em Proceedings of the Nuclear Physics Conference INPC 2001}, volume 610,
  page 746, Berkeley. California (USA). AIP.

\bibitem[Broglia et~al., 2004]{Broglia:04b}
Broglia, R.~A., Col{\'{o}}, G., Onida, G., and Roman, H.~E. (2004).
\newblock {\em Solid State Physics of Finite Systems: metal clusters,
  fullerenes, atomic wires}.
\newblock Springer Verlag, Berlin, Heidelberg.

\bibitem[Broglia et~al., 1973]{Broglia:73}
Broglia, R.~A., Hansen, O., and Riedel, C. (1973).
\newblock Two--neutron transfer reactions and the pairing model.
\newblock {\em Adv. Nucl. Phys.}, 6:287.

\bibitem[Broglia and Riedel, 1967]{Broglia:67}
Broglia, R.~A. and Riedel, C. (1967).
\newblock Pairing vibration and particle-hole states excited in the reaction
  $^{206}${P}b(t, p)$^{208}${P}b.
\newblock {\em Nucl. Phys.}, 92:145.

\bibitem[Broglia et~al., 2000]{Broglia:00}
Broglia, R.~A., Terasaki, J., and Giovanardi, N. (2000).
\newblock The {A}nderson--{G}oldstone--{N}ambu mode in finite and in infinite
  systems.
\newblock {\em Phys. Rep.}, 335:1.

\bibitem[Broglia and Winther, 2004]{Broglia:04a}
Broglia, R.~A. and Winther, A. (2004).
\newblock {\em Heavy Ion Reactions}.
\newblock Westview Press, Cambridge, MA.

\bibitem[Broglia and Winther, 2005]{Broglia:05c}
Broglia, R.~A. and Winther, A. (2005).
\newblock {\em Heavy Ion Reactions, 2nd ed.}
\newblock Westview Press, Perseus Books, Boulder.

\bibitem[Chabanat et~al., 1997]{Chabanat:97}
Chabanat, E., Bonche, P., Haensel, P., Meyer, J., and Schaeffer, R. (1997).
\newblock A {S}kyrme parametrization from subnuclear to neutron star densities.
\newblock {\em Nucl. Phys. A}, 627:710.

\bibitem[Charlton, 1976]{Charlton:76}
Charlton, L.~A. (1976).
\newblock Finite--range evaluation of {$(p-d$},{$d-t)$} with momentum space
  techniques.
\newblock {\em Phys. Rev. C}, 14:506.

\bibitem[Chatterjee et~al., 2008]{Chatterjee:08}
Chatterjee, A., Navin, A., Shrivastava, A., Bhattacharyya, S., Rejmund, M.,
  Keeley, N., Nanal, V., Nyberg, J., Pillay, R.~G., Ramachandran, K., Stefan,
  I., Bazin, D., Beaumel, D., Blumenfeld, Y., de~France, G., Gupta, D.,
  Labiche, M., Lemasson, A., Lemmon, R., Raabe, R., Scarpaci, J.~A., Simenel,
  C., and Timis, C. (2008).
\newblock 1n and 2n transfer with the borromean nucleus {$^6$}{H}e near the
  {C}oulomb barrier.
\newblock {\em Phys. Rev. Lett.}, 101:032701.

\bibitem[Cohen et~al., 1962]{Cohen:62}
Cohen, M.~H., Falicov, L.~M., and Phillips, J.~C. (1962).
\newblock Superconductive tunneling.
\newblock {\em Phys. Rev. Lett.}, 8:316.

\bibitem[Cooper, 1956]{Cooper:56}
Cooper, L.~N. (1956).
\newblock Bound electron pairs in a degenerate {F}ermi gas.
\newblock {\em Phys. Rev.}, 104:1189.

\bibitem[Cooper and Feldman, 2011]{Cooper:11}
Cooper, L.~N. and Feldman, D. (2011).
\newblock {\em BCS: 50 years}.
\newblock World Scientific, Singapore.

\bibitem[Coraggio et~al., 2004]{Coraggio:04}
Coraggio, L., Covello, A., Gargano, A., and Itaco, N. (2004).
\newblock Structure of particle-hole nuclei around $^{100}${S}n.
\newblock {\em Phys. Rev. C}, 70(3):034310.

\bibitem[Covello et~al., 1997]{Covello:97}
Covello, A., Andreozzi, F., Coraggio, L., Gargano, A., and Porrino, A. (1997).
\newblock Realistic shell-model calculations for sn isotopes.
\newblock In Pan, X.-W., Feng, D.~H., and Vallières, M., editors, {\em
  Contemporary Nuclear Shell Models}, volume 482 of {\em Lecture Notes in
  Physics}, pages 122--131. Springer Berlin / Heidelberg.

\bibitem[Duguet et~al., 2010]{Duguet:10}
Duguet, T., Lesinski, T., Hebeler, K., and Schwenk, A. (2010).
\newblock Lowest-order contribution of chiral three-nucleon interactions to
  pairing properties of nuclear ground states.
\newblock {\em Mod. Phys. Lett. A}, 25:1989.

\bibitem[Esaki, 1974]{Esaki:73}
Esaki, L. (1974).
\newblock Long journey into tunneling.
\newblock In {\em Le Prix Nobel en 1973}, page~64. PA Norstedt, and
  S{\"{o}}ner, Stockholm.

\bibitem[Esbensen et~al., 1997]{Esbensen:97}
Esbensen, H., Bertsch, G.~F., and Hencken, K. (1997).
\newblock {Application of contact interactions to Borromean halos}.
\newblock {\em Phys. Rev. C}, 56:3054.

\bibitem[Farine and Schuck, 1999]{Farine:99}
Farine, M. and Schuck, P. (1999).
\newblock Surface properties of nuclear pairing with the {G}ogny force in a
  simplified model.
\newblock {\em Phys. Lett. B}, 459:444.

\bibitem[Farine and Schuck, 2002]{Farine:02}
Farine, M. and Schuck, P. (2002).
\newblock Size dependence of pairing in finite fermi systems.
\newblock In Nazarewicz, W. and Vretenar, D., editors, {\em The Nuclear
  Many--Body Problem}, page 189, Amsterdam. Kluwer.

\bibitem[Flynn et~al., 1972]{Flynn:72}
Flynn, E.~R., Igo, G.~J., and Broglia, R.~A. (1972).
\newblock Three-phonon monopole and quadrupole pairing vibrational states in
  $^{206}${P}b.
\newblock {\em Phys. Lett. B}, 41:397.

\bibitem[Giaever, 1974]{Giaever:73}
Giaever, I. (1974).
\newblock Electron tunneling and superconductivity.
\newblock In {\em Le Prix Nobel en 1973}, page~84. PA Norstedt and S{\"{o}}ner,
  Stockholm.

\bibitem[Glendenning, 1963]{Glendenning:63}
Glendenning, N.~K. (1963).
\newblock {Nuclear stripping reactions}.
\newblock {\em Annu. Rev. Nucl. Sci.}, 13:191.

\bibitem[Glendenning, 1965]{Glendenning:65}
Glendenning, N.~K. (1965).
\newblock {Nuclear Spectroscopy with Two--Nucleon Transfer Reactions}.
\newblock {\em Phys. Rev.}, 137:B102.

\bibitem[Glendenning, 1968]{Glendenning:68}
Glendenning, N.~K. (1968).
\newblock {\em Tables of structure Amplitudes for $(p,t)$ Reaction}.
\newblock Lawrence Radiation Laboratory Report.

\bibitem[Golovkov et~al., 2009]{Golovkov:08}
Golovkov, M.~S., Grigorenko, L.~V., Ter-Akopian, G.~M., Fomichev, A.~S.,
  Oganessian, Y.~T., Gorshkov, V.~A., Krupko, S.~A., Rodin, A.~M., Sidorchuk,
  S.~I., Slepnev, R.~S., Stepantsov, S.~V., Wolski, R., Pang, D.~Y., Chudoba,
  V., Korsheninnikov, A.~A., Kuzmin, E.~A., Nikolskii, E.~Y., Novatskii, B.~G.,
  Stepanov, D.~N., Roussel-Chomaz, P., Mittig, W., Ninane, A., Hanappe, F.,
  Stuttge, L., Yukhimchuk, A.~A., Perevozchikov, V.~V., Vinogradov, Y.~I.,
  Grishechkin, S.~K., and Zlatoustovskiy, S.~V. (2009).
\newblock The $^{8}${H}e and $^{10}${H}e spectra studied in the $(t,p)$
  reaction.
\newblock {\em Phys. Lett. B}, 672:22.

\bibitem[Gori et~al., 2004]{Gori:04}
Gori, G., Barranco, F., Vigezzi, E., and Broglia, R.~A. (2004).
\newblock {Parity inversion and breakdown of shell closure in Be isotopes}.
\newblock {\em Phys. Rev. C}, 69:041302.

\bibitem[Guazzoni et~al., 1999]{Guazzoni:99}
Guazzoni, P., Jaskola, M., Zetta, L., Covello, A., Gargano, A., Eisermann, Y.,
  Graw, G., Hertenberger, R., Metz, A., Nuoffer, F., and Staudt, G. (1999).
\newblock {Level structure of $^{120}$Sn: High resolution ($p,t$) reaction and
  shell model description}.
\newblock {\em Phys. Rev. C}, 60:054603.

\bibitem[Guazzoni et~al., 2006]{Guazzoni:06}
Guazzoni, P., Zetta, L., Covello, A., Gargano, A., Bayman, B.~F., Graw, G.,
  Hertenberger, R., Wirth, H.-F., and Jaskola, M. (2006).
\newblock Spectroscopy of {$^{110}$}{S}n via the high-resolution
  {$^{112}$}{S}n{$(p,t)$} {$^{110}$}{S}n reaction.
\newblock {\em Phys. Rev. C}, 74:054605.

\bibitem[Gunnarsson, 1997]{Gunnarsson:97}
Gunnarsson, O. (1997).
\newblock Superconductivity in fullerides.
\newblock {\em Rev. Mod. Phys.}, 69:575.

\bibitem[Gunnarsson, 2004]{Gunnarsson:04}
Gunnarsson, O. (2004).
\newblock {\em Alkali-doped Fullerides: Narrow-band Solids with Unusual
  Properties}.
\newblock World Scientific, Singapore.

\bibitem[Hansen, 1996]{Hansen:96}
Hansen, P.~G. (1996).
\newblock Attack on a convoy of nucleons.
\newblock {\em Nature}, 384:415.

\bibitem[Hansen and Tostevin, 2003]{Hansen:03}
Hansen, P.~G. and Tostevin, J.~A. (2003).
\newblock Direct reactions with exotic nuclei.
\newblock {\em Annu. Rev. Nucl. Part. Sci.}, 53:219.

\bibitem[Hashimoto and Kawai, 1978]{Hashimoto:78}
Hashimoto, N. and Kawai, M. (1978).
\newblock The {$(p,d)$} {$(d,t)$} process in strong {$(p,t)$} reactions.
\newblock {\em Prog. Theor. Phys.}, 59:1245.

\bibitem[Hebeler et~al., 2009]{Hebeler:09}
Hebeler, K., Duguet, T., Lesinski, T., and Schwenk, A. (2009).
\newblock Non-empirical pairing energy functional in nuclear matter and finite
  nuclei.
\newblock {\em Phys. Rev. C}, 80:044321.

\bibitem[Heyde, 1997]{Heyde:98}
Heyde, K. (1997).
\newblock {\em {From Nucleons to the Atomic Nucleus}}.
\newblock Springer, Heidelberg.

\bibitem[Igarashi et~al., 1991]{Igarashi:91}
Igarashi, M., Kubo, K., and Yagi, K. (1991).
\newblock Two--nucleon transfer mechanisms.
\newblock {\em Physics Reports}, 199:1.

\bibitem[Josephson, 1969]{Josephson:69}
Josephson, B. (1969).
\newblock Weakly coupled superconductors.
\newblock In Parks, R.~D., editor, {\em Superconductivity}, volume~1, page 423,
  New York. Marcel Dekker, Inc.

\bibitem[Josephson, 1974]{Josephson:73}
Josephson, B. (1974).
\newblock The discovery of tunneling supercurrents.
\newblock In {\em Le Prix Nobel en 1973}, page 104. PA Norstedt and
  S{\"{o}}ner, Stockholm.

\bibitem[Josephson, 1962]{Josephson:62}
Josephson, B.~D. (1962).
\newblock Possible new effects in superconductive tunnelling.
\newblock {\em Phys. Lett.}, 1:251.

\bibitem[Keeley et~al., 2007a]{Keeley:07b}
Keeley, N., Raabe, R., Alamanos, N., and Sida, J. (2007a).
\newblock Fusion and direct reactions of halo nuclei at energies around the
  {C}oulomb barrier.
\newblock {\em Prog. Part. Nucl. Phys.}, 59:579.

\bibitem[Keeley et~al., 2007b]{Keeley:07a}
Keeley, N., Skaza, F., Lapoux, V., Alamanos, N., Auger, F., Beaumel, D.,
  Becheva, E., Blumenfeld, Y., Delaunay, F., Drouart, A., Gillibert, A., Giot,
  L., Kemper, K.~W., Nalpas, L., Pakou, A., Pollacco, E.~C., Raabe, R.,
  Roussel-Chomaz, P., Rusek, K., Scarpaci, J.-A., Sida, J.-L., Stepantsov, S.,
  and Wolski, R. (2007b).
\newblock Probing the {$^8$}{H}e ground state via the
  {$^8$}{H}e{$(p,t)$}{$^6$}{H}e reaction.
\newblock {\em Phys. Lett. B}, 646:222.

\bibitem[Khan et~al., 2004]{Khan:04}
Khan, E., Sandulescu, N., Giai, N.~V., and Grasso, M. (2004).
\newblock Two-neutron transfer in nuclei close to the drip line.
\newblock {\em Phys. Rev. C}, 69:014314.

\bibitem[Khoa and von Oertzen, 2004]{Khoa:04}
Khoa, D.~T. and von Oertzen, W. (2004).
\newblock Di-neutron elastic transfer in the
  $^{4}${H}e($^{6}${H}e,$^{6}${H}e)$^{4}${H}e reaction.
\newblock {\em Phys. Lett. B}, 595:193.

\bibitem[Kisslinger and Sorensen, 1963]{Kisslinger:63}
Kisslinger, L.~S. and Sorensen, R.~A. (1963).
\newblock Spherical nuclei with simple residual forces.
\newblock {\em Rev. Mod. Phys.}, 35:853.

\bibitem[Kobayashi, 1993]{Kobayashi:93}
Kobayashi, T. (1993).
\newblock {Nuclear structure experiments on $^{11}$Li}.
\newblock {\em Nucl. Phys. A}, 553:465.

\bibitem[Kobayashi et~al., 1989]{Kobayashi:89}
Kobayashi, T., Shimoura, S., Tanihata, I., Katori, K., Matsuta, K., Minamisono,
  T., Sugimoto, K., Müller, W., Olson, D.~L., Symons, T. J.~M., and Wieman, H.
  (1989).
\newblock Electromagnetic dissociation and soft giant dipole resonance of the
  neutron-dripline nucleus {$^{11}$Li}.
\newblock {\em Phys. Lett. B}, 232:51.

\bibitem[Kucharek et~al., 1989]{Kucharek:89}
Kucharek, H., Ring, P., Schuck, P., Bengtsson, R., and Girod, M. (1989).
\newblock Pairing properties of nuclear matter from the {G}ogny force.
\newblock {\em Phys. Lett. B}, 216:249.

\bibitem[Kuo et~al., 1971]{Kuo:71}
Kuo, T. T.~S., Lee, S.~Y., and Ratcliff, K.~F. (1971).
\newblock A folded-diagram expansion of the model-space effective hamiltonian.
\newblock {\em Nucl. Phys. A}, 176:65.

\bibitem[Landau and Lifshitz, 1981]{Landau:81}
Landau, L. and Lifshitz, L. (1981).
\newblock {\em Quantum Mechanics, 3rd ed.}
\newblock Butterworth-Heinemann.

\bibitem[Lauritzen et~al., 1993]{Lauritzen:93}
Lauritzen, B., Anselmino, A., Bortignon, P.~F., and Broglia, R.~A. (1993).
\newblock Pairing phase transition in small particles.
\newblock {\em Ann. Phys.}, 223:216.

\bibitem[Lenske and Schrieder, 1998]{Lenske:98}
Lenske, H. and Schrieder, G. (1998).
\newblock Probing the structure of exotic nuclei by transfer reactions.
\newblock {\em Eur. Phys. J. A}, 2:41.

\bibitem[Lilley, 1983]{Lilley:83}
Lilley, J.~S. (1983).
\newblock $^{208}${P}b(${}^{16}${O},${}^{18}${O})${}^{206}${P}b reaction.
\newblock private communication.

\bibitem[Machleidt, 2001]{Machleidt:01}
Machleidt, R. (2001).
\newblock High-precision, charge-dependent {B}onn nucleon-nucleon potential.
\newblock {\em Phys. Rev. C}, 63(2):024001.

\bibitem[Maglione et~al., 1985]{Maglione:85}
Maglione, E., Pollarolo, G., Vitturi, A., Broglia, R.~A., and Winther, A.
  (1985).
\newblock Absolute cross sections of two--nucleon transfer reactions induced by
  heavy ions.
\newblock {\em Phys. Lett. B}, 162:59.

\bibitem[Mahaux et~al., 1985]{Mahaux:85}
Mahaux, C., Bortignon, P.~F., Broglia, R.~A., and Dasso, C.~H. (1985).
\newblock Dynamics of the shell model.
\newblock {\em Phys. Rep.}, 120:1.

\bibitem[Matsuo and Serizawa, 2010]{Matsuo:10}
Matsuo, M. and Serizawa, Y. (2010).
\newblock Surface-enhanced pair transfer amplitude in quadrupole states of
  neutron-rich {S}n isotopes.
\newblock {\em Phys. Rev. C}, 82:024318.

\bibitem[Mc~Millan and Rowell, 1969]{McMillan:69}
Mc~Millan, W.~L. and Rowell, J.~M. (1969).
\newblock Tunneling and strong--coupling superconductivity.
\newblock In Parks, R.~D., editor, {\em Superconductivity}, volume~1, page 561,
  New York. Marcel Dekker, Inc.

\bibitem[Mottelson, 1976]{Mottelson:76}
Mottelson, B.~R. (1976).
\newblock {\em Elementary Modes of Excitation in Nuclei, Le Prix Nobel en
  1975}.
\newblock Imprimerie Royale Norstedts Tryckeri, Stockholm.
\newblock p. 80.

\bibitem[Nakamura et~al., 2006]{Nakamura:06}
Nakamura, T., Vinodkumar, A.~M., Sugimoto, T., Aoi, N., Baba, H., Bazin, D.,
  Fukuda, N., Gomi, T., Hasegawa, H., Imai, N., Ishihara, M., Kobayashi, T.,
  Kondo, Y., Kubo, T., Miura, M., Motobayashi, T., Otsu, H., Saito, A.,
  Sakurai, H., Shimoura, S., Watanabe, K., Watanabe, Y.~X., Yakushiji, T.,
  Yanagisawa, Y., and Yoneda, K. (2006).
\newblock {Observation of strong low-lying E1 strength in the two-neutron halo
  nucleus $^{11}$Li}.
\newblock {\em Phys. Rev. Lett.}, 96:252502.

\bibitem[Nambu, 1995]{Nambu:91}
Nambu, Y. (1995).
\newblock Dynamical symmetry breaking.
\newblock In Eguchi, T. and Nishijima, K., editors, {\em Broken Symmetry,
  Selected papers of Y. Nambu}, page 436. World Scientific, Singapore.

\bibitem[Nikam and Ring, 1987]{Nikam:87a}
Nikam, R.~S. and Ring, P. (1987).
\newblock Manifestation of the berry phase in diabolic pair transfer in
  rotating nuclei.
\newblock {\em Phys. Rev. Lett.}, 58:980.

\bibitem[Palstra et~al., 1995]{Palstra:95}
Palstra, T. T.~M., Zhou, O., Iwasa, Y., Sulewski, P.~E., Fleming, R.~M., and
  Zegarski, B.~R. (1995).
\newblock Superconductivity at 40{K} in {C}esium doped {C}$_{60}$.
\newblock {\em Solid State Comm.}, 93:327.

\bibitem[Pastore et~al., 2008]{Pastore:08}
Pastore, A., Barranco, F., Broglia, R.~A., and Vigezzi, E. (2008).
\newblock Microscopic calculation and local approximation of the spatial
  dependence of the pairing field with bare and induced interactions.
\newblock {\em Phys. Rev. C}, 78:024315.

\bibitem[Perenboom et~al., 1981]{Perenboom:81}
Perenboom, J. A. A.~J., Wyder, P., and Meier, F. (1981).
\newblock Electronic properties of small metallic particles.
\newblock {\em Phys. Rep.}, 78:173.

\bibitem[Pethick and Smith, 2009]{Pethick:09}
Pethick, C.~J. and Smith, H. (2009).
\newblock {\em Bose--Einstein Condensation in Dilute Gases}.
\newblock Cambridge University Press, Cambridge.

\bibitem[Potel et~al., 2010]{Potel:10}
Potel, G., Barranco, F., Vigezzi, E., and Broglia, R.~A. (2010).
\newblock {Evidence for phonon mediated pairing interaction in the halo of the
  nucleus $^{11}$Li}.
\newblock {\em Phy. Rev. Lett.}, 105:172502.

\bibitem[Raman et~al., 1987]{Raman:87}
Raman, S., Malarkey, C.~H., Milner, W.~T., Nestor, C.~W., and Stelson, P.~H.
  (1987).
\newblock {Transition probability, $B(E2)\uparrow$, from the ground to the
  first-excited $2^+$ state of even-even nuclides}.
\newblock {\em Atomic Data and Nuclear Data Tables}, 36:1.

\bibitem[Richter, 1993]{Richter:93}
Richter, A. (1993).
\newblock Trends in nuclear physics.
\newblock {\em Nucl. Phys. A}, 553:417c.

\bibitem[Ring and Schuck, 1980]{Ring:80}
Ring, P. and Schuck, P. (1980).
\newblock {\em The {N}uclear {M}any--{B}ody {P}roblem}.
\newblock Springer, Berlin.

\bibitem[Rodriguez-Gallardo et~al., 2008]{Rodriguez:09}
Rodriguez-Gallardo, M., Arias, J.~M., Gomez-Camacho, J., C., J.~R., Moro,
  A.~M., J, T.~I., and Tostevin, J.~A. (2008).
\newblock Four-body continuum-discretized coupled channels calculations using a
  transformed harmonic oscillator basis.
\newblock {\em Phys. Rev. C}, 77:064609.

\bibitem[Sackett et~al., 1993]{Sackett:93}
Sackett, D., Ieki, K., Galonsky, A., Bertulani, C.~A., Esbensen, H., Kruse,
  J.~J., Lynch, W.~G., Morrissey, D.~J., Orr, N.~A., Sherrill, B.~M., Schulz,
  H., Sustich, A., Winger, J.~A., De\'ak, F., Horv\'ath, A., Kiss, A., Seres,
  Z., Kolata, J.~J., Warner, R.~E., and Humphrey, D.~L. (1993).
\newblock {Electromagnetic excitation of $^{11}${L}i}.
\newblock {\em Phys. Rev. C}, 48:118.

\bibitem[Sagawa et~al., 1993]{Sagawa:93}
Sagawa, H., Brown, B.~A., and Esbensen, H. (1993).
\newblock {Parity inversion in the N=7 isotones and the pairing blocking
  effect}.
\newblock {\em Phys. Lett. B}, 309:1.

\bibitem[Satchler, 1980]{Satchler:80}
Satchler, G.~R. (1980).
\newblock {\em Introduction to Nuclear Reactions}.
\newblock Mc Millan, New York.

\bibitem[Scalapino, 1969]{Scalapino:69}
Scalapino, D.~J. (1969).
\newblock The electron--phonon interaction and strong coupling superconductors.
\newblock In Parks, R.~D., editor, {\em Superconductivity}, volume~1, page 449,
  New York. Marcel Dekker, Inc.

\bibitem[Schrieffer, 1964]{Schrieffer:64}
Schrieffer, J. (1964).
\newblock {\em Superconductivity}.
\newblock Benjamin, New York.

\bibitem[Shimizu et~al., 1989]{Shimizu:89}
Shimizu, Y.~R., Garrett, J.~D., Broglia, R.~A., Gallardo, M., and Vigezzi, E.
  (1989).
\newblock Pairing fluctuations in rapidly rotating nuclei.
\newblock {\em Rev. of Mod. Phys.}, 61:131.

\bibitem[Shulgina et~al., 2009]{Shulgina:09}
Shulgina, N., Jonson, B., and Zhukov, M.~V. (2009).
\newblock $^{11}$li structure from experimental data.
\newblock {\em Nucl. Phys. A}, 825:175.

\bibitem[Simon et~al., 1999]{Simon:99}
Simon, H., Aleksandrov, D., Aumann, T., Axelsson, L., Baumann, T., Borge, M.
  J.~G., Chulkov, L.~V., Collatz, R., Cub, J., Dostal, W., Eberlein, B., Elze,
  T.~W., Emling, H., Geissel, H., Gr\"unschloss, A., Hellstr\"om, M., Holeczek,
  J., Holzmann, R., Jonson, B., Kratz, J.~V., Kraus, G., Kulessa, R., Leifels,
  Y., Leistenschneider, A., Leth, T., Mukha, I., M\"unzenberg, G., Nickel, F.,
  Nilsson, T., Nyman, G., Petersen, B., Pf\"utzner, M., Richter, A., Riisager,
  K., Scheidenberger, C., Schrieder, G., Schwab, W., Smedberg, M.~H., Stroth,
  J., Surowiec, A., Tengblad, O., and Zhukov, M.~V. (1999).
\newblock {Direct Experimental Evidence for Strong Admixture of Different
  Parity States in $ ^{11}${L}i}.
\newblock {\em Phys. Rev. Lett.}, 83:496.

\bibitem[Smith et~al., 2008]{Smith:08}
Smith, M., Brodeur, M., Brunner, T., Ettenauer, S., Lapierre, A., Ringle, R.,
  Ryjkov, V.~L., Ames, F., Bricault, P., Drake, G. W.~F., Delheij, P., Lunney,
  D., Sarazin, F., and Dilling, J. (2008).
\newblock First {P}enning-trap mass measurement of the exotic halo nucleus
  {$^{11}$}{Li}.
\newblock {\em Phys. Rev. Lett.}, 101:202501.

\bibitem[Soloviev, 1965]{Soloviev:65}
Soloviev, V.~G. (1965).
\newblock Quasi--particle and collective structure of the states of even,
  strongly-deformed nuclei.
\newblock {\em Nucl. Phys.}, 69:1.

\bibitem[Suzuki and Lee, 1980]{Suzuki:80}
Suzuki, K. and Lee, S.~Y. (1980).
\newblock Convergent theory for effective interaction in nuclei.
\newblock {\em Prog. Theor. Phys.}, 64:2091.

\bibitem[Takemasa et~al., 1979]{Takemasa:79}
Takemasa, T., Tamura, T., and Udagawa, T. (1979).
\newblock Exact finite range calculations of light--ion induced two--neutron
  transfer reactions.
\newblock {\em Nucl. Phys. A}, 321:269.

\bibitem[Tamura et~al., 1970]{Tamura:70}
Tamura, T., B\`{e}s, D.~R., Broglia, R.~A., and Landowne, S. (1970).
\newblock Coupled-channel {B}orn-{A}pproximation calculation of two-nucleon
  transfer reactions in deformed nuclei.
\newblock {\em Phys. Rev. Lett.}, 25:1507.

\bibitem[Tang and Herndon, 1965]{Tang:65}
Tang, Y.~C. and Herndon, R.~C. (1965).
\newblock Form factors of {$^{3}$}{H} and {$^{4}$}{H}e with repulsive--core
  potentials.
\newblock {\em Phys. Lett.}, 18:42.

\bibitem[Tanihata, 1996]{Tanihata:96}
Tanihata, I. (1996).
\newblock {Neutron halo nuclei}.
\newblock {\em J. Phys. G}, 22:157.

\bibitem[Tanihata et~al., 2008]{Tanihata:08}
Tanihata, I., Alcorta, M., Bandyopadhyay, D., Bieri, R., Buchmann, L., Davids,
  B., Galinski, N., Howell, D., Mills, W., Mythili, S., Openshaw, R.,
  Padilla-Rodal, E., Ruprecht, G., Sheffer, G., Shotter, A.~C., Trinczek, M.,
  Walden, P., Savajols, H., Roger, T., Caamano, M., Mittig, W., Roussel-Chomaz,
  P., Kanungo, R., Gallant, A., Notani, M., Savard, G., and Thompson, I.~J.
  (2008).
\newblock Measurement of the two-halo neutron transfer reaction
  {$^1$H($^{11}$Li,$^{9}$Li)$^3$H} at {3A} {MeV}.
\newblock {\em Phys. Rev. Lett.}, 100:192502.

\bibitem[Terasaki et~al., 2002a]{Terasaki:02a}
Terasaki, J., Barranco, F., Broglia, R.~A., Vigezzi, E., and Bortignon, P.~F.
  (2002a).
\newblock Solution of the {D}yson equation for nucleons in the superfluid
  phase.
\newblock {\em Nucl. Phys. A}, 697:127.

\bibitem[Terasaki et~al., 2002b]{Terasaki:02b}
Terasaki, J., Barranco, F., Vigezzi, E., Broglia, R.~A., and Bortignon, P.~F.
  (2002b).
\newblock Effect of particle-phonon coupling on pairing correlations in finite
  systems -- the atomic nucleus --.
\newblock {\em Progr. Theor. Phys.}, 108:495.

\bibitem[Tostevin, 2007]{Tostevin:07}
Tostevin, J. (2007).
\newblock Nuclear reactions used to probe the structure of nuclei far from
  stability.
\newblock {\em Acta Phys. Pol. B}, 38:1195.

\bibitem[Tostevin et~al., 2004]{Tostevin:04}
Tostevin, J.~A., Podoly{{\'{a}}}k, G.~Brown, B.~A., and Hansen, P.~G. (2004).
\newblock Correlated two-nucleon stripping reactions.
\newblock {\em Phys. Rev. C}, 70:064602.

\bibitem[Van~der Sluys et~al., 1993]{VanderSluys:93}
Van~der Sluys, V., Van~Neck, D., Waroquier, M., and Ryckebusch, J. (1993).
\newblock Fragmentation of single-particle strength in spherical open-shell
  nuclei: Application to the spectral functions in $^{142}${N}d.
\newblock {\em Nucl. Phys. A}, 551:210.

\bibitem[Vinh~Mau, 1995]{Vinh:95}
Vinh~Mau, N. (1995).
\newblock Particle-vibration coupling in one neutron halo nuclei.
\newblock {\em Nucl. Phys. A}, 592:33.

\bibitem[von Oertzen et~al., 1983]{vonOertzen:83b}
von Oertzen, W., Brown, R.~E., Flynn, E.~R., Peng, J.~C., and Sunier, J.~W.
  (1983).
\newblock Pairing enhancement of two-neutron transfer in
  ($^{14}${C},$^{12}${C}) reactions.
\newblock {\em Z. Phys. A}, 313:371.

\bibitem[Wiringa et~al., 1984]{Wiringa:84}
Wiringa, R.~B., Smith, R.~A., and Ainsworth, T.~L. (1984).
\newblock {Nucleon-nucleon potentials with and without $\Delta{}(1232)$ degrees
  of freedom}.
\newblock {\em Phys. Rev. C}, 29:1207.

\bibitem[Yoshida, 1962]{Yoshida:62}
Yoshida, S. (1962).
\newblock Note on the two-nucleon stripping reaction.
\newblock {\em Nucl. Phys.}, 33:685.

\bibitem[Zinser et~al., 1995]{Zinser:95}
Zinser, M., Humbert, F., Nilsson, T., Schwab, W., Blaich, T., Borge, M. J.~G.,
  Chulkov, L.~V., Eickhoff, H., Elze, T.~W., Emling, H., Franzke, B.,
  Freiesleben, H., Geissel, H., Grimm, K., Guillemaud-Mueller, D., Hansen,
  P.~G., Holzmann, R., Irnich, H., Jonson, B., Keller, J.~G., Klepper, O.,
  Klingler, H., Kratz, J.~V., Kulessa, R., Lambrecht, D., Leifels, Y., Magel,
  A., Mohar, M., Mueller, A.~C., M\"unzenberg, G., Nickel, F., Nyman, G.,
  Richter, A., Riisager, K., Scheidenberger, C., Schrieder, G., Sherill, B.,
  Simon, H., Stelzer, K., Stroth, J., Tengblad, O., Trautmann, W., Wajda, E.,
  and Zude, E. (1995).
\newblock Study of the unstable nucleus $^{10}${Li} in stripping reactions of
  the radioactive projectiles $^{11}${Be} and $^{11}${Li}.
\newblock {\em Phys. Rev. Lett.}, 75:1719.

\bibitem[Zinser et~al., 1997]{Zinser:97}
Zinser, M., Humbert, F., Nilsson, T., Schwab, W., Simon, H., Aumann, T., Borge,
  M. J.~G., Chulkov, L.~V., Cub, J., Elze, T.~W., Emling, H., Geissel, H.,
  Guillemaud-Mueller, D., Hansen, P.~G., Holzmann, R., Irnich, H., Jonson, B.,
  Kratz, J.~V., Kulessa, R., Leifels, Y., Lenske, H., Magel, A., Mueller,
  A.~C., Münzenberg, G., Nickel, F., Nyman, G., Richter, A., Riisager, K.,
  Scheidenberger, C., Schrieder, G., Stelzer, K., Stroth, J., Surowiec, A.,
  Tengblad, O., Wajda, E., and Zude, E. (1997).
\newblock {Invariant-mass spectroscopy of $^{10}$Li and $^{11}$Li}.
\newblock {\em Nucl. Phys. A}, 619:151.

\end{thebibliography}

\end{document}